\documentclass[12pt]{article}
\usepackage{cite,setspace}
\usepackage{amsmath, amsthm, amssymb,slashed, bbm, dsfont,relsize}

\usepackage{ifpdf}
\ifpdf
  \usepackage[pdftex]{graphicx}
  \usepackage{epstopdf}
\else
  \usepackage[dvips]{graphicx}
\fi
\parskip=\baselineskip
\def\Bbb{\mathbb}
\def\Tr{{\rm Tr}}
\def\TT{{\sf T}}
\def\sH{{\eusm H}}
\def\veps{\varepsilon}
\def\st{{\mathrm{st}}}
\DeclareMathAlphabet{\pxitfont}{OML}{pxmi}{m}{it}
\DeclareMathAlphabet{\pxitfontn}{U}{pxmia}{m}{it}
\def\Qb{\pxitfont Q}
\def\I{{\mathcal I}}
\def\I{{\mathcal I}}
\def \bA {\mathlarger{\mathlarger{a}}}
\def \bB {\mathlarger{\mathlarger{b}}}

\def \sA {\hat{\mathcal A}}
\def \sB {\hat{\mathcal B}}
\def \sC {\hat{\mathcal C}}
\def \sM {\hat{\mathcal M}}
\def \sN {\hat{\mathcal N}}
\def \sP {\hat{\mathcal P}}
\def \bA {\mathlarger{\mathlarger{a}}}
\def \bB {\mathlarger{\mathlarger{b}}}
\def \lalpha {\mathlarger{\mathlarger{\mathlarger{\alpha}}}}
\def \lbeta {\mathlarger{\mathlarger{\mathlarger{\beta}}}}

\font\teneurm=eurm10 \font\seveneurm=eurm7 \font\eighteurm=eurm8 \font\fiveeurm=eurm5
\newfam\eurmfam
\textfont\eurmfam=\teneurm \scriptfont\eurmfam=\seveneurm
\scriptscriptfont\eurmfam=\fiveeurm
\def\eurm#1{{\fam\eurmfam\relax#1}}
\font\teneusm=eusm10 \font\seveneusm=eusm7 \font\fiveeusm=eusm5
\newfam\eusmfam
\textfont\eusmfam=\teneusm \scriptfont\eusmfam=\seveneusm
\scriptscriptfont\eusmfam=\fiveeusm
\def\eusm#1{{\fam\eusmfam\relax#1}}
\font\tencmmib=cmmib10 \skewchar\tencmmib='177
\font\sevencmmib=cmmib7 \skewchar\sevencmmib='177
\font\fivecmmib=cmmib5 \skewchar\fivecmmib='177
\newfam\cmmibfam
\textfont\cmmibfam=\tencmmib \scriptfont\cmmibfam=\sevencmmib
\scriptscriptfont\cmmibfam=\fivecmmib
\def\cmmib#1{{\fam\cmmibfam\relax#1}}
\def\H{{\mathcal H}}
\def\L{{\mathcal L}}
\def\g{\text{{\teneurm g}}}
\def\hH{H'}

\def\n{\text{{\teneurm n}}}
\def\sn{\text{{\eighteurm n}}}

\def\h{\text{{\teneurm h}}}
\def\Wp{{\mathcal W}_\parallel}
\def\sg{\text{{\eighteurm g}}}
\def\sh{\text{{\eighteurm h}}}
\def\ssg{\text{{\seveneurm g}}}
\def\5d{{\mathrm{5d}}}
\def\4d{{\mathrm{4d}}}
\def\neg{\negthinspace}
\def\O{{\mathcal O}}
\def\16{{\bf 16}}
\def\parz{\parallel}
\def\1{{\bf 1}}
\def\2{{\bf 2}}
\def\3{{\bf 3}}
\def\4{{\bf 4}}
\def\Mth{{\mathrm M}}
\def\U{{\mathcal U}}
\def\M{{\mathcal M}}
\def\st{{\mathrm{st}}}
\def\d{{\mathrm d}}
\def\J{{\mathcal J}}
\def\eJ{{\sf J}}
\def\JJ{{\eusm J}}
\def\eS{{\cmmib U}}
\def\Spin{{\mathrm{Spin}}}
\def\t{\widetilde}
\def\bar{\overline}
\def\tilde{\widetilde}
\def\SIgma{\Sigma}
\def\R{{\Bbb{R}}}\def\Z{{\Bbb{Z}}}
\def\eR{{\cmmib R}}
\def\cZ{{\eusm Z}}
\def\WW{{\sf W}}
\def\eV{{\sf V}}
\def\cQ{{\cmmib Q}}
\def\hsH{{\hat{\eusm H}}}
\def\RR{{\mathcal R}}
\def\ZZ{{\mathcal Z}}
\def\N{{\mathcal N}}
\def\hat{\widehat}
\def\D{{\mathcal D}}
\font\teneurm=eurm10 \font\seveneurm=eurm7 \font\fiveeurm=eurm5
\newfam\eurmfam
\textfont\eurmfam=\teneurm \scriptfont\eurmfam=\seveneurm
\scriptscriptfont\eurmfam=\fiveeurm
\def\eurm#1{{\fam\eurmfam\relax#1}}
\font\teneusm=eusm10 \font\seveneusm=eusm7 \font\fiveeusm=eusm5
\newfam\eusmfam
\textfont\eusmfam=\teneusm \scriptfont\eusmfam=\seveneusm
\scriptscriptfont\eusmfam=\fiveeusm
\def\eusm#1{{\fam\eusmfam\relax#1}}
\font\tencmmib=cmmib10 \skewchar\tencmmib='177
\font\sevencmmib=cmmib7 \skewchar\sevencmmib='177
\font\fivecmmib=cmmib5 \skewchar\fivecmmib='177
\newfam\cmmibfam
\textfont\cmmibfam=\tencmmib \scriptfont\cmmibfam=\sevencmmib
\scriptscriptfont\cmmibfam=\fivecmmib
\def\cmmib#1{{\fam\cmmibfam\relax#1}}
\numberwithin{equation}{section}
\def\neg{\negthinspace}
\def\d{\mathrm d}
\def\bar{\overline}
\def\C{{\Bbb C}}
\def\Z{{\Bbb Z}}

\def\F{{\mathcal F}}
\def\ad{{\mathrm{ad}}}
\def\W{{\mathcal W}}
\def\dA{\dot A }
\def\dB{\dot B }
\def\bar{\overline}
\begin{document}
\begin{titlepage}

\vskip 1.5in

\begin{center}
{\bf\Large{Some Details On The Gopakumar-Vafa and \vskip.05cm Ooguri-Vafa Formulas}}
\vskip0.5cm 
{Mykola Dedushenko$^1$ and Edward Witten$^2$} 
\vskip.5cm
 {\small{\textit{$^1$Joseph Henry Laboratories, Princeton University, Princeton NJ USA 08540}}}
 \vskip.2cm
 {\small{\textit{$^2$School of Natural Sciences, Institute for Advanced Study, Princeton NJ USA 08540 \\}}}
 {\small{\textit{and Department of Physics, University of Washington, Seattle WA 98195}}}
\end{center}
\vskip.5cm
\baselineskip 16pt
\begin{abstract}
The Gopakumar-Vafa (GV) formula expresses certain couplings that arise in Type IIA compactification to 
four dimensions on a Calabi-Yau
manifold in terms of a counting of BPS states in M-theory.  The couplings in question have applications 
to topological strings and supersymmetric
black holes.  In this paper, we reconsider the GV formula, taking a close look at the Schwinger-like computation that was suggested
in the original GV work. The goal is to understand the background that must be used in this computation, 
the role played by the extended supersymmetry
of this background, and how the computation gives a holomorphic result though superficially depending only on particle masses.
We also examine in a similar way the Ooguri-Vafa (OV) formula, which is an extension of the GV formula to include D4-branes.

\end{abstract}
\date{November, 2014}
\end{titlepage}
\def\Hom{\mathrm{Hom}}
\def\X{{\mathcal X}}

\renewcommand{\baselinestretch}{0.15}\normalsize
\tableofcontents
\renewcommand{\baselinestretch}{1.0}\normalsize 

\section{Introduction}

In supersymmetric field theories and string theories, terms in the effective potential that
are supersymmetric but cannot be written as integrals over all of superspace
often play a special role.  Such couplings are often loosely called $F$-terms, as opposed to $D$-terms that can be written as integrals over 
all of superspace.  

Type IIA superstring theory compactified on a Calabi-Yau manifold $Y$ provides an interesting example.  This theory has
$\N=2$ supersymmetry in four dimensions, with eight unbroken supercharges, so a $D$-term is an interaction that can be written
$\int\d^4x\,\d^4\theta\,\d^4\bar\theta\,S$ for some $S$ (the $\theta$'s and $\bar\theta$'s are superspace coordinates of negative or positive
chirality).  This theory has the unusual property that for every  integer $\g\geq0$, it has an $F$-term that can only appear precisely in $\g$-loop order, with
no higher order or nonperturbative corrections.   Somewhat schematically,  this interaction is
\begin{equation}\label{zell}I_\sg=-i\int_{\R^4} \d^4x\,\d^4\theta\,\F_\sg(\X ^\Lambda)(\W_{ A B}\W^{ A B})^{\sg},\end{equation}
where $\F_\sg$ is a holomorphic function of $\N=2$  chiral superfields 
$\X ^\Lambda=X^\Lambda+\dots$ associated to vector multiplets, and 
$\W_{ A B}=\W_{ B A}$ is a chiral superfield whose bottom component is the anti-selfdual part of the graviphoton field strength 
(here $ A, B=1,2$ are spinor indices of negative chirality).  
More details will be provided later.  These interactions, which were identified
in \cite{BCOV}, are believed
to have a special significance: they determine the entropy of a half-BPS black hole as a function of its charges \cite{OSV}.

Since the coupling $I_\sg$  can only arise in one order of perturbation theory, it can be computed exactly, in principle, for each $\g$.  But actually,
the functions $\F_\sg(\X )$ have an interesting simplifying property:  they are genus $\g$ topological string amplitudes,
which can be computed by counting, in a certain sense, holomorphic maps from a genus $\g$ Riemann surface $\Sigma$ (of unspecified complex
structure) to $Y$.  This relationship was proposed and partly justified in \cite{BCOV} and was confirmed in detail in
\cite{AGNT}. See also an analysis  in \cite{OVtwo} using the hybrid formalism. The ``counting'' of holomorphic maps from $\Sigma$ to $Y$ is not straightforward; there are many technicalities involving the fact
that $\Sigma$ might be singular and its map to $Y$ might not be an embedding.  The computation in \cite{AGNT} used the power of superconformal
field theory and  circumvented all the technicalities.  In the special case that none of the subtleties arise
($\Sigma$ is  smooth and isolated or rigid
 and its map to $Y$ is an embedding), a rather direct explanation of the relation of $\F_\sg(\X )$ to topological string theory
is possible; see section 5.2 of \cite{BW}.  For $\g=0$, the calculations relating $\F_0$ to the topological string essentially reduce to facts that were known earlier for the heterotic string \cite{DSWW}.  Some  aspects of the computation of the $I_\sg$'s have been re-examined more recently
\cite{OT}.
 
 \subsection{The Gopakumar-Vafa Formula}\label{gvform}
 
 In \cite{GV1,GV2}, Gopakumar and Vafa proposed another approach to the interactions $I_\sg$.  The basic idea was to lift the analysis
 from Type IIA superstring theory to M-theory.   Type IIA superstring theory on $\R^4\times Y$ is equivalent to M-theory on $\R^4\times S^1\times Y$,
 where $S^1$ is sometimes called the M-theory circle.
 A Type IIA superstring worldsheet wrapped on $\Sigma\subset \R^4\times Y$ lifts to an M2-brane wrapped on $\Sigma\times S^1\subset
 \R^4\times S^1\times Y$.  If the radius of the $S^1$ is large enough compared to the 11-dimensional Planck scale and the length scales of $Y$
 and of $\Sigma$, then we can ignore the internal structure of the M2-brane and just think of it as a point particle propagating around the $S^1$.
In this limit,  it must be possible to compute the $I_\sg$ by counting contributions due to M2-brane states propagating around the circle.
 An M2-brane state that is not BPS will generate an effective interaction that is an integral over all of superspace.  This may be intuitively obvious,
 and is explained in section \ref{collective}.   So it must be possible to compute the $I_\sg$'s, when the M-theory circle is large,
 by counting contributions of BPS states of M2-branes.  
 
 The region in which the M-theory approach is useful corresponds, in Type IIA superstring theory, to the limit that the string coupling constant $g_\st$ 
 and the  volume of $Y$ are  large.  So one might expect that the M-theory 
 approach could determine the $\F_\sg$'s only under those conditions.
 But it turns out that the complete answer can be deduced from what happens in this regime using the fact
 that $I_\sg$ has a known dependence on $g_\st$
 -- it arises precisely in $\g$-loop order
with $2\g-2$ Ramond-Ramond insertions and so is proportional to $g_\st^{4\sg-4}$ -- along with holomorphy of $\F_\sg(\X)$.
 
 \def\I{{\mathcal I}}
 Gopakumar and Vafa proposed that the $I_\sg$'s could be effectively understood by considering a supersymmetric background obtained by turning
 on an anti-selfdual graviphoton field on $\R^4$, in the presence of a chosen expectation value 
 for the bottom components $X^\Lambda$ of the superfields $\X ^\Lambda$.
  The graviphoton field is the bottom component of the superfield $\W_{ A B}$ that appears in the $I_\sg$, so this
 would be a supersymmetric background with expectation values for the bottom components of all the superfields that appear in the
 definition of $I_\sg$.  The idea was that the sum of the $I_\sg$'s, namely
 \begin{equation}\label{tonzo}\I =\sum_{\sg\geq 0} I_\sg=-i\int \d^4x\,\d^4\theta\, \sum_{\sg\geq 0}\F_\sg(\X ^\Lambda)(\W^2)^\sg,  \end{equation}
 is the superspace effective action for a supersymmetric background characterized by a constant anti-selfdual 
 graviphoton field on $\R^4$.  To be more precise, the right hand side
 of eqn. (\ref{tonzo}) vanishes after performing the $\d^4\theta$ integral if the background is really supersymmetric, so one should think of this
 as an effective action that describes the response to a small perturbation around a supersymmetric background.  (The perturbation may be made
 by giving a small spatial dependence to $X^\Lambda$ or to the graviphoton field, by turning on a small spatial curvature, and/or by coupling to fermions.)
 Gopakumar and Vafa further proposed that the effective action in the presence of the anti-selfdual graviphoton 
 field could be computed by adapting Schwinger's celebrated
 result for the effective action due to a charged particle propagating in a constant electromagnetic field.  
 
 On this basis, Gopakumar and Vafa proposed a formula --  the GV formula -- expressing the 
 quantity $\I$ as a sum of contributions of M-theory BPS states.  The formula depends only on the masses, charges, and spins
 of the   BPS states.  
 It has had many physical and mathematical applications (for a few of the early rigorous mathematical tests of the GV formula, see
 \cite{HST,FF,GG}, and for a recent treatment in a more general context of symplectic manifolds see \cite{IP}).   The aim of this paper is to understand
 more precisely the  derivation of this formula.
 
 \subsection{Some Questions}\label{questions}
 
We proceed  to an overview of some questions concerning the GV formula and its derivation.   For some of these questions,
the explanations we give in this introduction will hopefully be sufficient.  Some other points require more elaboration, which
will occupy most of this paper.

\subsubsection{Can The Answer Be Determined In Principle?}\label{detprinc}

Perhaps the most basic question about trying to determine the interactions $I_\sg$ by a lift to M-theory is whether we know
enough about M-theory to have a hope of success.  The low energy effective action of M-theory is largely unknown,
except for a few terms of low dimension.  Certainly
there is no known method to systematically compute the effective action on $\R^5$ that arises from Calabi-Yau compactification.

The interactions $I_\sg$ were defined in eqn. (\ref{zell}) as terms in a four-dimensional effective action.  We have to ask what contributions
to the $I_\sg$ can arise by classical dimensional reduction from five dimensions.   Any contributions that can arise that way can be 
predicted only to the extent that we know the effective action of M-theory in five dimensions.   A Schwinger-like calculation on 
$\R^4\times S^1$, or any other quantum computation that depends on compactifying on $S^1$,  is useful for computing terms in the four-dimensional effective action on $\R^4$  that do {\it not} come by classical dimensional reduction from a local interaction on $\R^5$.  Such a computation
cannot predict what was already present in the effective action on $\R^5$.

Luckily, with very limited exceptions, the $I_\sg$ do not come by classical dimensional
reduction from five dimensions.   As we explain in section 
\ref{classred}, only certain very special terms arise this way (and only for $\g=0,1$),
and one knows just enough about the M-theory effective action to determine them.

\subsubsection{The Background And Its Supersymmetry}\label{subac}
 From a field theory point of view, one can compute the effective action in an expansion 
around any background, whether or not the background is a classical solution and whether or  not it is supersymmetric.   But 
one generally cannot learn anything about $F$-terms in 	a supersymmetric  effective action by expanding around
a background that is not supersymmetric.  Consider, for example, a non-supersymmetric background in which there is a superfield $P$ such that $\int\d^4\bar\theta P\not=0$.  Then in expanding around that background, it is difficult to distinguish an 
$F$-term $\int \d^4 x\d^4\theta \F$ (for some chiral superfield $\F$) from a $D$-term $\int\d^4x \d^4\bar\theta\d^4\theta \,P\F$.  

In field theory, it is conceivable that one could compute $F$-terms by expanding around a background 
that is supersymmetric but is not a classical solution.
It is not clear that this makes sense in string/M-theory, which lack a satisfactory off-shell formulation.

In Type IIA superstring theory on $\R^4\times Y$, is it possible to turn on an anti-selfdual graviphoton field while preserving some supersymmetry?
Generic experience in supersymmetric gauge theories suggests that at least in linear order, a selfdual or anti-selfdual gauge field would
preserve half of the supersymmetries -- the ones of (say) positive chirality.

Beyond linear order, we have to consider back-reaction of the gauge fields on  other bosonic fields.  In Type IIA superstring theory
on $\R^4\times Y$, the fields that might exhibit  back-reaction are the metric and  the scalar fields. 

In general, in four-dimensional gauge theory, a selfdual or anti-selfdual gauge field does not produce gravitational back-reaction, since a duality
condition ensures vanishing of  the Maxwell
stress tensor $T_{\mu \nu}=\frac{1}{2}\left(F_{\mu\alpha}F_{\nu}^\alpha-\allowbreak \frac{1}{4}\eta_{\mu\nu}F_{\alpha\beta}F^{\alpha\beta}\right)$.   But generically, a selfdual or anti-selfdual gauge field does produce back-reaction on spin 0 fields,
which couple to  scalar densities $F_{\mu\nu}F^{\mu\nu}$ and $F_{\mu\nu}\t F^{\mu\nu}$ (where $\t F=\star F$ is the dual of $F$) 
that are
nonzero even if $F$ is selfdual or anti-selfdual.

In our problem, we write $b_2$ for the second Betti number of $Y$ (thus in general $b_2\geq 1$).  At low energies,
Type IIA superstring theory on $\R^4\times Y$ has gauge group $U(1)^{b_2+1}$ (in fact, $b_2$ abelian gauge fields arise as modes of the RR three-form field $C$, and one
is the RR 1-form).  Of the  $U(1)$ gauge fields, one linear combination
is the graviphoton, which is in the supergravity multiplet (meaning that it is related by supersymmetry to the graviton) and the
others are in $b_2$ vector multiplets.  The subtlety is that the specific linear combination of gauge fields that is the graviphoton
depends on the scalar fields $X^\Lambda$ of the vector multiplets.  

In Type IIA superstring theory on $\R^4\times Y$, a general background characterized by $U(1)^{b_2+1}$ gauge fields on $\R^4$,
 whether or not the field strength is selfdual or anti-selfdual,  does produce scalar back-reaction.  However, precisely in case the gauge field that is turned on is the graviphoton (with constant field strength), 
this back-reaction vanishes. If in addition the graviphoton field strength is selfdual or anti-selfdual, it produces no gravitational back-reaction and
we get a supersymmetric background.\footnote{The fact that a graviphoton field produces no scalar back-reaction is also important
in the attractor mechanism for half-BPS black holes \cite{Attractor}.  (This was pointed out to us by J. Maldacena. A related remark was made in
\cite{BeS}.)  
As we noted in section \ref{gvform}, the entropy of these half-BPS black holes is believed to be governed by the same supersymmetric
interactions that appear in the GV formula \cite{OSV}.  The graviphoton field in a half-BPS black  hole is neither selfdual nor anti-selfdual, 
so it produces gravitational back-reaction (and the metric of the black hole is not flat), even though it produces no scalar back-reaction.}  

The existence of this supersymmetric background is a very nice
fact that is essential in making the GV calculation possible.   This background has been studied in 
the literature (mostly for the special case $b_2=1$)
from several points of view \cite{GGHPR,BeS,OVtwo}, and  a remarkable property has been found.
Naively, an anti-selfdual graviphoton background would be expected to preserve at most 
half of the supersymmetry -- four of the eight supercharges
of $\N=2$ supersymmetry -- namely the supersymmetries of just one chirality.
It turns out, however, that the background obtained by turning on a (constant)
anti-selfdual graviphoton on a flat $\R^4$ preserves eight supersymmetries, not just four.  Turning on the  graviphoton 
 ``deforms'' the supersymmetries
of the opposite chirality and changes the supersymmetry algebra, but it does not 
change the total number of unbroken supersymmetries.  We will call
the extra four supersymmetries the extended supersymmetry of the problem.

The importance of the extended supersymmetry is as follows.  
In performing the Schwinger-like calculation that leads to the GV formula, one needs to know not just the masses and
electric charges but also the magnetic moments of the BPS particles.  In specific cases,
one can compute the magnetic moments, but to get the GV formula, one needs a universal answer,
which follows from the extended supersymmetry. 

The anti-selfdual graviphoton background has one more important and unusual property:  this background is not real.  
To understand this, as in \cite{GGHPR}, we can consider this background as a solution of minimal five-dimensional
supergravity, in which the bosonic fields are the five-dimensional metric and a single $U(1)$ gauge field.
(This theory arises from compactification on a Calabi-Yau manifold $Y$ that has $b_2=1$.  The resulting solution can be embedded in the
supergravity that is relevant for any $b_2$.)
This theory has a real solution in Lorentz signature that is called a supersymmetric G\"{o}del solution in \cite{GGHPR} because of its peculiar causal properties (closed timelike curves).  The solution has time-translation symmetry, and one can compactify the time direction, giving a solution
on $ \R^4\times S^1$, still with Lorentz signature.  (The solution has translation symmetry in the $\R^4$ directions, but these
translations do not commute; their commutators generate the rotation of $S^1$.)  This solution is real, but it does not have a simple relation to
Type IIA superstring theory since the $S^1$ direction is timelike, while the relation of  
M-theory to Type IIA involves compactification on a spacelike circle.  
One can ``Wick rotate'' the solution to make the $S^1$ direction spacelike, but then if the metric is to be real, the gauge fields are no longer real.  This is the relevant
background for deriving the GV formula.

The fact that the gauge fields are not real actually poses no problem.  
In fact, Schwinger's method of computing 
the one-loop effective action due to a charged particle in an external electromagnetic field with constant
field strength does not require the background field to be real.  In his original calculation,
Schwinger considered both electric and magnetic fields in Lorentz signature, which is
somewhat like considering a complex-valued magnetic field in Euclidean signature.  
It turns out, as we explain at the end of  
section \ref{pf}, that to resolve  some puzzles concerning the derivation of the GV formula, one needs to know
 that the pertinent graviphoton background is not real.

Likewise, the fact that the Lorentz signature version of the graviphoton background has closed
timelike curves is no problem.  We will explain this in three related ways.  First, by rotating to
Euclidean signature, we eliminate the closed timelike curves in favor of an imaginary gauge field,
which poses no problem as explained in the last paragraph.  Second, the superparticle action that
we actually use for the computation in section \ref{schwpart} is physically sensible.
Third, the GV formula describes a series of interactions $I_\sg$ each of which is perturbative in the graviphoton
field; the Schwinger-like computation that determines the sum of these interactions is in principle just a way
to conveniently organize perturbation theory.  In perturbation theory in the graviphoton field $\TT$, one does not see closed timelike
curves; their radius is of order $1/\sqrt{\TT}$.

\subsubsection{Particles, Instantons, And Fields}\label{pf}

Quantum mechanics has wave/particle duality, but the range of validity of a computation based on particles can
be quite different from the range of validity of a computation based on fields.  Is the Schwinger calculation that leads to the GV
formula supposed to be a computation based on particles or on fields?

We understand M-theory on $\R^5\times Y$ -- to the extent that we do understand it -- in the regime that the radius of $Y$ is large
compared to the eleven-dimensional Planck length.  Under these conditions, a massive BPS state that arises from an M2-brane wrapped
on a two-manifold $\Sigma\subset Y$
is much heavier than the Planck mass.  It is much more natural to study such a heavy object as a particle rather than as a field.
In fact, upon replacing $\R^5\times Y$ with $\R^4\times S^1\times Y$ with the $S^1$ assumed to be large, there is no problem at all in doing the relevant computation.
We simply consider an M2-brane wrapped on the volume-minimizing submanifold
 $p\times S^1\times \Sigma$ (where $p$ is a point in $\R^4$, which we treat as a collective
coordinate for the M2-brane).  Such a Euclidean wrapped M2-brane can be thought of as an instanton in M-theory on 
$\R^4\times S^1\times Y$.  Instantons arising in this way from wrapped branes have been widely studied, originally in
\cite{bbs}.  To obtain the GV formula, or at least its contribution from a given BPS state, we simply have to compute a product of determinants
associated to small fluctuations around the wrapped M2-brane.  This computation gives the expected Schwinger-like formula in a straightforward
fashion (one uses the deformed supersymmetry algebra to determine the necessary couplings, as explained in section \ref{subac}).  As in many
such problems, a
one-loop computation is enough, because holomorphy ensures that higher order corrections that arise in expanding around the
instanton cannot contribute to the functions $\F_\sg(\X ^\Lambda)$.

The instanton obtained by wrapping an M2-brane on $p\times S^1\times \Sigma$ can be understood in five-dimensional terms as
a particle whose worldline is $p\times S^1$.
Thus we treat the massive BPS states as particles, not fields.  In the limit that the $S^1$ is large (so that the M-theory
description is useful), we isolate the contribution in which these particles travel a long distance -- all the way around the $S^1$.  Because
this distance can be assumed large, we can ignore the internal structure of the BPS states and treat them as point particles.  That 
makes the one-loop computation very simple.

  Actually, 
massive BPS states that arise from wrapped M2-branes can have arbitrarily large spin.  Entirely apart from their trans-Planckian masses,
one would have trouble treating these objects via fields because one does not have a reasonable field theory of massive particles
of high spin coupled to background gauge and gravitational fields.   We do not need such a field theory because we are in a regime in
which particle theory is more appropriate.

M-theory on $\R^5\times Y$ also has massless BPS states that arise from compactification on $Y$ of  eleven-dimensional
supergravity.  These states have spin at most 2 and of course we do have a sensible field theory description of 
them -- namely five-dimensional supergravity.  By contrast, it would be tricky to give a useful particle description of these states. (The starting point would have to be a relativistic description of a massless superparticle in five dimensions; a useful one is hard to come by.
There would be no useful semiclassical  description involving an expansion around a classical
particle orbit since there is no natural lightlike geodesic to expand around.\footnote{One might be tempted to use 
a particle description for four-dimensional
mass eigenstates  arising in Kaluza-Klein reduction on $S^1$. (These states are wrapped D2-branes coupled to D0-branes.)  Such a description makes little sense for individual
mass eigenstates because from a $d=4$ point of view,
the distance traveled by the D2-branes is not large and there is no basis for treating them as elementary particles running
around a loop.  One does not usually 
consider a loop diagram with a D-brane running around the loop, unless the loop is a large loop in spacetime, just as one does not usually perform
a Schwinger calculation with deuterons or DNA molecules running around a loop.  A better route to a somewhat similar formalism
is described in the next paragraph: start with the fact that field theory is valid for a massless (or light)  supermultiplet in five 
dimensions, and then make a Kaluza-Klein
expansion of the resulting field theory formulas.})  Thus we determine the
contribution to the GV formula of five-dimensional massless BPS states by a 1-loop computation in massless field theory on
$\R^4\times S^1$.  (We perform such a computation in detail for a massless hypermultiplet in section \ref{hypercalc}
and determine the contributions of other
massless supermultiplets by using the fact that a non-BPS supermultiplet does not contribute to the interactions $I_\sg$.)  

Since the $\R^4\times S^1$ background that is relevant in deriving the GV formula is invariant under rotations of the $S^1$, 
the field theory computation in this background  can be expressed in terms of a sum of contributions of four-dimensional mass
eigenstates\footnote{We do not need to know if these mass eigenstates are precisely BPS. If some of them are not BPS (a possibility
discussed in section \ref{tenc}), this means that some 4d supermultiplets whose net contribution to the GV formula is 0 combine to 
non-BPS 4d supermultiplets.} that arise in Kaluza-Klein reduction.  The mass eigenstates are states of definite momentum around $S^1$.
  As in \cite{GV1,GV2}, a Poisson resummation
re-expresses the sum of contributions of states of definite momentum around $S^1$ as a sum of contributions 
of configurations with definite winding number.
The winding number sum is generally the most useful version
 of the GV formula.  For massive 5d BPS states, the particle-based treatment gives directly the winding number sum, with no 
 Poisson resummation.
 
One might be disappointed at the thought that field theory is 
 only useful for the very few BPS states whose mass vanishes in five dimensions.  Actually, this is not quite the case.  Field theory
 can also be used for hypermultiplets or vector multiplets that are anomalously light.
Generically, when $Y$ is large enough so that M-theory is a useful description, the massive BPS states are much heavier than
the Planck mass, and cannot be reasonably treated in field theory.  However, when varying the Kahler moduli of $Y$, it is possible \cite{wittenphase} to reach a critical point at which a massive charged BPS hypermultiplet
goes to zero mass (this happens when a copy of $\Bbb{CP}^1\subset Y$ collapses to a point; the light BPS hypermultiplet arises
from an M2-brane wrapped on this $\Bbb{CP}^1$).  
Near such a critical point, the hypermultiplet in question is light enough that it makes sense to treat it by field theory,
and of course we do have a sensible field theory for massive 5d hypermultiplets.   Likewise, when the Kahler moduli of $Y$ are varied so that $Y$ develops
a curve of ADE singularities, it is possible for a vector multiplet to become light, so that again a field theory description of this multiplet makes sense.

So in deriving the GV formula, there is a situation in which
it makes sense to perform a field theory computation for a 5d massive hypermultiplet or vector multiplet. Happily these are multiplets for which a satisfactory
field theory does exist.   When the field theory  computation is performed
in terms of a sum over 4d mass eigenstates that arise from Kaluza-Klein reduction, it has the following property.  Let $\cZ  $ be the central
charge of the four-dimensional $\N=2$ algebra.  It is a holomorphic function of the complex fields $X^\Lambda$, and 
the mass of a 4d BPS state is $m=|\cZ  |$.  Schwinger's one-loop calculation is naturally
expressed in terms of the magnetic field $F$ and the mass, so it looks like the answer will be a function of $|\cZ  |$, contradicting the fact
that the functions $\F_\sg(X^\Lambda)$ are supposed to be holomorphic in the $X^\Lambda$.  This puzzle was actually one of the reasons
for writing the present paper. The resolution of the puzzle depends upon the
fact that the background that is used in deriving the GV formula is not real.  In fact, the relevant magnetic field $F$ in this background\footnote{The $F$ that is relevant here is a linear combination of the $U(1)^{b_2+1}$ field strengths,
with coefficients that depend on the charges of the BPS state considered.  See eqn.  (\ref{effield4d}).}
 is proportional to $\bar \cZ  $,
and the dimensionless ratio $F/m^2\sim 1/\cZ  $ that appears in the Schwinger formula is holomorphic.

Can one similarly vary the Kahler moduli of $Y$ to reach a point at which a BPS state of high spin becomes light and hence should be treated
in field theory?  This would be problematical, since we do not know a suitable field theory for a state of high spin.
An M2-brane state of high spin becomes light if a curve in $Y$ of high genus collapses to a point.  However, a curve of positive genus cannot collapse to a point in $Y$ except when a whole divisor, for instance
a copy of $D=\Bbb{CP}^2\subset Y$, collapses to a point.  In this case, curves in $D$ of arbitrarily high genus are collapsing, so BPS states of arbitrarily
high spin go to zero mass,  and in addition there is a BPS string
whose tension goes to zero (it comes from an M5-brane wrapped on $D$).  It is believed that M-theory develops a non-trivial infrared
critical point when a single divisor $D$ shrinks to a point \cite{wittenphase,MS}.  One does not have a description of such a 
critical point that would be useful for a field theory computation of the behavior of the $\F_\sg$'s in this regime.

What happens when a whole family of divisors collapses is described next.

\subsubsection{Duality With The Heterotic String}\label{dualhet}

\def\K3{{\mathrm{K3}}}
Though there is no useful field theory description of massive particles of high spin, there sometimes is a useful weakly-coupled string theory
description of such particles.  Under certain conditions, M-theory on $\R^5\times Y$ has an alternative description in terms of the heterotic
string on $\R^5\times \K3\times S^1$.  In this description, there are two kinds of BPS states:  elementary string states
and states that arise from fivebranes wrapped on $\K3\times S^1$.  Both types of BPS state contribute to the $I_\sg$'s, but the contributions
from the elementary string states are particularly simple:  they arise precisely at 1-loop order in perturbation theory.  These 1-loop contributions
were analyzed in \cite{AGNTtwo,MoMa,Serone}.  This provided part of the background to the work of Gopakumar and Vafa, 
where it was argued that
in a more general sense, even when a perturbative string formalism is not available, the $I_\sg$'s can be computed as a sum of
1-loop contributions from BPS states.  

To find a heterotic string description of M-theory on $\R^5\times Y$, we proceed via F-theory \cite{Evidence}.  We first assume that $Y$ is elliptically
fibered, which means that there is a holomorphic map $Y\to B$, where $B$ is a del Pezzo surface, possibly with orbifold singularities,
and the generic fiber is an elliptic curve $E$:
\begin{equation}\label{zomo}\begin{matrix} E & \to &Y\cr && \downarrow \cr && B.\end{matrix} \end{equation} 
Given this, M-theory on $Y$ is equivalent to F-theory on $S^1\times Y$ (or in other words to Type IIB superstring theory on
$S^1\times B$, with a coupling parameter that is controlled by the map $Y\to B$).  We would like to convert this
to a heterotic string description, using the fact that F-theory on K3 is equivalent to the heterotic string on $T^2$.  To use
this duality, we have to assume that $Y$ is fibered by K3 surfaces,\footnote{Duality of Type II on a Calabi-Yau threefold to the heterotic string
was first discovered in \cite{KV}. The fact that the relevant Calabi-Yau threefolds are K3 fibrations was discovered
 in \cite{KLM}.} meaning that there is a holomorphic map $Y\to S\cong \Bbb{CP}^1$
whose generic fiber is a K3 surface:
\begin{equation}\label{zomox}\begin{matrix} \K3 & \to &Y\cr && \downarrow \cr && S.\end{matrix} \end{equation}
The two fibrations (\ref{zomo}) and (\ref{zomox}) are compatible if the $\K3$ fibers of $Y\to S$ are themselves elliptically fibered, now
over $S'\cong\Bbb{CP}^1$:
\begin{equation}\label{zomorz}\begin{matrix} E & \to &\K3\cr && \downarrow \cr && S'.\end{matrix} \end{equation}  
Then $B$ is a ``rational ruled surface'': it admits a holomorphic map to $S\cong \Bbb{CP}^1$ with generic fiber $S'\cong\Bbb{CP}^1$:
\begin{equation}\label{zomotz}\begin{matrix} S' & \to &B\cr && \downarrow \cr && S.\end{matrix} \end{equation} 
(The map $Y\to S$ in (\ref{zomox}) is the composition of the map $Y\to B$ in (\ref{zomo}) with the map $B\to S$ in (\ref{zomotz}).)
Duality with the heterotic string is a consequence of the fibration (\ref{zomox}).  To get a heterotic string description, we use the duality
between F-theory on K3 and the heterotic string on $E'\cong T^2$.  Applying this duality to every fiber in (\ref{zomox}), we arrive
at an elliptically-fibered K3 surface, which we call $\K3'$:
\begin{equation}\label{zomoxic}\begin{matrix} E' & \to &\K3'\cr && \downarrow \cr && S.\end{matrix} \end{equation}
F-theory on $Y$ is equivalent to the heterotic string on $\K3'$, and M-theory on $Y$ is equivalent to the heterotic string on $\K3'\times S^1$.

In this situation, an M-theory BPS state that arises from an M2-brane wrapped on an embedded Riemann surface
$C\subset Y$ corresponds to an elementary
heterotic string state if and only if $C$ is contained in one of  the K3 fibers of the map $Y\to S$ (equivalently, $C$ projects
to a point in $S$).  Ordinary heterotic string perturbation
theory suffices to compute the contributions of these states to the $I_\sg$'s, and considerations of supersymmetry and holomorphy
ensure that there are no contributions beyond 1-loop order.  Everything can be computed directly, with no need for considerations
such as those of the present paper. The GV formula asserts that a
similar answer holds in general, even when there is no weakly coupled description to justify a 1-loop calculation.  

\subsubsection{The Role Of The Holomorphic Anomaly}\label{ha}

The claim that the amplitudes (\ref{zell}) or (\ref{tonzo}) can be compared to topological string theory requires a clarification, as follows.

Let $t^I=\langle X^I/X^0\rangle$ be the holomorphic
Kahler moduli in compactification of Type IIA superstring theory on the Calabi-Yau manifold $Y$, and let $\t t^I$ be the corresponding
antiholomorphic moduli.  In a physically sensible compactification of string theory, $\t t^I$ is simply the complex conjugate of 
$t^I$.   A perturbation to the moduli is made, in physical string theory, by adding to the worldsheet action a chiral interaction term, 
to vary the $t^I$, along with its hermitian conjugate, to vary the $\t t^I$.  In topological field theory, 
one does not necessarily require the action to be real, so the $t^I$ can be varied independently of the
$\t t^I$.  That is indeed why we have written $\t t^I$ rather than $\bar t^I$ 
for the antiholomorphic variables that naively are the complex conjugates of the $t^I$.

Formally, one can use worldsheet supersymmetry to show that topological string amplitudes are holomorphic in 
$t^I$ and independent of $\t t^I$,
but in reality this is not so; there is a holomorphic anomaly \cite{BCOV}.  To define topological string amplitudes, one has to pick a ``basepoint''
-- that is, a value of the $\t t^I$.  
Then one varies the $t^I$ (away from the physical value, which would be the complex conjugate of $\t t^I$), keeping $\t t^I$ fixed.  The resulting function of the $t^I$ is holomorphic.   Its
dependence on the $\t t^I$ is given by the holomorphic anomaly equation.   Although topological string amplitudes thus depend on $\t t^I$,
there is actually a standard choice, which is roughly $\t t^I=\infty$.  
To be more precise, the basepoint is defined by taking the
$\t t^I$ to infinity within the Kahler cone of $Y$.  
 One summarizes
this by saying that the basepoint is at infinite volume.
It turns out that it does not matter in exactly what direction one goes to infinity in the Kahler cone
(as long as one stays away from the boundaries of the cone). 

This has an analog in physical string theory.  Formally, one can use spacetime supersymmetry to show that the interactions (\ref{zell})
or (\ref{tonzo}) are holomorphic in the variables $Z^I=X^I/X^0$, whose expectation values are the moduli $t^I$.  Actually, this reasoning
applies to a Wilsonian effective action, obtained by integrating out massive fields only.  
If the interactions (\ref{zell}) and (\ref{tonzo})
are understood as low energy effective actions (whose tree level matrix elements are supposed to give directly the low energy limits
of the appropriate scattering amplitudes), then actually they are not holomorphic; they have a dependence on the complex conjugates
of the $X^I$'s that comes from loops of massless particles and 
precisely matches the dependence of the topological string amplitudes on the $\t t^I$'s.   
The claim \cite{BCOV,AGNT} that the interactions (\ref{zell}) can be computed in topological string theory refers to low energy effective
actions, not Wilsonian ones, and  this claim involves a match between the nonholomorphic dependence on the two sides.
If we want
the interactions (\ref{zell}) and (\ref{tonzo}) to agree with standard topological string amplitudes, which are defined with a basepoint at infinite volume,
then we have to similarly define the interactions (\ref{zell}) and (\ref{tonzo}) in terms of an expansion near infinite volume.

The GV formula  asserts that a certain sum of contributions of BPS states in M-theory reproduces the physical
string amplitudes (\ref{zell}) or (\ref{tonzo}).   With what basepoint should the physical string amplitudes
 be computed in order to make the GV formula true?  
The answer to this question is clear from the fact that the GV formula is obtained by a computation in 
M-theory on $\R^4\times S^1\times Y$, where the radius  of $S^1$ is taken to be large.  M-theory 
in this situation corresponds
to Type IIA superstring theory on $\R^4\times Y$ with a large volume for $Y$ (this
statement follows from elementary considerations that are explained in \cite{cadavid} and also in section \ref{omix} below).  Thus the
derivation of the GV formula involves an expansion around large volume, and the corresponding topological string amplitudes are the standard
ones with a basepoint at infinity.   

Among other things, this resolves the following small puzzle.  The notion of ``large volume'' in Type II superstring theory is not completely
natural, because in general the Kahler moduli space of a Calabi-Yau manifold $Y$ contains a variety of large volume limits;
one  corresponds to large volume on $Y$
and the others correspond to 
large volume on  additional Calabi-Yau manifolds $Y_\alpha$ that are birationally equivalent to $Y$.  However,
in M-theory, the $Y_\alpha$'s are separated from each other and from $Y$ by phase transitions \cite{wittenphase}. (Near such a phase
transition, a charged BPS hypermultiplet becomes light, and the scope of field theory is enlarged, as already noted in section \ref{pf}.) Built into the M-theory
analysis is  a choice of a particular
$Y$ (within its birational equivalence class) and thus a particular large volume limit in Type IIA superstring theory.

For a detailed re-evaluation of the full effective action of Type II superstring theory on a Calabi-Yau manifold, with a thorough study of 
many questions involving the holomorphic anomaly and the relation to the topological string, see \cite{dewit}.

\subsubsection{BPS States In Ten And Eleven Dimensions}\label{tenc}

We can construct a supersymmetric theory on $\R^5$ by compactifying M-theory on a Calabi-Yau 
manifold $Y$, and similarly we can construct a supersymmetric
theory on $\R^4$ by compactifying Type IIA superstring theory on $Y$.  BPS states can be defined in 
either of these two theories.  What is the relation between
the BPS states defined in M-theory and the BPS states defined in Type IIA superstring theory?  And which set of 
BPS states enters the GV formula?

The answer to  the second question should already be clear.  The GV formula arises by lifting from Type IIA superstring theory on
$\R^4\times Y$ to M-theory on $\R^4\times S^1\times Y$, and reinterpreting an elementary string worldsheet 
$\Sigma\subset \R^4\times Y$ in terms
of a membrane state, wrapped on $\SIgma$, that propagates around the $S^1$.   Thus the states that are 
counted in the GV formula are the BPS states in M-theory.

Given this, the first question is less relevant for us, but we discuss it anyway for completeness.  Naively,
 if we start with an M-theory BPS state of mass $m$
and compactify on a circle of circumference $2\pi R$, then  in Type IIA we get a whole Kaluza-Klein 
tower of BPS states with a momentum $p=n/R$
around the circle, for some integer $n$,
and  a mass \begin{equation}\label{melf} m_n=\sqrt{m^2+(n/R)^2}.\end{equation}  But in general, 
this claim is oversimplified: the states in question have mass
approximately $\sqrt{m^2+(n/R)^2}$ if $R$ is large, but this formula may not be exact.  If it is not,
then these states are 
not BPS states in Type IIA.
So in general BPS states do not remain BPS after compactification on a circle and 
there is no simple way to deduce the spectrum of BPS states in Type IIA from a knowledge of the BPS states in M-theory.

It is actually natural to lose BPS states in compactification on a circle, because in compactification on a circle, some of the 
rotation symmetry is lost.
In five dimensions, massive states can be classified according to how they transform under the rotation group $SO(4)$ 
or more precisely its double
cover $\Spin(4)$.  This group  is isomorphic to a product $SU(2)_\ell\times SU(2)_r$ of two $SU(2)$ subgroups, 
so a massive state in five dimensions has spins
$j_\ell,j_r\in \Z/2$.  If orientations are chosen properly, the  BPS supermultiplets contributing
to the GV formula have the property that all states in a such a supermultiplet have the same $j_r$ (but not the same $j_\ell$), and 
the contribution of the multiplet
to the GV formula depends on $j_\ell$ but not $j_r$.  Reduction from five to four dimensions by compactifying one of the 
spatial directions on a circle
breaks the $SU(2)_\ell\times SU(2)_r$ symmetry to a diagonal $SU(2)$ subgroup.  Because of this reduction 
in symmetry, it can happen that BPS multiplets
that are protected in five dimensions (they cannot be deformed into multiplets that are not BPS) 
are no longer protected after compactification to four dimensions.\footnote{For instance, 
let $W_\ell$ be the representation $2(0,0)\oplus (1/2,0)$ of $SU(2)_\ell\times SU(2)_r$, and $W_r$ the representation
$2(0,0)\oplus (0,1/2)$.  In five dimensions, $W_r\otimes W_r$ is protected but $W_\ell\otimes W_r$ is not (see section \ref{prelim}).
After reduction to four dimensions, $W_\ell$ and $W_r$ become equivalent and $W_r\otimes W_r$ ceases to be protected.}
In this case, it is very natural that the mass
formula (\ref{melf}) is not exact and that (at least for some values of the momentum around the circle -- the integer $n$ in eqn. (\ref{melf}))
compactification on a circle does not convert five-dimensional BPS states to four-dimensional ones.

Conversely, given BPS states in four dimensions, from a knowledge of their $SU(2)$ quantum numbers, one cannot in general 
reconstruct the $SU(2)_\ell\times SU(2)_r$
quantum numbers of underlying BPS states in five dimensions, and hence one cannot reconstruct their contribution to the GV formula.

The following example is one in which 5d BPS states may be lifted in reduction to four dimensions,\footnote{If there is any example in which
5d BPS states are lifted in reduction to four dimensions, we would expect this to happen in any fairly generic example.  But a generic
example of a family of curves in a Calabi-Yau manifold is difficult to study.  The example given in the text is meant to be a relatively
simple one that emphasizes the strong coupling issues that probably are relevant to the general question. It is known \cite{GG} that in this type of example,
topological string amplitudes can be expressed via the GV formula in terms of states that presumably are the M-theory BPS states. (It is difficult to determine
the M-theory BPS states in this example because the theory of parallel M2-branes is strongly coupled.)} and also shows the difficulty
in reconstructing the 5d spectrum from what happens in four dimensions.
To understand this example would
  involve more work and, probably, tools that are not presently available.   The rest of the paper does not depend on this example.
   In Type IIA superstring theory, suppose that the Calabi-Yau
  manifold $Y$ contains an isolated embedded curve $\Sigma$ of genus $\g$.  The normal bundle $N$ to $\Sigma$ has degree $2\g-2$.  The assumption
  that $\Sigma$ is isolated means that $H^0(\Sigma,N)=0$ (so $\Sigma\subset X$ has no infinitesimal deformations); for example, if $N$ is a direct
  sum of line bundles, this condition implies that each has degree $\g-1$.  Wrap two D2-branes on $\Sigma$, with a $U(2)$ Chan-Paton
  bundle $E$.  We write $\ad(E)$ for the adjoint bundle derived from $E$.  The world-volume theory of the D2-branes contains scalar
  fields $\phi$ forming a section of $N\otimes \ad(E)$. If $H^0(\Sigma,N\otimes \ad(E))\not=0$, it may be possible to give an expectation value to
  $\phi$, in which case a D2-brane doubly wrapped on $\Sigma$ is replaced by a D2-brane singly-wrapped on a curve $\Sigma'\subset Y$ of genus
   $\g'>\g$  that is a ``spectral cover'' of $\Sigma$ in the sense of Hitchin.  However, for generic $E$, $H^0(\Sigma,N\otimes \ad(E))=0$.  We focus
   on that situation.
  Let $M_E$ be the moduli space  that parametrizes deformations of $E$.  The space of BPS states in this situation is $H^*(M_E)$, the de Rham
  cohomology\footnote{The statement that the space of BPS states is $H^*(M_E)$ is rather formal.  To make a precise statement, one
  must understand how to treat singularities in $M_E$ (including the locus in $M_E$ along which $H^0(\Sigma,N\otimes \ad(E))\not=0$; along this
  locus, extra zero-modes appear in the D-brane quantization, so this should
  be regarded as a locus of singularities).   At least naively, for $E$ of rank 2, $H^*(M_E)$ depends  on whether the degree of $E$ is even
  or odd (it should only depend on the degree mod 2, since tensoring $E$ by a fixed line bundle $\mathcal L\to \Sigma$ can shift this degree by
  any even integer).  As this degree is the momentum around the M-theory circle, this gives an example in which the spectrum of 4d BPS states appears
   likely
  to depend on that momentum.  If so, this certainly shows that the spectrum of 4d BPS states does not descend directly from five dimensions.} of $M_E$.  In this type of example, the $SU(2)$ that corresponds to the spin of the BPS states is the Lefschetz
  $SU(2)$ that acts on the cohomology of any Kahler manifold, in this case $M_E$.   Given the spectrum of 4d BPS states of this system, 
  there appears to be no simple way to determine the corresponding space of 5d BPS states that arise in wrapping two M2-branes on $\Sigma$
  in M-theory.   For example, does a 4d multiplet of spin $j$ descend from spin $(j,0)$ or $(0,j)$ in 5d? There is no simple way to answer,
  since the theory of two parallel M2-branes is strongly coupled.  For reducible representations of $SU(2)$, there are more general questions
  of this sort; for example, does spin 0 plus spin 1 in four dimensions come from $(1/2,1/2)$ or from $(0,0)\oplus (0,1)$ in five dimensions?  But actually,
  it may be necessary to add additional states to make the appropriate 5d supermultiplets, since 
 part of the BPS spectrum  in five dimensions may be lost in reduction in four dimensions.
  
There are, however, some situations in which 5d BPS states are definitely {\it not} lifted in compactification to four dimensions.  
That is the case for BPS states that arise from an M2-brane singly wrapped on a smooth, isolated curve $\Sigma\subset Y$.  This is clear
from explicit quantization of the M2-brane and its D2-brane counterpart in string theory.   This statement generalizes under some
conditions  when $\Sigma$ has moduli
(but still with single wrapping) \cite{GV2,HST}.
For another kind of example,  two-dimensional
conformal field theory tells us that a BPS perturbative string state remains BPS after further compactification on a circle.  So M-theory BPS
states that are dual to perturbative heterotic string states, as described in section \ref{dualhet}, are not lifted upon reduction to four dimensions.

\subsubsection{Trivial And Non-Trivial $F$-Terms}\label{trivial}

In understanding the derivation of the GV and OV formula, it helps to make a certain elementary distinction.

$D$-terms are a subspace of all possible effective couplings, namely those that can be written as $\int \d^4x\d^4\theta\d^4\bar\theta\,S$ for
some $S$.  But any $D$-term can also be written as an $F$-term, since $\int\d^4x\d^4\theta\d^4\bar\theta\,S=\int\d^4x\d^4\theta\, U$
with $U=\int\d^4\bar\theta\,S$.  We regard an $F$-term of this form as trivial.  When one speaks loosely of ``the space of $F$-terms,''
one really means ``the space of $F$-terms modulo trivial ones that can be written as $D$-terms.''  The space of $F$-terms in this sense is
really not a subspace of the space of all interactions but a quotient.

The interactions (\ref{zell}) that enter the GV formula  are non-trivial
$F$-terms, and moreover they generate the space of $F$-terms that can be constructed from these particular chiral superfields, modulo trivial ones.
(In the OV case, the corresponding interactions (\ref{mell}) are non-trivial $F$-terms but do not generate the space of $F$-terms modulo trivial ones;
there are additional $F$-terms that can be constructed from the same fields but are not determined by the OV formula.)  In deriving the GV or OV
formula by evaluating an instanton amplitude, one is only interested in a certain leading term, since the subleading corrections are $D$-terms or
equivalently trivial $F$-terms.

 \subsection{The Ooguri-Vafa Formula}\label{oov}
 
 The Gopakumar-Vafa formula has a close analog with D4-branes included \cite{OV}.  We will call this analog the
 Ooguri-Vafa (OV) formula.    Here in Type IIA superstring theory on $\R^4\times Y$, we pick a linear subspace $\R^2\subset
 \R^4$ and a special Lagrangian submanifold
 $L\subset Y$ and, for some $N\geq 0$, we add $N$ D4-branes wrapped on $\R^2\times L$.  Or more generally,
 we pick several special Lagrangian submanifolds $L_i\subset Y$, and we wrap $N_i$ D4-branes on $\R^2\times L_i$.
For the derivation of the OV formula, all D4-branes must be
 supported on the same $\R^2\subset\R^4$,
 though they can be supported on different Lagrangian submanifolds $L_i\subset Y$.  (Otherwise, as we explain in section
 \ref{ovover}, some of the supersymmetry that is needed in the derivation is lost.)
 
 In this situation, string perturbation  theory is modified in a familiar way.  A string worldsheet $\Sigma$ is still oriented, but
 it may have boundaries, each component of which must be mapped to one of the $L_i$.  Contributions in string
 perturbation theory in which $\Sigma$ has genus $\g$ with $\h$ holes or boundary components and hence Euler
 characteristic $\chi=2-2\g-\h$ are as usual weighted in string perturbation theory with a factor $g_\st^{-\chi}=g_\st^{2\sg+\sh-2}$.  (A Ramond-Ramond insertion comes with an extra factor of $g_\st$.)
 
 In general, in such a problem, the low energy effective action  has terms that are supported on the worldvolumes of
 the branes, which here means on  $\R^2\subset\R^4$.  As explained
 in \cite{OV}, for the particular case described here,  the interactions
 $I_\sg$ that enter the GV formula have analogs supported on the branes:
 \begin{equation}\label{mell}J_\sn=\int_{\R^2}\d^2x \,\d^2\theta \,\RR_\sn(\X ^\Lambda;\U^\sigma) \W_\parz^\sn. ~~\n\geq 0. \end{equation}
 Here the $\X ^\Lambda$ are the same chiral superfields that appear in eqn. (\ref{zell}), except that now, since the D4-branes explicitly
 break half of the supersymmetry, we integrate over only half as many $\theta$'s (the ones that 
 correspond to the unbroken supersymmetry) and
 we restrict $\X ^\Lambda$ to depend only on those $\theta$'s. Thus the $\X ^\Lambda$ are now viewed as chiral superfields in a theory with
 $(2,2)$ supersymmetry on $\R^2$.  For $\n\geq 0$, $\RR_\sn$ is a holomorphic function of the $\X ^\Lambda$.  
 Also, in a sense that will be explained in
 section \ref{fourov}, $\W_\parz$ is the ``parallel'' component of the graviphoton superfield $\W_{ A B}$ that appears in 
 eqn. (\ref{zell}); and $\U^\sigma$, $\sigma=1,\dots,
 b_1(L)$, are chiral superfields associated to the moduli of $L$.
 
 Like the bulk interactions $I_\sg$ that appear in eqn. (\ref{zell}), the particular $F$-terms $J_\sn$
 arise only from particular
 terms in string perturbation theory.  In fact, the interaction $J_\sn$ is only generated by worldsheets $\Sigma$ with
 \begin{equation}\label{merx} 2\g+\h-1 = \n. \end{equation}  Consequently, $J_\sn$ has a known dependence on $g_\st$.
 Like the $I_\sg$'s, the $J_\sn$'s can be interpreted as topological string amplitudes, in this case, amplitudes for open and closed 
 topological strings.
 
 The idea of Ooguri and Vafa \cite{OV} was simply that like the closed-string interactions $I_\sg$,  
 the open-string interactions $J_\sn$  can be determined
 from a lift to M-theory on $\R^4\times S^1\times Y$.  In this lift, the D4-branes supported on $\R^2\times L_i$
 become M5-branes supported on $\R^2\times S^1\times L_i$.
 A fundamental string worldsheet
 $\Sigma$ (possibly with boundaries on the D4-branes) is replaced by an M2-brane with 
 worldvolume $S^1\times \Sigma$ (possibly with boundaries on the
 M5-branes).  When the $S^1$ is large, the calculation can be performed in terms of M2-brane
 states propagating around the $S^1$.  It may be intuitively obvious, and will be clear in the 
 technical derivation in section \ref{fourov}, that only BPS
 states contribute to $F$-terms.  Here, to get a contribution to the Type IIA effective action supported on $\R^2$,
  the relevant BPS states are states in M-theory on $\R^5\times Y$ that do have boundaries, so that they propagate only along the M5-branes,
 that is, only along $\R^3\times L_i\subset \R^5\times Y$.   From a low energy point of view, one suppresses $L_i$ and 
 $Y$ and says simply that these states propagate along $\R^3\subset \R^5$.
  
 It must  be possible to compute the interactions $J_\sn$ as a sum of contributions of these BPS states.  
 Ooguri and Vafa argued that, as in the closed
 string case, instead of the individual $J_\sn$, one should consider the sum
\begin{equation}\label{elx}\J=\sum_{\sn=0}^\infty J_\sn=\sum_{\sn=0}^\infty\int_{\R^2}\d^2x \,\d^2\theta \, \RR_\sn(\X ^\Lambda;\U^\sigma)\W_\parz^\sn.\end{equation}
Their proposal was that this sum can be interpreted as a superspace effective interaction for a background in 
which an anti-selfdual graviphoton 
is turned on in the presence of the M5- or D4-branes, and that it can be computed by a Schwinger computation, 
now for charged BPS states propagating on $\R^3\subset \R^5$.  
The required background is  the same one that is relevant to the original GV formula, but now with the branes included.  
The facts summarized in section \ref{subac} are again essential:
in particular, to get a universal formula for the contributions of BPS states to the function $\J$, one needs the extended supersymmetry of
the graviphoton background.   We explain details of this computation, from our point of view, in section \ref{fourov}.   We should remark
that although the computations are straightforward, there are a number of details that we have found difficult to understand.  These are
mainly associated to infrared subtleties that can occur in three spacetime dimensions. In generalizing
the OV formula beyond the type of example considered in the original
work, one is likely to have to grapple with these infrared questions.

In practice (starting with the original paper \cite{OV}), the OV formula is often 
combined with additional arguments that involve among other things 
a geometric transition \cite{GVother,Vafa} between different Calabi-Yau manifolds in which D-branes are replaced by fluxes.  
In particular, this has led to many results about knot theory \cite{LMV,GSV}.
 
 \section{The Background And Its Supersymmetry}\label{backsup}
 
 \def\T{{\mathcal T}}
 \def\d{{\mathrm d}}
 \subsection{The Background In Five Dimensions}\label{fivedback}
 \subsubsection{The Supersymmetric G\"{o}del Solution}\label{supergodel}
 
 The bosonic fields of minimal supergravity in five dimensions are the metric tensor $g$ and a $U(1)$ gauge field $V$, whose
 field strength is the 5d graviphoton $\TT=\d V$.  
 To describe the supersymmetric G\"{o}del solution \cite{GGHPR},
we parametrize $\R^5$ with coordinates $t$ and $x^\mu$, $\mu=1,\dots,4$.  The desired solution has the property that $\TT$ has
no component in the $t$ direction, and its components in the $x^\mu$ directions are constant and anti-selfdual.  We set $V_\nu=\frac{1}{2}\TT^-_{\mu\nu}x^\mu$, where
 $\TT^-_{\mu\nu}$ is constant (independent of $t$ and the $x^\mu$), antisymmetric, and anti-selfdual in the four-dimensional sense. 
 We take the metric to be 
 \begin{equation}\label{miro}\d s^2=-(\d t-V_\mu \d x^\mu)^2+\sum_{\mu=1}^4 (\d x^\mu)^2. \end{equation}
 For real $\TT^-$, this is a real and supersymmetric solution of 5d supergravity
 in Lorentz signature.    It has the special property that 
 the 5d graviphoton  can also be viewed as the field strength of  a ``Kaluza-Klein'' gauge field.
 
This is actually not a physically sensible solution, 
since a large circle in the hyperplane $t=0$ can be a closed timelike curve.   For our purposes, we would like to compactify
 the $t$ direction to a circle, and moreover we want this circle to be spacelike, so that it can be interpreted as the M-theory circle.  
 To make the circle spacelike, we will set $t$ to be a multiple of $-i y$, where $y$ will be a real variable of period $2\pi$.
 To give the circle an arbitrary circumference $2\pi e^\sigma$, we take the relation between $t$ and $y$ to be $t=-iye^\sigma$.
The  solution (\ref{miro}) can then be written
 \begin{equation}\label{iro} \d s^2=e^{2\sigma}\left(\d y+B_\mu \d x^\mu\right)^2+
  \sum_\mu (\d x^\mu)^2, \end{equation}  
 where we have defined
 \begin{equation}\label{morox}B_\mu=-ie^{-\sigma}V_\mu. \end{equation} 
 This compactified solution can be generalized in an obvious way to depend on another real parameter:  we 
 give a constant expectation value to $V_y$,
 the component  in the $y$ direction of the gauge field $V$.   

Clearly, to make the metric in eqn. (\ref{iro}) real, we have to take $V_\mu$ and $\TT^-$ to be imaginary.
This is not really troublesome, since a Schwinger-like calculation in a constant magnetic field  still makes sense if the magnetic
field is imaginary.  (An imaginary magnetic field in Euclidean signature is somewhat analogous to a constant electric field in Lorentz signature, which was one of the original
cases studied by Schwinger.)

 The 4d interpretation of the 5d metric (\ref{iro}) requires some care.
 The 4d metric in Einstein frame is not $g_{\mu\nu}=\delta_{\mu\nu}$, which we would read off from (\ref{iro}), but rather is
 \begin{equation}\label{zorox} g^E_{\mu\nu}= e^{\sigma}\delta_{\mu\nu}. \end{equation}
  It is also convenient to define 
 \begin{equation}\label{plorox} \WW^-_{\mu\nu}=4e^{\sigma/2}\TT^-_{\mu\nu},\end{equation}
 which turns out to be the 4d graviphoton.  Thus
 \begin{equation}\label{belz}B_\nu=-i\frac{e^{-3\sigma/2}}{8}\WW^-_{\mu\nu}x^\mu,~~~V_{\nu}=\frac{1}{2}\TT^-_{\mu\nu}x^\mu . \end{equation} 
 We also write $\WW^-$ as the curvature of a 4d gauge field
 \begin{equation}\WW^-_{\mu\nu}=\partial_\mu U_\nu-\partial_\nu U_\mu,~~ U_\mu =4e^{\sigma/2}V_\mu. \end{equation}
 $\TT^-$, $\WW^-$,  $V$ and $U$ will be imaginary and $B$ real.   We define the 5d scalar quantity
 \begin{equation}\label{zory}(\TT^-)^2 = \delta^{\mu\mu'}\delta^{\nu\nu'}\TT^-_{\mu\mu'}\TT^-_{\nu\nu'}, \end{equation} raising and lowering
 indices using the 5d metric (\ref{iro}).
 But in defining a corresponding 4d scalar quantity $(\WW^-)^2$,
 we raise and lower indices using the 4d Einstein frame metric:
 \begin{equation}\label{welz} (\WW^-)^2=g^{E\mu\mu'}g^{E\nu\nu'}\WW_{\mu\mu'}\WW_{\nu\nu'}=16 e^{-\sigma}(\TT^-)^2.\end{equation} 
 
We have described the  basic five-dimensional solution that is used in the computation leading to the GV formula, along with its reduction
to four dimensions.
This solution has two properties that are important in deriving the GV formula: {\it (i)} it preserves all of the supersymmetry, not just half of it, as one might expect
for a solution with  an anti-selfdual graviphoton;
{\it (ii)} it generalizes straightforwardly to the case that an arbitrary number of vector multiplets are included.   We describe these two
properties in sections \ref{extended} and \ref{genany}.

\subsubsection{Extended Supersymmetry}\label{extended}

The supersymmetry algebra of the supersymmetric G\"{o}del solution (\ref{miro}) can be described as follows. In describing spinors, we use the obvious orthonormal
frame field
\begin{equation}\label{morf}e^t=\d t-V_\mu\d x^\mu,~~~e^\mu=\d x^\mu, ~~\mu=1,\dots,4,\end{equation}
or the dual vector fields
\begin{equation}\label{norfo} v_t=\frac{\partial}{\partial t},~~ v_\mu=\frac{\partial}{\partial x^\mu}+V_\mu\frac{\partial}{\partial t}.\end{equation}
The spinor representation of $SO(1,4)$ is four-dimensional and pseudoreal.  Since it is pseudoreal, the supersymmetry generators in minimal 5d supergravity
are actually a pair of spinors, which we denote $\epsilon^{\alpha i}$, where $\alpha=1,\dots, 4$ is an $SO(1,4)$ spinor index and $i=1,2$ reflects the doubling needed to make
the supersymmetry generator real (note
that no symmetry acting on this index is assumed).  Indices
are raised and lowered using the $SO(1,4)$-invariant antisymmetric tensor $C_{\alpha\beta}$ (sometimes called the charge conjugation matrix) and a $2\times 2$ antisymmetric tensor $\veps_{ij}$.
In five-dimensional Minkowski spacetime, the supersymmetry algebra is 
\begin{equation}\label{mordox}\{Q_{\alpha i},Q_{\beta j}\}=-i\Gamma^M_{\alpha\beta}\veps_{ij}P_M +C_{\alpha\beta}\veps_{ij}\zeta,\end{equation}
where $P_M$, $M=0,\dots,4$ are the momentum generators, 
$(\Gamma^M)^\delta_\beta$ are  Dirac gamma-matrices, $\Gamma^M_{\alpha\beta}=(\Gamma^M)^\delta_\beta C_{\delta\alpha}$, and we include the 5d  central charge $\zeta$.

Since the graviphoton field breaks $SO(1,4)$, it is convenient to write everything in terms of a $4+1$-dimensional split with coordinates $x^\mu$, $\mu=1,\dots,4$ and $t$.  For this, we introduce four-dimensional gamma-matrices $\gamma^\mu$ with chirality matrix $\gamma_5=-i\Gamma_0$, decompose $Q_{\alpha i}$ in terms of spinors $Q_{Ai}$ and $Q_{\dA i}$, $A,\dA=1,2$ of negative and positive chirality, and
we write the momentum generators as $H=-P_0$ and $P_\mu$, $\mu=1,\dots,4$.
In 5d Minkowski spacetime, the supersymmetry algebra now reads
\begin{align}\label{mofo} \{Q_{Ai},Q_{Bj}\}&=\veps_{AB}\veps_{ij}( H+\zeta) \cr
                                        \{Q_{\dA i}, Q_{\dB j}\}&=\veps_{\dA\dB}\veps_{ij}(H-\zeta) \cr
                                         \{Q_{A i},Q_{\dB j}\}&= -i\Gamma^\mu_{A\dot B}\veps_{ij}P_\mu. \end{align}

Now let us discuss what happens to the supersymmetry algebra when the graviphoton field is turned on.  The Killing spinor equation for a supersymmetry
generator $\epsilon$ implies that it is independent of $t$ and obeys the four-dimensional equation
\begin{equation}\label{omfo}\partial_\mu\epsilon-\frac{1}{4}\TT^-_{\nu\rho}\gamma^{\nu\rho}\gamma_\mu\epsilon =0. \end{equation}
Since $\gamma_\mu$ reverses the chirality and $\TT^-_{\nu\rho}\gamma^{\nu\rho}$ annihilates spinors of positive chirality, this
equation is trivially satisfied  for any constant spinor $\eta_{Ai}$  of negative chirality by
\begin{equation}\label{merg} \epsilon_{Ai}=\eta_{Ai}, ~~~\epsilon_{\dot Ai}=0.\end{equation}
This is enough to maintain half the supersymmetry.  But somewhat  less trivially,
if $\eta_{\dot A i}$ is a constant spinor of positive chirality, the equation can also be solved by
\begin{equation}\label{zerg}\epsilon_{\dot A i}=\eta_{\dot A i},~~~\epsilon_{A i}=\TT^-_{\mu\nu}x^\mu\gamma^\nu_{A\dot A}\eta^{\dot A}_i, \end{equation}
so a constant anti-selfdual graviphoton actually preserves all of the supersymmetry.

The extended supersymmetry is certainly surprising, but if one looks more closely, there is a surprise hidden even in the more trivial-looking supersymmetries (\ref{merg}).  In gauge theory, a background with anti-selfdual field strength $F_{A B}$ preserves the
supersymmetries of positive chirality.   (Anti-selfduality means that $F_{\dA\dB}=0$, so the transformation of the gluino field $\lambda$ associated to a positive chirality supersymmetry generator $\epsilon^{\dA}$ is 
$\delta\lambda_{\dA}=F_{\dA \dB}\epsilon^{\dB}=0$.) But
the ``trivial'' supersymmetries in an anti-selfdual graviphoton background have {\it negative} chirality.

Since the anticommutator of two supersymmetries will be a bosonic symmetry,  we have to understand
the bosonic symmetries of this spacetime in order to understand the supersymmetry algebra.  The Killing vector fields associated to the generators $H$ and $P_\mu$
of translation symmetries are
\begin{align}\label{zibo} h & =-\frac{\partial}{\partial t} \cr
                                       p_\mu& =\frac{\partial}{\partial x^\mu}-V_\mu\frac{\partial}{\partial t}. \end{align}
 Note the contribution to $p_\mu$ that is proportional to $V^\mu$; it reflects the fact that the graviphoton background is translation-invariant in the 
 $x^\mu$ directions only up to
 a time translation.  Because of this contribution, the translation generators  do not commute:
 \begin{align}\label{ilbo} [p_\mu,p_\nu] & =\TT^-_{\mu\nu}h \cr
                                        [p_\mu,h]& =0. \end{align}
(As discussed below, the commutator of the conserved charges $P_\mu$ corresponding to $p_\mu$ also contains a central term that is not seen in the commutator of the $p_\mu$.)
                                        
We also must consider rotation symmetries.  Without the graviphoton field, we would have a full action of $\mathrm{Spin}(4)\cong SU(2)_\ell\times SU(2)_r$,
with $SU(2)_\ell$ rotating spinor indices $A,B$ of negative chirality and $SU(2)_r$ rotating spinor indices $\dA,\dB$ of positive chirality.  A constant
anti-selfdual graviphoton
field breaks $SU(2)_\ell\times SU(2)_r$ to $U(1)_\ell\times SU(2)_r$.  The Killing vector fields that generate the unbroken rotation symmetries are unchanged
from what they would be at $\TT^-=0$.  The $SU(2)_r$ generators do not appear in the anticommutators of two supersymmetries (or of the other bosonic symmetry
generators).  Thus $SU(2)_r$ can be viewed as a group of outer automorphisms of the supersymmetry algebra.  However, as we discuss momentarily, the
$U(1)_\ell$ generator does appear on the right hand side of the supersymmetry algebra.   The generator of $U(1)_\ell$ is associated to the Killing vector field
\begin{equation}\label{moxic} j=4V^\mu\frac{\partial}{\partial x^\mu}. \end{equation}
We also express $j$ in terms of standard angular momentum generators $j_{\mu\nu}$:
\begin{equation}\label{noxic}j=\TT^{-\mu\nu}j_{\mu\nu},~~~   j_{\mu \nu}
=x_\mu\frac{\partial}{\partial x^\nu}-x_\nu\frac{\partial}{\partial x^\mu}. \end{equation} 
It is convenient to write $J_{\mu\nu}$ for the conserved angular momentum 
corresponding
to $j_{\mu\nu}$, and set
\begin{equation}\label{mixt}\eJ=\TT^{-\,\mu\nu}J_{\mu\nu}. \end{equation}
We also want an analogous quantity for the theory compactified to four dimensions, but here we should be careful.
In more detail $\eJ=\TT^-{}_{\mu\nu}\left( \delta^{\nu\sigma} x^\mu\frac{\partial}{\partial x^\sigma}-\delta^{\mu\sigma}x^\nu\frac{\partial}{\partial x^\sigma}\right)$, where $\delta^{\nu\sigma}$ is the standard flat metric on $\R^4$.   In four dimensions, we want to make a similar
definition using the 4d graviphoton field $\WW^-$ (eqn. (\ref{plorox})) and  the Einstein metric
$g^E$  (eqn. (\ref{zorox})), so we define
\begin{equation}\label{fixt}\JJ= \WW^-_{\mu\nu}\left( g^{E\nu\sigma} x^\mu\frac{\partial}{\partial x^\sigma}-g^{E\mu\sigma}x^\nu\frac{\partial}{\partial x^\sigma}\right),\end{equation}
obeying
\begin{equation}\label{lixt} \eJ=\frac{e^{\sigma/2}}{4}\JJ.\end{equation}

In discussing the supersymmetry algebra, just as at $\TT^-=0$, we write $Q_{Ai}$ and $Q_{\dA i}$ for   supersymmetries
whose generators are parametrized by the negative and positive chirality spinors  $\eta_{Ai}$ and $\eta_{\dA i}$ that appear in eqns. (\ref{merg}) and (\ref{zerg}) above.
Turning on the constant anti-selfdual graviphoton field modifies the supersymmetry algebra in two ways. 
 The most obvious change is that because
the generators (\ref{zerg}) of positive chirality supersymmetries have a contribution linear in the $x^\mu$, the $Q_{\dA i}$ do not commute with the $P_\mu$:
\begin{equation} \label{elbox}[P_\mu, Q_{\dA i}]=\TT^-_{\mu\nu}\Gamma^\nu_{\dA A} Q^A_i. \end{equation}
Given this, there must be a correction to the anticommutator $\{Q_{\dA i},Q_{\dB j}\}$, to 
avoid a problem with the $Q_{\dA i}\cdot Q_{\dB j}\cdot Q_{A k}$
Jacobi identity.  To compute what happens, all one has to know is that, in five-dimensional 
notation, if $\epsilon_{\alpha i}$ and $\epsilon'_{\beta j}$ are
two Killing spinor fields, then, up to possible central terms,
 the anticommutator of the corresponding supersymmetries is associated to the Killing vector 
 field\footnote{If we use constant gamma-matrices $\Gamma^m$ referred to the local Lorentz frame (\ref{morf}), 
 this formula will give the components
of the vector field $u^m$ in the dual basis (\ref{norfo}).} $u^m=\varepsilon^{ij}\bar\epsilon'_i
\Gamma^m\epsilon_j$.  The graviphoton field produces no correction to the anticommutator $\{Q_{A i},Q_{B j}\}$ of
negative chirality supersymmetries, since eqn. (\ref{merg}) asserts that (in the local Lorentz frame (\ref{morf})), 
there is no $\TT^-$-dependent contribution
to the generators of these supersymmetries.  In computing $\{Q_{Ai},Q_{\dA j}\}$, we do have to take into 
account the $\TT^-$-dependent contribution
to the generator of $Q_{\dot A j}$.  But this just goes into building up the $\TT^-$-dependent part of the Killing vector field $p_\mu$ (eqn. (\ref{zibo})),
leaving no $\TT^-$-dependent correction to the usual relation $\{Q_{Ai},Q_{\dA j}\}=-i
\varepsilon_{ij}\sum_{\mu=1}^4 P_\mu \Gamma^\mu_{A\dA}.$
Where one does find a correction is in the anticommutator $\{Q_{\dot Ai},Q_{\dot Bj}\}$, which acquires a term $\varepsilon_{\dot A\dot B}\varepsilon_{ij}\eJ$.  

The central charge $\zeta$ that appears in the $\{Q,Q\}$ anticommutator (eqn. (\ref{mofo})) also appears in $[P_\mu,P_\nu]$.
This should come as no surprise; it reflects the fact that in the presence of a constant magnetic field 
on $\R^4$ -- in our case the graviphoton field -- 
translations only commute up to a gauge transformation. To evaluate $[P_\mu,P_\nu]$, we use 
the $P_\mu \cdot Q_{A i} \cdot Q_{\dB j}$ Jacobi identity. Since
\begin{equation}
\label{peqQQ}
P_\nu = -\frac{i}{4}\Gamma_\nu^{\dA A}\veps^{ij}\{Q_{A i}, Q_{\dA j}\},
\end{equation}
the commutator $[P_\mu,P_\nu]$ can be simply computed using  $[P_\mu,Q_{A i}]=0$, eqn. 
(\ref{elbox}) for $[P_\mu, Q_{\dA j}]$, and eqn. (\ref{mofo}) for
$\{Q_{A i}, Q_{B j}\}$. We find that $[P_\mu,P_\nu]$ is proportional to $H+\zeta$ (and not to $H$, as one might have supposed
from eqn. (\ref{ilbo}) for $[p_\mu,p_\nu]$).

Putting the pieces together, the supersymmetry algebra is
\begin{align}\label{susy5d}[P_\mu,P_\nu]& = -i\TT^-_{\mu\nu}(H+\zeta) \cr
[\eJ, P_\mu] &=2i\TT^-_{\mu\nu}P^\nu \cr
[\eJ, Q_{Ai}] &=-\frac{i}{2}\TT^-_{\mu\nu}\Gamma^{\mu\nu}_{AB}Q^B_i \cr
[P_\mu,Q_{\dA i}]& =    \TT^-_{\mu\nu}\Gamma^\nu_{\dA B} Q^B_{i}, \cr
\{Q_{A i},Q_{B j}\}& = \varepsilon_{A B}\varepsilon_{ij}(H+\zeta) \cr
\{Q_{Ai},Q_{\dA j}\}& = -i\Gamma^\mu_{A\dA}\varepsilon_{ij}P_\mu\cr
\{Q_{\dA i},Q_{\dB j}\}&=\varepsilon_{\dA\dB}\varepsilon_{ij}(H-\zeta+\eJ),
                                                    \end{align}
with other commutators and anticommutators vanishing.  

  This 5d supersymmetry algebra  will be our starting point in the
particle-based computation in section \ref{schwpart}.  
In section \ref{schwfields}, we will perform a field theory computation that is conveniently expressed
in terms of Kaluza-Klein reduction to four dimensions.  
 For this, we will want a 4d version of the above algebra that arises after rotation to Euclidean time and compactifying the time direction on a circle of radius $e^\sigma$. We write the metric as in eqn. (\ref{iro}), and  use a rescaled graviphoton field $\WW^-_{\mu\nu}$ as in eqn. (\ref{plorox}).    In going to four dimensions, the 
 gamma-matrices are scaled by $e^{\sigma/2}$ to refer them to the Einstein frame because of the $e^\sigma$ in the definition of the 4d Einstein metric in (\ref{zorox}), and accordingly to keep the supersymmetry
 algebra in a standard form, the supersymmetry generators must be scaled by $e^{\sigma/4}$.
 Thus, we introduce 4d supersymmetry generators $\Qb_{Ai}$, $\Qb_{\dot A j}$, defined by
 \begin{equation}\label{melof}\Qb_{Ai}=e^{-\sigma/4}Q_{Ai},~~~\Qb_{\dot Aj}=e^{-\sigma/4}Q_{\dot Aj}.\end{equation}
 Rotation to Euclidean time causes $H$ to be accompanied by an extra factor of $-i$ (assuming one wishes $H$ to be remain hermitian).  After compactification, $H$ becomes a central charge in the 4d sense.
  The full 4d central charge is \begin{equation}\label{murkox} \cZ  =e^{-\sigma/2}(\zeta+ i H)\end{equation} and the 
  supersymmetry algebra in four dimensions is 
  \begin{align}\label{morfic} [P_\mu,P_\nu]& = -\frac{i}{4}\WW^-_{\mu\nu}\bar{\cZ  } \cr
[\JJ,P_\mu]&=2i\WW^-_{\mu\nu}P^\nu,\cr
[\JJ,\Qb_{A i}]&=-\frac{i}{2}\WW^-_{\mu\nu}\gamma^{\mu\nu}_{AB}\Qb_i^B,\cr
[P_\mu,\Qb_{\dA i}]& =\frac{1}{4}\WW^-_{\mu\nu}\gamma^\nu_{\dA B} \Qb^B_i, \cr
\{\Qb_{A i},\Qb_{B j}\}& = \varepsilon_{A B}\varepsilon_{ij}\bar{\cZ  } \cr
\{\Qb_{Ai},\Qb_{\dA j}\}& = -i\gamma^\mu_{A\dA}\varepsilon_{ij}P_\mu\cr
\{\Qb_{\dA i},\Qb_{\dB j}\}&=-\varepsilon_{\dA\dB}\varepsilon_{ij}(\cZ   - \frac{1}{4}\JJ)
                                                    \end{align}                                     
As always, the moduli of the compactification do not appear explicitly in the algebra, but they affect the possible values of the central
charge.  For example, if the 5d gauge fields have holonomies around the M-theory circle, 
this affects the values of $H$ and therefore of $\cZ  $.

\def\CC{{\mathcal C}}
\def\V{{\mathcal V}}
\subsubsection{Generalization To Any $b_2(Y)$}\label{genany}
So far, we have considered the supersymmetric G\"{o}del solution in pure supergravity, but we will need its generalization to include vector multiplets.
For this, we could proceed abstractly, but it is convenient to consider
the motivating example of compactification of M-theory to five dimensions on a Calabi-Yau manifold $Y$  with second Betti number $b_2$.
We introduce a basis $\omega_I$, $I=1,\dots, b_2$ of $H^2(Y,\Z)$, and define
\begin{equation}\label{omex} \CC_{IJK}=\frac{1}{6}\int_Y\omega_I\wedge\omega_J\wedge\omega_K. \end{equation}
The Kahler class $\omega$ of $Y$ can be expanded as a linear combination of the $\omega_I$:
\begin{equation}\label{refo}\omega=\sum_{I=1}^{b_2} v^I\omega_I .\end{equation}
The $v^I$ are interpreted as scalar fields in five dimensions (they take values in a certain Kahler cone)
and their expectation values are moduli of the compactification.  It turns out that only $b_2-1$ combinations
of the $b_2$ fields $v^I$ are in 5d vector multiplets.  The volume of $Y$, which is
\begin{equation}\label{uyt}\V=\CC_{IJK}v^Iv^Jv^K,\end{equation}
is part of a hypermultiplet (sometimes called the universal hypermultiplet).  
The remaining $b_2-1$ combinations of the $v^I$ are in vector multiplets.  It is convenient to define
these combinations by setting
\begin{equation}\label{buyt}h^I=\frac{v^I}{v},~~v=\V^{1/3},\end{equation}
 so that
\begin{equation}\label{norf}\CC_{IJK}h^Ih^Jh^K=1.\end{equation}
The $h^I$, with this constraint, parametrize the vector multiplet moduli space in five dimensions.  

In five dimensions, a vector multiplet contains a real scalar field and a $U(1)$ gauge field.
To find the gauge fields, we make a Kaluza-Klein expansion of the M-theory three-form field $C$:
\begin{equation}\label{monoxic} C=\sum_I V^I\omega_I. \end{equation}
The $V^I$ are abelian gauge fields in five dimensions, with field strengths $F^I=\d V^I$.  
One linear combination of the $V^I$, namely $V=\sum_I h_I V^I$, with $h_I=\CC_{IJK}h^Jh^K$, is in the supergravity multiplet.  This is the 
the 5d graviphoton field.  To be more exact, the graviphoton field strength is $\TT=\sum_I h_I\d V^I$; $\d V$ is not gauge-invariant unless the $h_I$ are
constant.
The orthogonal linear combinations of the $V^I$ are in vector multiplets,
together with the $h^I$.  To describe orthogonal linear combinations of the $V^I$, it is useful to introduce vectors $h^I_x$, $x=1,\dots,b_2-1$ 
tangent to the hypersurface (\ref{norf}), i.e.
obeying $ h_Ih^I_x=0$.  These can be defined as $h^I_x=\partial h^I/\partial \phi^x$, where $\phi^x$ are local coordinates on the 
hypersurface (\ref{norf}).  The linear combinations of the $V^I$ that are orthogonal to the graviphoton field\footnote{\label{kelter} This orthogonality is
in the natural metric $a_{IJ}=\frac{1}{4}\int_Y\omega_I\wedge *\omega_J$  on the Kahler cone. The hypersurface metric 
$g_{xy}=h^I_x h^J_y a_{IJ}$ is induced from this.}
are  $V_x=\sum_{IJK}\CC_{IJK}h^Ih^J_xV^K$.   To be more precise, the gauge-invariant 
field strengths $F_x =\sum_{IJK}\CC_{IJK}h^Ih_x^J\d V^K$ are
in vector multiplets.

The precise meaning of the statement that $\TT=\sum_I h_I\d  V^I$ is in the supergravity multiplet is that it appears in the supersymmetry transformation
of the spin 3/2 gravitino field:
\begin{equation}\label{murf} \delta\Psi_M = \nabla_M\epsilon + \frac{i}{8}\TT_{NP}\left(\Gamma_M^{~NP}-4\delta_M^N\Gamma^P\right)\epsilon+\dots,\end{equation}
where the ellipsis represents fermionic terms. By contrast, the $V_x$ appear along with derivatives of scalars in supersymmetry
transformations of spin 1/2 fermi fields that are in vector multiplets. Let $\lambda^x$ be fermionic fields related to the $\phi^x$  by supersymmetry. Then
the precise meaning of the statement that $F_x$ is in a vector multiplet is that 
\begin{equation}\label{urf} \delta\lambda^x = \frac{i}{2}\partial_M\phi^x \Gamma^M\epsilon + \frac{1}{4}F^x_{MN}\Gamma^{MN}\epsilon +\dots,\end{equation}
where again the ellipsis represents fermionic terms, and the index $x$ in $F^x$ was raised using the metric defined in footnote \ref{kelter}.

Now it should be clear how to embed the original supersymmetric G\"{o}del 
solution (\ref{miro}), which corresponds to the case $b_2=1$ (no vector multiplets), in the theory with an arbitrary
number of vector multiplets.  Using the same $V$ and the same metric as before, we simply take the $h^I$ to be arbitrary constants,  the $V_x$ to vanish, and $V^I=h^IV$.  
This will ensure the vanishing of the right hand side of eqn.
(\ref{urf}) and all desired properties are satisfied.
Similarly, the compactified version of the solution is 
 \begin{equation}\label{irom} \d s^2=e^{2\sigma}\left(\d y+B_\mu \d x^\mu\right)^2+
 e^{-\sigma} \sum_\mu (\d x^\mu)^2,~~~~~~B_\mu=-i\frac{e^{-\sigma}}{2}\TT^-_{\nu\mu}x^\nu,~~~V^I_\mu=\frac{h^I}{2}\TT^-_{\nu\mu}x^{\nu} ,\end{equation}
again with
 \begin{equation}\label{ploroxt} \TT^-_{\mu\nu}=\frac{e^{-\sigma/2}}{4}\WW^-_{\mu\nu}.\end{equation}
  Again we take $\TT^-_{\mu\nu}$ and $V^I_\mu$ to be imaginary
 and the metric to be real.    If $y$ is understood to be a periodic variable (with period $2\pi$), 
 we can slightly generalize this solution by giving nonzero constant values to $V_y^I$,
 the components of the fields $V^I$ in the $y$ direction. 
 
 Each gauge field $V^I$, for $I=1,\dots, b_2$, couples to a conserved charge $Q_I$.  The $Q_I$ are components of the homology class of
 an M2-brane wrapped in $Y$.   A wrapped M2-brane with world volume $\Sigma$ is an eigenstate of $Q_I$ with eigenvalue
 \begin{equation}\label{moff}q_I=\int_\Sigma \omega_I. \end{equation}
 The central charge in the 5d supersymmetry algebra is $\zeta=\sum_I h^IQ_I$ (see eqn. \ref{erf}).  Its values for a BPS particle with charges $Q_I=q_I$ is $\zeta(\vec q)=\sum_I h^Iq_I$.
A BPS particle with those charges couples to the linear combination $V(\vec q)=\sum_J q_J V^J$ of the $V^I$.  In the background (\ref{irom}),
 the field strength of $V(\vec q)$ is
 \begin{equation}\label{zobel} F^{\{\vec q\}}_{\mu\nu}=\sum_I q_I h^I \TT^-_{\mu\nu}=\zeta(\vec q) \TT^-_{\mu\nu}. \end{equation}
So for each set of charges $\vec q=\{q_1,q_2,\dots,q_{b_2}\}$, we will do a Schwinger calculation with background field $F^{\{\vec q\}}_{\mu\nu}=
\zeta(\vec q)\TT^-_{\mu\nu}$.   Part of the reason that a simple answer emerges is that the mass $m(\vec q)$ of a BPS particle with charges $\vec q$ is also proportional to 
$\zeta(\vec q)$, so that the ratio $F^{\vec q}/m(\vec q)$ depends only on $\TT^-_{\mu\nu}$ and not on the vector multiplet moduli.

 \subsection{The Background From A 4d Point Of View}\label{fourdpoint}

 \subsubsection{Duality-Invariant Formalism}\label{dualinv}
 
 Here, we will describe the same background more fully from a 4d point of view.\footnote{Some original supergravity references are
  \cite{AA,BB,CC,DD}.  Our conventions are those of \cite{EE}.}  We primarily work in Einstein frame (which is natural in supergravity) rather
 than the string frame.  
 As we have already explained, M-theory compactification on a Calabi-Yau manifold $Y$ gives a theory in five dimensions with $b_2(Y)$ abelian gauge fields,
 of which $b_2-1$ linear combinations are in vector multiplets and one is the graviphoton.  Upon further compactification on a circle, we get one more vector
 multiplet, which comes from Kaluza-Klein reduction of the 5d metric on a circle.  Thus in four dimensions, there are $b_2+1$ abelian gauge fields, of which
 $b_2$ linear combinations are in vector multiplets and one linear combination is the graviphoton.  
 
 In five dimensions, a vector multiplet contains a real scalar field; we described the scalars in $b_2-1$ vector multiplets via $b_2$ scalar fields $h^I$ that obey
 a constraint (\ref{norf}).  In four dimensions, a vector multiplet contains a {\it complex} scalar field.  It is convenient to describe the scalar fields in $b_2$ vector
 multiplets via $b_2+1$ complex scalar fields $X^\Lambda$, $\Lambda=0,\dots,b_2$  that obey  a gauge-invariance
 \begin{equation}\label{monko}X^\Lambda \to \lambda X^\Lambda, ~~\lambda\in U(1),\end{equation}
 and a constraint
 \begin{equation}\label{yonko}N_{\Lambda\SIgma}X^\Lambda\bar X{}^\SIgma=-1,\end{equation}
 where $N_{\Lambda\Sigma}$ will be defined later.    We will also eventually impose a condition to fix the $U(1)$ gauge-invariance.  Alternatively,
 to emphasize the complex structure of the vector multiplet moduli space, one can replace the constraint (\ref{yonko}) by an inequality
 $N_{\Lambda\SIgma}X^\Lambda \bar X{}^\Sigma<0$ and replace the $U(1)$ gauge-invariance with a $\C^*$ gauge-invariance $X^\Lambda
 \to \lambda X^\lambda$, $\lambda\in\C^*$.
 
 The $X^\Lambda$ are the bottom components of superfields $\X^\Lambda$ that also contain fermion fields $\Omega^\Lambda$ and the field strengths $F^\Lambda_{\mu\nu}$ of
 the $U(1)$ gauge fields, which appear in the combinations:
 \begin{equation}
 \label{fcal}
 \F_{\mu\nu}^{\Lambda,-}=F_{\mu\nu}^{\Lambda,-}-\frac{1}{2}\bar{X}^\Lambda \WW^-_{\mu\nu}+{\rm fermions}.
 \end{equation}
 Here the 4d graviphoton field strength $\WW^-_{\mu\nu}$ will be defined later and the fermionic terms are not important to us. Here and elsewhere in this paper, if $\F$ is
 a two-form then $\F^-$ is its anti-selfdual part. The superfields $\X^\Lambda$ have expansions
 \begin{equation}\label{onkort}\X^\Lambda=X^\Lambda+\bar\theta^i \Omega_i^\Lambda +\frac{1}{2}\varepsilon_{ij}\bar\theta^i\sigma^{\mu\nu}\theta^j \F^{\Lambda,-}_{\mu\nu}+\dots -\frac{1}{6}(\varepsilon_{ij}\bar\theta^i\sigma^{\mu\nu}\theta^j)^2 \Delta \bar X^\Lambda, 
 ~~\Lambda=0,\dots,b_2, \end{equation}
 where the $\theta$'s are superspace coordinates of negative chirality and $\Delta=D_\mu D^\mu$ is the Laplacian.  Under the scaling (\ref{monko}), the field strengths $F^\Lambda_{\mu\nu}$
 are invariant (Dirac quantization of magnetic flux gives a natural normalization of these field strengths, so they should not be rescaled), 
 so the $\theta$'s transform as $\theta\to \lambda^{1/2}\theta$ and hence
 the chiral superspace measure transforms as
 \begin{equation}\label{itono}\d^4\theta\to \lambda^{-2}\d^4\theta. \end{equation}
 The reader will note that although the $X^\Lambda$ parametrize a complex manifold, as is manifest in the description in which
 they satisfy a $\C^*$ gauge invariance and an inequality $N_{\Lambda\Sigma}X^\Lambda \bar X^\SIgma<0$, 
  there are non-holomorphic terms in the expansion (\ref{onkort}).  Part of the reason for this is that 
 the superfields $\X^\Lambda$ and their superspace derivatives
obey a linear nonholomorphic
constraint.

 The kinetic energy of the vector multiplets comes from a holomorphic coupling
 \begin{equation}\label{bronoko}I_0=-i\int \d^4x\d^4\theta\,\F_0(\X^\Lambda). \end{equation}
 For consistency with the scaling (\ref{monko}) and (\ref{itono}), the function $\F_0$, which is called the prepotential, must be homogeneous in the $\X^\Lambda$ of
 degree 2.  The interaction $I_0$ is in fact the case $\g=0$ of the interactions $I_\sg$ (defined in eqn. (\ref{zell})) that are described by the GV formula.  
 
 \def\Sp{{\mathrm{Sp}}}
The vector multiplets can be conveniently described in a $T$-duality invariant language.  This makes some formulas we will need more transparent,
even though, since there is no $T$-duality in M-theory, $T$-duality is not important in the derivation of the GV formula.   One introduces the fields
 \begin{equation}\label{monox}\hat F_\Lambda=\frac{\partial \F_0}{\partial X^\Lambda},\end{equation}
 which transform as $\hat F_\Lambda\to \lambda \hat F_\Lambda$, just like the $X^\Lambda$. One furthermore introduces the symplectic
 form $\Upsilon=\sum_\Lambda \d X^\Lambda \wedge\d \hat F_\Lambda$ and the group $\Sp(2b_2+2,\Z)$
 of integer-valued linear transformations of the whole set of fields 
 \begin{equation}\label{onox} \begin{pmatrix} X^\Lambda\cr \hat F_\Lambda\end{pmatrix} \end{equation}
 that preserve this symplectic form.  From this point of view, the equation (\ref{monox}) can
 be described more symmetrically by saying that the vector multiplets parametrize (the quotient by $\C^*$ of) 
 a $\C^*$-invariant Lagrangian submanifold of $\C^{2b_2+2}$, which we view as a complex symplectic manifold with holomorphic symplectic form $\Upsilon$. One  defines
 \begin{align}\label{yrf}N_{\Lambda\Sigma}&=2 {\rm Im}\,\hat F_{\Lambda\Sigma},~~\hat F_{\Lambda\Sigma}=\frac{\partial^2 \F_0}{\partial X^\Lambda \partial X^\Sigma}\cr
 \mathcal{N}_{\Lambda\Sigma}&=\bar{\hat F}_{\Lambda\Sigma} + i\frac{(NX)_\Lambda(NX)_\Sigma}{(X,NX)}.\end{align}
 Here $(NX)_\Lambda=N_{\Lambda\Sigma}X^\Sigma$ and $(X,NX)=N_{\Lambda\Sigma}X^\Lambda X^\Sigma$. 
These objects \ appear in the kinetic energy of the fields $\X^\Lambda$ after performing $\theta$ integrals. The constraint (\ref{yonko}) can be written in a manifestly symplectic-invariant way:
 \begin{equation}
 \label{symcon}
 i(X^\Lambda\bar{\hat  F_\Lambda} - \hat F_\Lambda\bar X^\Lambda)=-1.
 \end{equation}
 
 The action of $\Sp(2b_2+2,\Z)$ on
 the fields $X^\Lambda,\hat F_\Lambda$ has to be accompanied by linear transformations of the field strengths $F^\Lambda_{\mu\nu}$ and their duals.
 In fact, $\Sp(2b_2 +2,\Z)$ acts linearly on 
  \begin{equation}\label{blonox} \begin{pmatrix} F_{\mu\nu}^{\Lambda+}\cr G^+_{\Lambda\,\mu\nu}\end{pmatrix} \end{equation}
 where the $G_{\Lambda\,\mu\nu}$ are the duals of $F^\Lambda_{\mu\nu}$ which can be defined by\footnote{If $\mathcal{L}$ is the
  Lagrangian density, one can define  $G_\Lambda$ by $G^{-\mu\nu}_\Lambda=2i\frac{\partial \mathcal{L}}{\partial F^{\Lambda-}_{\mu\nu}}$.}:
 \begin{equation}
 G^+_{\Lambda\mu\nu}=\mathcal{N}_{\Lambda\Sigma}F^{\Sigma+}_{\mu\nu}.
 \end{equation}
  
  One advantage of the redundant description via pairs of fields $X^\Lambda,\,\hat F_\Lambda$ and also $F_{\mu \nu}^\Lambda,\,G_{\Lambda\,\mu\nu}$ is that
  this makes it possible to describe the graviphoton field in a manifestly holomorphic and duality-invariant fashion.  The anti-selfdual part of the 4d graviphoton field
  is
  \begin{equation}\label{zobo}\WW^-_{\mu\nu}=2(X^\Lambda G^-_{\Lambda\,\mu\nu}-\hat F_\Lambda F^{\Lambda\,-}_{\mu\nu}).\end{equation} 
  In Lorentz signature, the selfdual part of the graviphoton field is the complex conjugate of this or
    \begin{equation}\label{zobort}\WW^+_{\mu\nu}=2(\bar X^\Lambda G^+_{\Lambda\,\mu\nu}-\overline{\hat F}_\Lambda F^{\Lambda\,+}_{\mu\nu}).\end{equation}
    In Euclidean signature, $\WW^+_{\mu\nu}$ is not the complex conjugate of $\WW^-_{\mu\nu}$, but these formulas remain valid.   Of course, we are interested
    in an anti-selfdual graviphoton background in which $\WW^+_{\mu\nu}=0$.

It is convenient to describe the anti-selfdual and selfdual parts of the graviphoton field using spinor indices, defined by
\begin{equation}\label{moffit} \WW^-_{ A B}=\gamma^{\mu\nu}_{ A B}\WW^-_{\mu\nu},~~ \WW^+_{\dA \dB}=\gamma^{\mu\nu}_{\dA \dB}\WW^+_{\mu\nu},~~ \gamma^{\mu\nu}
=\frac{1}{2}[\gamma^\mu,\gamma^\nu]. \end{equation}    
Here $A,B=1,2$ and $\dA, \dB=1,2$ are respectively spinor indices of negative and positive chirality.
Eqn. (\ref{zobo}) shows that $\WW^-_{ A B}$ scales with degree 1 under the scaling $X^\Lambda\to\lambda X^\Lambda$. More precisely,
$\WW^-_{ A B}$ is an anti-selfdual two-form valued in the pullback to spacetime of a holomorphic line bundle $\L$ over the vector multiplet moduli space;
$\L$ is characterized by the fact that $\WW^-_{AB}$ transforms with charge 1 under scaling. In terms
of superfields,  $\WW^-_{ A B}$ is
the bottom component of a chiral superfield $\W_{ A B}$ that likewise transforms with charge 1 under the equivalence (\ref{monko}), and  
$\W^2=\W_{ A B}\W^{ A B}$ is similarly a chiral
superfield of charge $2$.   To make the charges balance, in  the interaction $I_\sg$
\begin{equation}\label{ozell}I_\sg=-i\int_{\R^4} \d^4x\,\d^4\theta\,\F_\sg(\X ^\Lambda)(\W_{ A B}\W^{ A B})^{\sg}\end{equation}
that enters the GV formula, the functions $\F_\sg$ must be homogeneous of degree $2-2\g$.

\subsubsection{Background Gauge Fields In $d=4$}\label{dfour}

Next we  explain the 4d analogs of some observations that were made in section \ref{genany} for $d=5$.

In the supersymmetric graviphoton background,  the linear combinations of 
field strengths $\F^\Lambda_{\mu\nu}$ defined in (\ref{fcal}) and appearing as components of the vector superfields $\X^\Lambda$ must vanish.
This gives the very important relation
\begin{equation}
\label{vec4dback}
F^\Lambda_{\mu\nu} = \frac{1}{2}\bar{X}^\Lambda \WW^-_{\mu\nu},
\end{equation}
which is the 4d analog of the corresponding 5d statement $F^I_{\mu\nu}=h^I\TT^-_{\mu\nu}$.

A 4d particle with charges $\vec q=q_0,\dots, q_{b_2}$  
couples to the effective gauge field given by the linear combination $A_\mu(\vec q) =q_\Lambda A^\Lambda_\mu$. The field strength of $A_\mu(\vec q)$ in the
graviphoton background is 
\begin{equation}
\label{effield4d}
F_{\mu\nu}(\vec q)=\frac{1}{2}q_\Lambda \bar X^\Lambda \WW^-_{\mu\nu}=\frac{1}{4}\bar{\cZ  }
(\vec q)\WW^-_{\mu\nu}.
\end{equation}
Thus, the effective field strength seen by such a particle in the graviphoton background is proportional to $\bar \cZ  (\vec q)$, the complex conjugate
of the central charge $\cZ  (\vec q)=2\sum_\Lambda q_\Lambda X^\Lambda$.  This statement is the generalization of the fact (eqn. (\ref{zobel})) that in $d=5$, the effective field strength is proportional
to the 5d central charge $\zeta$.  

Now, the mass of a BPS particle in $d=4$ is $m(\vec q)=|\cZ  |$.  This means that the dimensionless ratio $F(\vec q)/m(\vec q)^2$
that appears in the Schwinger calculation is proportional to $1/\cZ  $ and in particular is holomorphic.  This is part of the mechanism by which a holomorphic
answer emerges from the Schwinger
calculation that we perform in terms of 4d fields in section \ref{schwfields}, even though the particle masses are certainly not holomorphic in $\cZ  $.

\subsubsection{Comparison To Perturbation Theory}\label{compper}

Now let us explain why in Type IIA superstring perturbation theory, $\F_\sg$ is generated only in genus $\g$.  We practice first with the case $\g=0$,
corresponding to the classical approximation.  We have written the above formulas in 4d Einstein frame, in which the dilaton (which is in a hypermultiplet)
does not couple directly to the $\F_\sg$'s, which govern vector multiplets.  To compare to string perturbation theory, we must transform to the string frame, which we do
by a Weyl transformation of the metric $g_{\mu\nu}\to e^{-2\phi} \hat g_{\mu\nu}$, where $\hat g_{\mu\nu}$ is the metric in string frame, 
$\phi$ is the four-dimensional dilaton, and the string coupling constant is
$g_{st}=e^\phi$.  The Einstein-Hilbert action $\frac{1}{2\kappa_4^2}\int \d^4x\sqrt g R$ becomes
\begin{equation}\label{menz}\frac{1}{2\kappa_4^2}\int \d^4x \sqrt{\hat g}e^{-2\phi}R(\hat g)\end{equation}
and is generated in string theory in genus 0.
More generally, any genus $\g$ contribution to the effective action for external fields from the 
Neveu-Schwarz (NS) sector only is proportional to $g_{st}^{2\sg-2}=\exp((2\g-2)\phi)$.
But a genus $\g$ contribution to the effective action that has in addition $k$ external Ramond-Ramond (RR) gauge field strengths (normalized in the standard way
to satisfy ordinary Dirac quantization and standard Bianchi identities) is proportional to $g_{st}^{2\sg-2+k}=\exp((2\g-2+k)\phi)$.  Let us see how this
works for $\F_0$.  Performing the $\theta$ integrals in (\ref{bronoko}) gives a variety of terms, among them
\begin{equation}\label{ifonko}I_0=\int \d^4x \sqrt{\hat g}\left(e^{-2\phi}\frac{\partial \F_0}{\partial X^\Lambda}\hat\Delta \bar X^\Lambda+\frac{\partial^2\F_0}{\partial
X^\Lambda\partial X^\SIgma}\F^{\Lambda\,-}_{\mu\nu} \F^{\Sigma\,-\,\mu\nu}+\dots\right),\end{equation}
where $\hat\Delta=\hat g^{\mu\nu}\hat D_\mu\hat D_\nu$ is the Laplacian defined with the string metric.  
Bearing in mind that $X^\Lambda$ is described by an NS-NS vertex operator and  $\F^{\Lambda\,-}=F^{\Lambda\,-}-\frac{1}{2}\bar{X}^\Lambda \WW^-+\dots$ is described by an  RR vertex operator, we see that in Type IIA superstring theory, such interactions can be generated only in genus 0.  The analog of this for $\F_\sg$ is immediate.
All we have to know is that $\WW^-_{\mu\nu}\WW^{-\,\mu\nu}=g^{\mu\mu'}g^{\nu\nu'}\WW^-_{\mu\nu}\WW^-_{\mu'\nu'}$ acquires a factor of $e^{4\phi}$ in transforming to the
string frame and that $\WW^-_{\mu\nu}$ comes from the Ramond-Ramond sector.
Performing $\theta$ integrals in the definition of $I_\sg$ gives a variety of terms such as
\begin{equation}\label{ifonkor}I_\sg=\int \d^4x \sqrt{\hat g}\left(\left(e^{-2\phi}\frac{\partial \F_\sg}{\partial X^\Lambda}\hat\Delta \bar X^\Lambda+\frac{\partial^2\F_\sg}{\partial
X^\Lambda\partial X^\SIgma}\F^{\Lambda\,-}_{\lambda\sigma}\F^{\Sigma\,-\,\lambda\sigma}+\dots\right)e^{4\sg\phi}(\WW^{-\mu \nu}\WW_{-\mu\nu})^\sg+\dots\right),\end{equation}
and we see the expected scaling for interactions that in superstring perturbation theory are generated only in genus $\g$.

\subsubsection{Shift Symmetries}\label{shiftsym}

The full power of the duality-invariant formalism sketched in section \ref{dualinv} is not really needed for our problem.    The reason is that
reduction on a circle from five to four dimensions gives a natural duality frame, defined by the fields that arise in classical dimensional
reduction from the classical gauge and gravitational fields in five dimensions.  Moreover, among the $U(1)$ gauge fields in four dimensions,
there is a distinguished one  $A^0_\mu=-B_\mu$, which arises from the components $g_{\mu 5}$ of the five-dimensional metric. The
other 4d gauge fields  arise in Kaluza-Klein reduction of the 5d gauge fields $V^I$:
\begin{equation}\label{melx}V^I=\sum_{\mu=1}^4 A_\mu^I\d x^\mu+ \alpha^I \left(\d y+ B_\mu \d x^\mu\right),~~I=1,\dots, b_2. \end{equation}
Here the $A_\mu^I$ are gauge fields in four dimensions, and the $\alpha^I$ are scalars.
The holonomy of $V^I$ around the Kaluza-Klein circle is $\exp(2\pi i\alpha^I)$, so we expect a symmetry 
\begin{equation}\label{solfo}\alpha^I\to \alpha^I+n^I. \end{equation}
This shift in $\alpha^I$ is generated by a gauge transformation  $\exp(in^I y)$ of $A^I$ 
 together with a redefinition of the gauge fields
\begin{equation}\label{zelx} A_\mu^I\to A_\mu^I+n^I A^0_\mu=A_\mu^I-n^IB_\mu, ~~n^I\in\Z.  \end{equation}
Thus although the definition of $A^0_\mu$ is completely natural, $A_\mu^I$ is only well-defined up to an integer multiple of $A^0_\mu$.

The gauge fields $A^0_\mu$ and $A_\mu^I$, or rather their field strengths $\d A^0$ and $\d A^I$, appear in the superfields
$\X^0=X^0+\dots +\theta^2 \d A^0_\mu+\dots$ and $\X^I=X^I+\dots+\theta^2 \d A^I+\dots$.   The symmetries (\ref{zelx}) extend
to symmetries of the superfields
\begin{equation}\label{omelx} \X^I\to \X^I+n^I \X^0,~~ n^I\in \Z.\end{equation}
These transformations (accompanied by corresponding transformations of the derivatives $\partial\F_0/\partial\X^\Lambda$)
are the only  $\Sp(2b_2+2,\Z)$ duality transformations 
that are important in the derivation of the GV formula.  

\def\S{{\mathcal S}}
The ratios $\ZZ^I=\X^I/\X^0$ are invariant under scaling of the homogeneous coordinates 
$\X^\Lambda$ and parametrize the vector multiplet moduli space. They transform simply under (\ref{omelx}):
\begin{equation}\label{ofgo}\ZZ^I\to \ZZ^I+n^I,~~n^I\in\Z. \end{equation}
Of course, these shifts really act only on the bottom components of the $\ZZ^I$, which we call $Z^I$:
\begin{equation}\label{wofgo}Z^I\to Z^I+n^I, ~~~n^I\in\Z. \end{equation}
The $Z^I$ have  a simple interpretation.  Consider an M2-brane wrapped on $p\times S^1\times\Sigma\subset \R^4\times S^1\times Y$,
where $p$ is a point in $\R^4$ and $\Sigma\subset Y$ is a holomorphic curve.  Such an M2-brane is a supersymmetric cycle,
and its action must be a holomorphic function of the vector multiplet moduli.  The charges of the wrapped M2-brane are
$q_I=\int_\Sigma\omega_I$ (the $\omega_I$ were introduced in eqn. (\ref{omex})),
and its mass is given\footnote{We work in units in which the M2-brane tension is 1.}  by the central 
charge $\zeta(\vec q)=\sum_I v^Iq_I$ (which will be positive for a supersymmetric cycle).
The real part of the action is simply the mass times the circumference of the 
Kaluza-Klein circle.  If we write the metric of $\R^4\times S^1\times Y$ in the M-theory description as
\begin{equation}\label{mth} \d s^2_{\Mth}= \d s^2_{10}+e^{2\gamma}(\d y + B_\mu dx^\mu)^2. \end{equation}
then the circumference is $2\pi e^\gamma$, so the real part of the M2-brane action is
$2\pi e^\gamma \sum_I v^Iq_I$.
On the other hand, the imaginary part of the action is just the $C$-field period 
$-\int_{p\times S^1\times \Sigma}C$.  Recalling the definition
(\ref{monoxic}) of the 5d gauge fields $V^I$, we see that this period is $2\pi \alpha^Iq_I$.  
So the Euclidean action of the wrapped M2-brane is
\begin{equation}\label{nofgo} S(\vec q)=2\pi\sum_I q_I \left(e^\gamma v^I-i\alpha^I\right). \end{equation}

It is convenient to re-express this formula in terms of the circumference of the M-theory circle defined in 5d supergravity.  
The reason this is not the same as the circumference $2\pi e^\gamma$ measured in the 11d metric
 is that compactification from 11 dimensions to 5 dimensions on the Calabi-Yau manifold $Y$ gives a 5d gravitational action
$\int \d^5x \sqrt g_\Mth \V R(g_\Mth)$, where $\V$ is the volume of $Y$ in the eleven-dimensional description, $g_\Mth$ is the 5d metric
in that description, and $R(g_\Mth)$ is the Ricci scalar.  In 5d supergravity, it is convenient and
usual to make a Weyl transformation to Einstein frame, replacing $g_\Mth$ by a 5d-metric via 
$g_\Mth=\V^{-2/3}g_{\5d}$.  The relation of $e^\gamma$
to the radius $e^\sigma$ of the circle in the 5d description is thus $e^\gamma=e^\sigma/v$, where 
$v=\V^{1/3}$.  Thus we rewrite (\ref{nofgo}) in 5d terms:
\begin{equation}\label{pofgo} S(\vec q)=2\pi\sum_I q_I \left(e^\sigma h^I-i\alpha^I\right),~~~ h^I=\frac{v^I}{v}. \end{equation}
The coefficients $e^\sigma h^I-i\alpha^I$ in $\S(\vec q)$ must be holomorphic functions of the vector multiplet moduli, or in other words
of the $Z^I$.  Comparing the transformations (\ref{solfo}) and (\ref{ofgo}), we find the relationship
\begin{equation}\label{merot}Z^I=\alpha^I+ie^\sigma h^I, \end{equation}
which describes the background values of the superfield $\ZZ^I$.  The action is thus $S(\vec q)=-2\pi i \sum_Iq_I Z^I$.  However,
 it is more convenient to introduce a superfield $\S(\vec q)$ whose bottom component is $S(\vec q)$:
 \begin{equation}\label{merzo}\S(\vec q)=-2\pi i \sum_I q_I \ZZ^I. \end{equation}
 This is more convenient because when we perform an actual computation in section \ref{schwpart}, 
 a particle propagating around the circle
 has fermionic as well as bosonic collective coordinates.  Writing the action as a superfield is an easy way to incorporate the fermionic
 collective coordinates.

In the 5d description, we write the real part of the action as $2\pi e^\sigma m(\vec q)$, with $m(\vec q)$ 
the mass of the wrapped M2-brane in that
description:
\begin{equation}\label{merf} m(\vec q)=\sum_I q_I h^I. \end{equation}
Since the wrapped M2-brane is BPS, this mass also equals the central charge in the 5d description:
\begin{equation}\label{erf}\zeta(\vec q)=\sum_I q_I h^I. \end{equation}

Eqn. (\ref{merot}) states in particular that $\mathrm{Im}\,Z^I=h^Ie^\sigma$.  
Recalling the constraint (\ref{norf}), this implies the useful relation
\begin{equation}\label{zerf}\CC_{IJK}\mathrm{Im}\,Z^I\,\mathrm{Im}\,Z^J\,\mathrm{Im}\,Z^K=e^{3\sigma}.\end{equation}

\subsubsection{Validity Of The Calculation}\label{omix}

In M-theory on $\R^4\times S^1\times Y$, we will perform a computation involving M2-branes wrapped on $S^1\times \Sigma$ where $\Sigma$
is  a non-trivial cycle in $Y$.    Our aim here is to describe the range of validity of the computation, and explain why this suffices to determine
the full answer.

For M-theory to be a reasonable description, we would like $Y$ not to be sub-Planckian, so we can ask for its volume $\V_\Mth$ in M-theory units
not to be sub-Planckian.  If we are not too close to a boundary of the Kahler cone of $Y$, then a wrapped M2-brane has a size of order $\V_\Mth^{1/6}$.
To justify a calculation in which wrapped M2-branes propagating around $S^1$ are treated as elementary particles, we would like the $S^1$
to be much larger than the size of the M2-branes, which will be generically of order $\V_\Mth^{1/6}$.  So we want 
\begin{equation}\label{wanted} e^\gamma>> \V_\Mth^{1/6}\gtrsim 1,\end{equation}
where as in (\ref{mth}), $e^\gamma$ is the radius of the M-theory circle in M-theory units.

\def\IIA{{\mathrm{IIA}}}
When we relate M-theory on $\R^4\times S^1\times Y$ to Type IIA superstring theory on $\R^4\times Y$, the ten-dimensional string coupling constant
$g_{10}=e^\phi$ is related to $\gamma$ by \cite{wittenold} 
\begin{equation}\label{zemilo} g_{10}=e^{3\gamma/2}. \end{equation} 
Moreover, the metric of $\R^4\times Y$ in the Type IIA  description is
\begin{equation}\label{rth}\d s^2_{\IIA}=e^\gamma \d s^2_{10}.\end{equation}
In particular, the volume of $Y$ in the string theory description is (recall that $e^\gamma=e^\sigma/v$)
\begin{equation}\label{ytro}\V_\IIA =e^{3\gamma}\V_\Mth=e^{3\sigma}. \end{equation}
Eqns. (\ref{wanted}), (\ref{zemilo}), and (\ref{ytro}) show that in the region in which our computation is valid, $g_{10}$ and $V_\IIA$ are both large.
In particular, the fact that $g_{10}$ is large means that, as expected, string perturbation theory is not useful in the region in which our calculation is valid.
 Moreover, as explained in section \ref{ha}, the fact that $V_\IIA$ is large 
means that we will not encounter the holomorphic anomaly.
Notice from (\ref{zemilo}) and (\ref{ytro}) that
\begin{equation}
\label{g_10_4}
g_{10}=e^{3\sigma/2}\frac{1}{\sqrt{\V_M}}=e^{3\sigma/2}g_\st,
\end{equation}
where $g_\st$ is a 4-dimensional string coupling introduced in section \ref{compper}.

The interactions $\F_\sg(\X)$, when expressed in string frame with Kahler moduli (and hence $\V_{\mathrm{IIA}}$)  held fixed, 
have a known dependence on $g_\st$, as explained in section \ref{compper}. Due to (\ref{g_10_4}), they have a known dependence on $g_{10}$ as well.  So a calculation
that is only valid for $g_{10}>>1$ can suffice to determine them.

Because of holomorphy, the same is true of a calculation that is only valid for large $\V_\IIA$.  To explain this, we use the homogeneity of $\F_\sg$ to write
$\F_\sg(\X)=(\X^0)^{2-2\sg}\Phi(\ZZ^1,\dots,\ZZ^{b_2})$. To avoid clutter, in the following argument, we take $b_2=1$, so there is only one $\ZZ$.
 Also we write $\ZZ=\alpha+i\beta$
where $\beta$ is defined in eqn. (\ref{merot}).  The shift symmetry $\ZZ\to \ZZ+n$, $n\in\Z$, implies that the general form of $\Phi(\ZZ)$ is
$\Phi(\ZZ)=\sum_{n\in \Z}c_n \exp(2\pi i n \ZZ)$, with constants $c_n$.  (Moreover, these constants vanish for $n<0$, 
since an exponential blowup for large volume would contradict what we know from supergravity.)
We can write
$\Phi(\ZZ)=\sum_{n=0}^\infty f_n(\beta)\exp(2\pi in\alpha)$, where $f_n(\beta)=c_n\exp(-2\pi n\beta)$.  Since each $f_n(\beta)$ has a known dependence on $\beta$,
if we can compute these functions for large $\beta$, this will suffice to determine the whole function $\Phi(\ZZ)$.  But large $\beta$ is precisely the large
volume region in which the Schwinger-like computation  is valid, so  that computation can suffice to determine
$\F_\sg(\X)$.

\subsubsection{Classical Reduction From Five Dimensions}\label{classred}

As explained in section \ref{detprinc} of the introduction, it is important to know to what extent the $I_\sg$'s or equivalently the 
$\F_\sg$'s  can arise by classical dimensional
reduction from five dimensions.  We will  describe the two contributions that are known and  explain why they are the only ones.

After performing the $\theta$ integrals, $I_0$ contributes a term to the effective action with two derivatives, so a classical contribution to $I_0$
  must come from the
two-derivative part of the effective action in five dimension, or in other words from the minimal supergravity action with vector multiplets. 
This contributes the much-studied classical prepotential
\begin{equation}\label{omox}\F_0^{\mathrm{cl}}(\X)=-\frac{1}{2}\sum_{IJK}\CC_{IJK}\frac{\X^I\X^J\X^K}{\X^0}.\end{equation}
With this prepotential,  and with the help of eqn. (\ref{zerf}), one finds that 
the constraint (\ref{yonko}) implies that $|X^0|=e^{-3\sigma/2}/2$.  We choose the phase so that 
\begin{equation}
\label{x0val}
X^0=-\frac{i}{2}e^{-3\sigma/2}.
\end{equation}
After performing $\theta$ integrals,  the four-dimensional  action that follows from  $\F^{\mathrm{cl}}_0(\X)$
includes  kinetic terms for the gauge fields:
\begin{equation}\label{tomow}-\frac{i}{4\pi}\int\d^4x \sqrt g \mathcal{N}_{\Lambda\Sigma}g^{\mu\mu'}g^{\nu\nu'}F^{\Lambda+}_{\mu\nu}F^{\Sigma+}_{\mu'\nu'}+c.c.
\end{equation}
Using (\ref{yrf}) for $\mathcal{N}_{\Lambda\Sigma}$ and (\ref{merot}) for the ratios $X^K/X^0$, we find a parity-violating part of  the kinetic term
\begin{equation}\label{omow}\mathcal I^-= -\frac{3}{2\pi}\sum_{IJK}\CC_{IJK}\int_{\R^4} \alpha^K (F^I-\alpha^I F^0)\wedge (F^J-\alpha^J F^0)\end{equation}
and a parity-conserving part
\begin{equation}\label{lomow}\mathcal I^+=-\frac{1}{2\pi}\sum_{IJ}\int_{\R^4}  e^\sigma a_{IJ} (F^I-\alpha^I F^0)\wedge \star (F^J-\alpha^J F^0)-\frac{1}{4\pi}\int_{\R^4}e^{3\sigma}F^0\wedge \star F^0, \end{equation}
where
\begin{equation}
\label{aIJ}
a_{IJ}=-3\CC_{IJK}h^K + \frac{9}{2}h_I h_J
\end{equation}
is the metric on the Kahler cone defined in footnote \ref{kelter}.
The parity-violating contribution descends from a Chern-Simons interaction
\begin{equation}
\label{mixl} -\frac{1}{(2\pi)^2}\sum_{IJK} \CC_{IJK}\int V^I\wedge \d V^J\wedge \d V^K\end{equation}
in five dimensions.
Notice that although $\mathcal I^-$ is not left fixed by the shift symmetries $\alpha^I\to \alpha^I+n^I$,
it changes only by a topological invariant,  so its contribution to the classical equations of motion does respect the shift symmetries.
As is usual in such problems, at the classical level, the shift symmetries are continuous symmetries with no restriction for the
 $n^I$ to be integers.  (Quantum mechanically, the shift symmetries are broken
to discrete symmetries by M2-brane instanton effects that will be studied in section \ref{schwpart}.)
The parity-conserving term $\mathcal I^+$  descends from the gauge theory kinetic energy $-\frac{1}{2}\int a_{IJ} \d V^I\wedge \star \d V^J$ in five dimensions,
along with the Einstein-Hilbert action, which contributes to the kinetic energy of $A^0$.

What about 
$I_1$?  This interaction contributes four-derivative terms to the effective action in four dimensions, so a classical contribution to $I_1$ comes from a term in the
five-dimensional effective action with four derivatives.  In eleven-dimensional M-theory, essentially only one multi-derivative
correction to the minimal two-derivative supergravity action is known.  This is a term \begin{equation}\label{dofos}\Delta I=\frac{1}{(2\pi)^4}\int C\wedge \Big[\frac{1}{768}(\Tr\,R^2)^2 - \frac{1}{192}\Tr\,R^4\Big]\end{equation}
(where $R$ is the Riemann tensor, which is viewed as a matrix-valued two-form in defining the trace) that was originally found \cite{duff} by its role
in anomaly cancellation in the field of an M5-brane.\footnote{The full supersymmetric completion of this coupling is not known.  
We normalize the $C$ field so that the periods of its curvature $G=\d C$ differ by integer multiples of $2\pi$.   See
 \cite{deAlwis:1996ez, deAlwis:1996hr} for detailed formulas with other conventions.} In compactification on $M_5\times Y$ (where for us $Y$ will be a Calabi-Yau three-fold and $M_5=\R^4\times S^1$), 
we consider a contribution to $\Delta I$ with two factors of $R$ tangent to $Y$ and the other two tangent to $\R^4\times S^1$.   This contribution
generates a Chern-Simons interaction in five dimensions $\frac{1}{16\cdot 24\pi^2}\sum_I c_{2,I}\int V^I\wedge \Tr \,R\wedge R$, where 
\begin{equation}\label{tefron} c_{2,I}=\frac{1}{16\pi^2}\int_Y \omega_I\Tr\,R\wedge R \end{equation}
are the coefficients of the second Chern class $c_2(Y)$ in a basis dual to the $\omega_I$.  Upon reduction to four dimensions, that Chern-Simons
coupling becomes
\begin{equation}\label{efron}\frac{1}{16\cdot 12\pi}\sum_I c_{2,I}\int \alpha^I\Tr\,R\wedge R. \end{equation}
Again, this possesses the shift symmetry, modulo a topological invariant.  This four-dimensional interaction can be derived from 
\begin{equation}\label{zefron}\F_1(\X)=-\frac{i}{64\cdot 12\pi}\sum_I c_{2,I}\frac{\X^I}{\X^0} =-\frac{i}{64\cdot 12\pi}\sum_I c_{2,I}\ZZ^I.  \end{equation}

Could there be other classical contributions to $\F_1(\X)$, apart from this known contribution?  Since $\F_1(\X)$ is invariant under scaling,
it is a function of the ratios $\ZZ^I=\X^I/\X^0$.  A term in $\F_1$ that is quadratic or higher order in the $\ZZ^I$ would violate the classical shift symmetries.
We already know about the linear terms.
What about a constant contribution?  
Depending on whether the constant is real or imaginary, it would contribute a parity-violating interaction $\int\, \Tr\,R\wedge R$ or a 
parity-conserving one $\int\, \Tr\,R\wedge \star R$.  The parity-violating contribution must be absent, since M-theory conserves parity.  To 
generate a parity-conserving $R^2$ interaction by classical dimensional reduction, we would have to start with $\int \Tr\,R\wedge\star R$ in five 
dimensions, but reduction of this to four dimensions gives $\int\,e^\sigma\Tr\,R\wedge\star R$, with an unwanted factor of $e^\sigma$.  This 
factor  is  absent in the four-dimensional effective interaction associated to a constant $\F_1$.  
Thus,  there is no way to generate classically a constant contribution to $\F_1$.

What about $\F_\sg$ for $\g>1$?  No classical contributions are known and we claim that there are none.  This actually almost follows from the shift symmetries.
The shift symmetries imply that $\F_\sg$ must be independent of the ratios $\ZZ^I$ and hence must be $d_\sg(\X^0)^{2-2\sg}$, for some constant $d_\sg$.

To show the vanishing of these constants, one may use a scaling argument
similar to the one used above for $\F_1$.  
For this, one observes that for $\g>1$,
$I_\sg$ generates among other things a  4d coupling $R^2F^{2\sg-2}$ in which indices are contracted using only the metric tensor and not the 4d 
Levi-Civita tensor.\footnote{The case $\g=1$ is exceptional partly because this statement fails for $\g=1$ if $\F_1$ is a real constant.} Such couplings can be lifted to $R^2F^{2\sg-2}$
couplings in five dimensions, but as we found for $\g=1$, the dimensional reduction of those couplings  to four dimensions does not give the 
power of $e^\sigma$ that is needed to match $I_\sg$.  Alternatively, one may  argue as follows using the fact that the 4d couplings derived from $I_\sg$ with $\g>1$ also include terms that
are of odd order in the  Levi-Civita tensor
$\veps_{\mu\nu\alpha\beta}$, and depend only on the scalars $X^\Lambda$, the field strengths $F^I$, and the Riemann tensor $R$ (and not their derivatives).   We can see as follows that these terms do not arise by reduction of covariant, gauge-invariant couplings in five dimensions.
To get such terms by reduction, the starting point should be 5d interactions
 that are local, generally covariant, and gauge-invariant, and odd order in the 5d Levi-Civita tensor (so that their reduction to $d=4$ will
be odd order in the 4d Levi-Civita tensor).
A 5d interaction that is covariant and odd order in the Levi-Civita tensor  cannot be the integral of a polynomial 
in the field strengths $F^I_{\mu\nu}$ and the Riemann tensor $R_{\mu\nu\alpha\beta}$;
such a polynomial would have an even number of indices and there would be no way to contract 
them with the help of any number of copies of the metric
tensor and an odd number of Levi-Civita tensors. We do not want to use covariant derivatives 
of $F$ or $R$, since then we will get covariant derivatives in
$d=4$.   So we have to start with a 5d interaction that  is gauge-invariant and local (meaning that its variation is a gauge-invariant local function) but is not the integral of a gauge-invariant
local density.  The only such interactions are the standard 
Chern-Simons interactions.\footnote{This fact is widely used but not always explained.  Let us say that a generally covariant interaction $U$ has weight $n$ if it scales as $U\to e^{n\lambda}U$
under a global Weyl transformation $g_{\mu\nu}\to e^\lambda g_{\mu\nu}$ (with constant $\lambda$).  Any generally covariant interaction is a linear combination of interactions of definite weight.  If $U$
has non-zero weight $n$ and its variation is a local gauge-invariant functional of the metric and other fields, then $U$ is the integral of a gauge-invariant local density, since
$U=(1/n)\int\d^5x\sqrt g \, g_{\mu\nu}\frac{\delta}{\delta g_{\mu\nu}}U$.  Interactions of weight 0 are very few (in any odd dimension and in particular in dimension 5, they are all parity-odd and proportional to the Levi-Civita tensor).  It is not
hard to see by hand that the standard Chern-Simons functions are the only interactions $U$ of weight 0 
 whose variations are local and gauge-invariant but that are not themselves
integrals of locally-defined gauge-invariant densities.  One can approach this last question by first classifying the possible variations of $U$; in five dimensions,
these must be bilinear in the gauge field strength $F$ or the Riemann tensor $R$.} We have already  analyzed their contributions.

In short, only some  very special and known contributions to $\F_0$ and $\F_1$ can arise by classical reduction from five dimensions.
Everything else can be determined by a Schwinger-like calculation.

 \section{The Schwinger Calculation With Particles}\label{schwpart}
 
 \subsection{Overview}
 
 In this section, we come finally to the Schwinger-like calculation for massive BPS states in five dimensions.  We view an M2-brane wrapped on
 $p\times S^1\times \Sigma\subset \R^4\times S^1\times Y$ (where $p$ is a point in $\R^4$ and  $\Sigma$ is a holomorphic curve in $Y$) as a supersymmetric
 instanton.  We work in the regime that
 the radius of $S^1$ is large, so the M2-brane can be treated as a point particle coupled to 5d supergravity. The particle action is uniquely determined
 by supersymmetry modulo irrelevant terms of higher dimension (once the graviphoton is turned on, this assertion depends on the extended supersymmetry of the graviphoton background), and this
 makes our considerations simple.
 In appendix \ref{holc}, we show explicitly how the supersymmetric particle action emerges from an underlying M2-brane action.  
   
In our computation, we consider only the leading order approximation to the particle action, ignoring all sorts of couplings of higher dimension,
and we integrate over fluctuations around the classical particle orbit in a one-loop approximation.  This computation gives a result with the correct
dependence on the radius $e^\sigma$ of the $S^1$ to  contribute to the chiral interactions $\F_\sg(\X)$.
The terms we neglect all have extra powers of $e^{-\sigma}$ and so cannot contribute to those couplings.

In section \ref{hyperplet}, we consider the basic example of a massive BPS hypermultiplet that arises from wrapping an M2-brane on an isolated genus 0 holomorphic
curve in $\Sigma$.  We evaluate the contribution to the GV formula for the case that
 (as tacitly assumed above and in the introduction) the BPS state wraps just once around the $S^1$.
In section \ref{winding}, we explain what multiple winding means in this context and thereby get the full contribution of the hypermultiplet to the GV formula.
In section \ref{moregenmass}, we evaluate the contribution of arbitrary massive BPS states to the GV formula.
This generalization is rather simple because of the extended supersymmetry of the graviphoton background. 

As explained in the introduction, the treatment of BPS states that are {\it massless} in five dimensions requires a different approach, based on fields rather than particles.
This is presented in section \ref{schwfields}.

\subsubsection{A Detail}

We perform our calculation in terms of the variables $Z^I$ and $\sigma$, and use the classical  constraint $X^0=-(i/2)e^{-3\sigma/2}$ (eqn. (\ref{x0val}))
to express the results in a manifestly supersymmetric fashion in terms of superfields $\X^\Lambda$.  But  that classical constraint equation  is derived
from the classical prepotential $\F_0$.  Since  we will be computing in particular
instanton corrections to $\F_0$, the constraint equation actually has instanton corrections.  So supersymmetry actually implies
that, when the effective action is expressed in terms of $Z^I$ and $\sigma$, it has  multi-instanton corrections and also instanton/anti-instanton corrections
(that is, it has terms whose $Z$-dependence corresponds to effects of multiple BPS particles and/or antiparticles).  These terms cannot be confused
with 1-instanton or multi-instanton corrections to the $\F_\sg$'s, because they have the wrong dependence on $\sigma$ (they vanish too rapidly for large $\sigma$).
But are there multi-instanton corrections to the $\F_\sg$'s, as opposed to the 1-instanton contributions that we will evaluate?  One expects 
the answer to this question to be ``no,'' because the interactions among massive
BPS particles are irrelevant at long distances, as we will explain for a related reason in section \ref{winding}.

 \subsection{Massive Hypermultiplet}\label{hyperplet}
 
 \subsubsection{The Free Action}\label{thaction}

The central charge $\zeta$ is real in five dimensions.  
The mass $M$ of a BPS particle is $M= |\zeta|$, 
and there are two types of massive BPS particles, with $\zeta>0$ or $\zeta<0$.  They arise from M2-branes and their antibranes wrapped on a holomorphic
curve $\Sigma\subset Y$ (equivalently, they arise from M2-branes wrapped on $\Sigma$ with positive or negative orientation).  
We will consider the case $\zeta>0$,
and we define what we mean by M2-branes (as opposed to antibranes) by saying that this case corresponds to wrapped M2-branes.  

Let us consider a supermultiplet consisting of  BPS particles of mass $M=\zeta$ at rest.  The supersymmetry algebra reduces to
\begin{equation}\label{moxicx}\{Q_{Ai},Q_{Bj}\}=2M\varepsilon_{ij}\varepsilon_{AB},~~\{Q_{\dA i},Q_{\dB j}\}=0=\{Q_{A i},Q_{\dA j}\}. \end{equation}
In a unitary theory, the vanishing of $\{Q_{\dA i},Q_{\dB j}\}$ implies that the operators $Q_{\dA i}$ annihilate the whole supermultiplet.  On
the other hand, the operators $Q_{A i}/\sqrt {M}$ generate a Clifford algebra.  The irreducible representation of this Clifford algebra consists
of two bosonic states transforming as $(0,0)$ under $SU(2)_\ell\times SU(2)_r\cong \Spin(4)$ and two fermionic states transforming as $(1/2,0)$.  
These four states make up a massive BPS hypermultiplet (of positive central charge).   In M-theory compactified to five dimensions on $Y$, such a massive hypermultiplet
arises from an M2-brane wrapped on an isolated genus 0 curve $\Sigma\subset Y$.  If by ``supermultiplet,'' we mean a set of states that provide an irreducible representation of the full superalgebra of
spacetime symmetries (including the rotation group $\Spin(4)\cong SU(2)_\ell\times SU(2)_r$), then a general BPS supermultiplet at rest consists of the tensor product of the states of a massive hypermultiplet with some
representation $(j_\ell,j_r)$ of $SU(2)_\ell\times SU(2)_r$.  We consider a hypermultiplet here, and analyze the contribution to the GV formula of a general BPS supermultiplet in section \ref{moregenmass}.

We ultimately will perform a one-loop calculation involving small
fluctuations around a particle trajectory of the form
$p\times S^1\subset \R^4\times S^1$, for some $p\in\R^4$.  In this one-loop approximation, the BPS particle is nearly at rest, meaning that it can
be treated nonrelativistically.  So we can approximate the Hamiltonian as $H=M+\hH$, where $\hH=\sum_{\mu=1}^4P_\mu^2/2M$ is the nonrelativistic
Hamiltonian; here $P_\mu,\,\mu=1,\dots,4$ is the momentum. We replace the supersymmetry
algebra (\ref{mofo}) with its nonrelativistic limit
\begin{align}\label{tofo} \{Q_{Ai},Q_{Bj}\}&=2M\veps_{AB}\veps_{ij} \cr
                                        \{Q_{\dA i}, Q_{\dB j}\}&=\veps_{\dA\dB}\veps_{ij}\frac{P^2}{2M} \cr
                                         \{Q_{A i},Q_{\dA j}\}&= -i\Gamma^\mu_{A\dot A}\veps_{ij}P_\mu. \end{align}
The momentum $P_\mu$ commutes with all supersymmetry generators, as does the nonrelativistic Hamiltonian $\hH$:
\begin{align}\label{rofo} [\hH,Q_{Ai}]=[\hH,Q_{\dA j}]=[\hH,P_\mu]=0. \end{align}

Eqn. (\ref{tofo}) tells us that $\psi_{Ai}=Q_{Ai}/M\sqrt 2$ (the normalization will be convenient)
obeys fermion
 anticommutation relations $\{\psi_{Ai},\psi_{Bj}\}=M^{-1}\veps_{AB}\veps_{ij}$.   Eqn. (\ref{rofo}) tells us further
 that the $\psi_{Ai}$ commute with the Hamiltonian and thus obey $\dot\psi_{Ai}=0$.  To derive this equation of motion
along with the anticommutation relations from an effective action, the action must be $\int\d t \frac{M}{2}i\veps^{AB}\veps^{ij}\psi_{Ai}\dot \psi_{Bj}$.  As for the bosonic coordinates 
$x^\mu$ that represent the motion of the center of mass, the fact that the translation generators $P_\mu$ are conserved and that the Hamiltonian is $\hH=P^2/2M$ tells us that
up to an additive constant, the action is a free particle action $\frac{M}{2}\int \d t \,\dot x^2$.  In Lorentz signature, the constant is minus the rest energy or $-M$ and thus the particle
action in this approximation is
\begin{equation}\label{uft}I=\int\d t\left(-M+\frac{M}{2} \dot x^\mu\dot x_\mu +\frac{iM}{2} \veps^{AB}\veps^{ij} \psi_{Ai}\dot\psi_{Bj}\right). \end{equation}  One can think of the
first two terms $-M+\frac{1}{2}M\dot x^2$ as a nonrelativistic approximation to a covariant action \begin{equation}\label{doolittle}I_{\mathrm{cov}} =-M\int \d \tau\sqrt{-g_{MN}\frac{\d x^M}{\d \tau}\frac{\d x^N}{\d \tau}},\end{equation}
where here $\tau$ is an arbitrary parameter along the particle path and $g_{MN}$ is the full 5d metric.
This action is
valid for any particle orbit with a large radius of curvature.

This action (\ref{uft}) satisfies the expected supersymmetry algebra, with 
\begin{equation}\label{dorfic}Q_{Ai}=M\sqrt 2\psi_{Ai},~~~Q_{\dA i}=-i\frac{M}{\sqrt 2}\frac{\d x^\mu}{\d t}\Gamma_{\mu A \dA}\psi^A_i,~~~P^\mu=M\frac{\d x^\mu}{\d t},~~~
\hH=\frac{P^2}{2M}.\end{equation}

What really uniquely determines the action  (\ref{uft}) is that it gives a minimal realization of the translation 
symmetries and supersymmetries that are spontaneously
broken by the choice of superparticle trajectory.  The conserved charges $Q_{Ai}=M\sqrt 2\psi_{Ai}$ and $P^\mu=M\dot x^\mu$, being linear in $\psi_{Ai}$ and 
$\dot x^\mu$, generate constant shifts of $\psi_{Ai}$ and $x^\mu$, which can be viewed as Goldstone fields for spontaneously broken symmetries.
As usual, the low energy action for the Goldstone fields is uniquely determined.

\subsubsection{Collective Coordinates}\label{collective}

In the instanton calculation, the zero-modes of $x^\mu(\tau)$ and $\psi_{Ai}(\tau)$ 
will be collective coordinates that parametrize the choice of the
superparticle orbit; we will denote those collective coordinates as $x^\mu$ and $\psi^{(0)}_{Ai}$.  By integrating over all non-zero modes while 
keeping the zero-modes fixed, we will generate an effective action $\int \d^4x \d^4\psi^{(0)}(\dots)$.  Up to an elementary factor that is computed
shortly,
the $\psi^{(0)}_{Ai}$ can be identified with the fermionic coordinates $\theta_{Ai}$ that are used in writing superspace effective actions
in four dimensions, so $\int\d^4x\d^4\psi^{(0)}(\dots)$ is a chiral effective action
$\int \d^4x \d^4\theta(\dots)$.  Such an interaction is potentially non-trivial in the sense explained in section \ref{trivial}; that is,
it may not be possible to write it as a $D$-term. A 5d BPS particle with $\zeta<0$ would have fermionic collective
coordinates  of opposite chirality and could similarly generate an anti-chiral interaction $\int\d^4x\d^4\bar\theta$.
 A superparticle that is not BPS spontaneously
breaks all supersymmetries, so it can be described with eight fermionic collective coordinates and can only generate $D$-terms, 
that is non-chiral interactions $\int \d^4x\,\d^4\theta\,\d^4\bar\theta\,(\dots)$.  

Let us determine the normalization of the  
measure for integration over the collective coordinates.  First of all, this measure is independent of $M$.  In fact, $M$
can be removed from the action by absorbing a factor of $\sqrt{M}$ in both $x^\mu$ and $\psi_{Ai}$, as we will do later (eqn. (\ref{murtz})).
Because the bosons $x^\mu$ and the fermions $\psi_{Ai}$ both have four components, this rescaling affects neither the Gaussian integral
for the non-zero modes nor the zero-mode measure $\d^4x \d^4\psi^{(0)}$.  

However, it is fairly natural to factor the zero-mode measure as $M^2\d^4 x\cdot M^{-2}\d^4\psi^{(0)}$.  This is based on the following observation.
With the action being proportional to $M$, the Gaussian integral over any non-zero bosonic mode gives a factor of $1/M^{1/2}$, and the integral
over any non-zero fermionic mode gives a factor of $M^{1/2}$.  To compensate for this, in defining the path integral measure, one includes
a factor of $M^{1/2}$ for every bosonic mode and a factor $M^{-1/2}$ for every fermionic mode.  So the bosonic and fermionic zero-mode
measures, up to constants, are $M^2\d^4x$ and $M^{-2}\d^4\psi^{(0)}$.  

To find the normalization of the  measure for fermion zero-modes, we compare a matrix element computed by integrating over collective
coordinates to the same matrix element computed in a Hamiltonian approach. Quantization of the four fermions $\psi_{Ai}$ gives a 
four-dimensional  Hilbert space $\sH$,
consisting of two spin 0 bosonic states and two fermionic states of spin $(1/2,0)$.  If $(-1)^F$ is the operator that distinguishes
bosons from fermions, then the anticommutation relations can be used to show that
\begin{equation}\label{muromo}\Tr_{\sH}(-1)^F \psi_{A1}\psi_{B1}\psi_{C2}\psi_{D2}=\frac{1}{M^2}\veps_{AB}\veps_{CD}. \end{equation}
Now recall that such a trace can be computed by a path integral on a circle with periodic boundary conditions for the fermions.
 The integral  $M^{-2}\int\d^4\psi^{(0)}\psi^{(0)}_{A1}\psi^{(0)}_{B1}\psi^{(0)}_{C2}\psi^{(0)}_{D2}$  over collective coordinates
should reproduce the formula (\ref{muromo}), so we want  
\begin{equation}\label{urmo}\int \d^4\psi^{(0)} \,\psi^{(0)}_{A1}\psi^{(0)}_{B1}\psi^{(0)}_{C2}\psi^{(0)}_{D2}=\veps_{AB}\veps_{CD}. 
  \end{equation}
  
Now we can compare the zero-mode measure $\d^4x\,\d^4\psi^{(0)}$ to the usual measure $\d^4x\d^4\theta\sqrt{g^E}$ of a four-dimensional
supersymmetric action.  This comparison involves a few steps.  First, with $Q_{Ai}=M\sqrt 2\psi_{Ai}$ and $\{\psi_{Ai},\psi_{Bj}\}=M^{-1}\veps_{AB}\veps_{ij}$,
we have $\{Q_{Ai},\psi_{Bj}\}= \sqrt 2\veps_{AB}\veps_{ij}$.  So $Q_{Ai}$ acts on the fermionic collective coordinates as $\sqrt 2{\partial}/{\partial
\psi^{(0)Ai}}$.   The 4d supersymmetry generators are $\Qb_{Ai}=e^{-\sigma/4}Q_{Ai}$ (eqn. (\ref{melf})) and the fermionic coordinates $\theta_{Ai}$
of superspace are usually normalized so that $\Qb_{Ai}$ acts on them as ${\partial}/{\partial\theta^{Ai}}$.  So we should set
$\sqrt 2\partial/\partial{\psi^{(0)\,Ai}}=e^{\sigma/4}\partial/\partial{\theta^{Ai}}$, or $\psi^{(0)}_{Ai}=\sqrt 2 e^{-\sigma/4}\theta_{Ai}$.  Hence 
\begin{equation}\label{zormo} \d^4\psi^{(0)}=\frac{e^\sigma}{4}\d^4\theta, \end{equation}
where $\d^4\theta$ is defined so that
\begin{equation}\label{wormox}\int\d^4\theta \,\theta_{A1}\theta_{B1}\theta_{C2}\theta_{D2}=\veps_{AB}\veps_{CD}. \end{equation}
We further write $\d^4x =\d^4x\sqrt{g^E} e^{-2\sigma}$, since $\sqrt{g^E}=e^{2\sigma}$ according to eqn. (\ref{zorox}).  So finally
the zero-mode measure is
\begin{equation}\label{tormox}\d^4x\d^4\psi^{(0)}=\d^4x\d^4\theta\sqrt{g^E} \frac{e^{-\sigma}}{4}.\end{equation}

Although we normalized the fermion zero-mode measure by comparison to a Hamiltonian
calculation, we have not yet done the same for the bosons.  This will be done in section \ref{thecomp}, by comparing to a counting of quantum states.

\subsubsection{The Action In A Graviphoton Background}\label{graviback}

Now let us turn on a graviphoton field.  A particle of charge $q$ couples to an abelian gauge field $A$ with a coupling 
\begin{equation}\label{zz} \int \d \tau\,qA_M \frac{\d x^M}{\d \tau}\end{equation}
(and possible non-minimal couplings involving magnetic moments, etc.), where $A_M$ has time component
$A_0$ and spatial components $A_\mu$.    In the case of a superparticle
coupled with charges $\vec q$ to the gauge fields  of 5d supergravity, and assuming that the background gauge field is precisely the graviphoton $\TT^-$,
we found in eqn. (\ref{zobel}) that the effective magnetic field is $\zeta(\vec q)\TT^-$, where $\zeta(\vec q)$ is the central charge.  
For a BPS superparticle of $M=\zeta(\vec q)$,
this means that we should replace $qA_\mu$ in eqn. (\ref{zz}) with $MV_\mu$, where 
$\TT^-_{\mu\nu}=\partial_\mu V_\nu-\partial_\nu V_\mu$.  A convenient gauge
choice  is $V_\nu=\frac{1}{2}\TT^-_{\mu\nu}x^\mu$.  
There is an important detail here, however.  In a 5d covariant form, the action for a charged point particle coupled to $V_\mu$,
on an orbit with a large radius
of curvature, is
\begin{equation}-M\int \d\tau \sqrt{-g_{MN}\frac{\d x^M}{\d \tau}\frac{\d x^N}{\d \tau}}+\int \d\tau\, V_M\frac{\d x^M}{\d\tau}.\end{equation}  To apply this
to the graviphoton background, we have to use the supersymmetric G\"odel metric  (\ref{miro}), which depends on $V_\mu$.  When we expand the square root taking
this into account, the effect is to double the coefficient of the $V_\mu \dot x^\mu$ coupling.
The action of the superparticle in the graviphoton background is thus, in some approximation,
\begin{equation}\label{hurmex}I=M\int\d t\left(-1+\frac{1}{2}\dot x^\mu\dot x_\mu+\frac{i}{2}\veps^{AB}\veps^{ij}\psi_{Ai}\frac{\d}{\d t}\psi_{Bj} +
\TT^-_{\mu\nu}
x^\mu\dot x^\nu\right). \end{equation}

Are there additional terms that should be included in this action?  The spontaneously broken supersymmetries $Q_{Ai}$ remain valid
symmetries
when the graviphoton field is turned on.  So they commute with the exact Hamiltonian $\hH$ that describes the superparticle, and hence the
fields $\psi_{Ai}=Q_{Ai}/M\sqrt 2$ are time-independent.  Hence we should not add to (\ref{hurmex}) a magnetic moment coupling
\begin{equation}\label{urmox}\int\d t \,\TT^-_{AB}i\veps^{ij}\psi_{Ai}\psi_{Bj},\end{equation}
as this will give a time-dependence to $\psi_{Ai}$.
(By contrast, we will encounter such magnetic moment couplings  in section \ref{moregenmass} for other fermion fields along the particle worldline.)
Other interactions that might be added to (\ref{hurmex}) are irrelevant in the limit that the circumference of the circle is large.  The precise scaling
argument behind that statement is explained at the end of this section.\footnote{\label{tumox} In determining the effective action (\ref{hurmex}) 
for a BPS hypermultiplet, we have not
needed the fact that the graviphoton background preserves eight supersymmetries; 
the four conserved supercharges $Q_{Ai}$ were enough.  In analyzing more general BPS supermultiplets
in section \ref{moregenmass}, we will need the full supersymmetry algebra to determine magnetic moments.}

The momentum conjugate to $x^\mu$ is $\pi_\mu=\delta I/\delta \dot x^\mu=M(\dot x_\mu-\TT^-_{\mu\nu}x^\nu)$.   Of course, it obeys
$[\pi_\mu,\pi_\nu]=0$, $[\pi_\mu,x^\nu]=-i\delta_\mu^\nu$.  By contrast, the conserved momentum that generates spatial translations is 
\begin{equation}\label{puft}P^\mu=M\left(\frac{\d x^\mu}{\d t}-2\TT^{-\,\mu\nu}x_\nu\right)=\pi^\mu-M\TT^{-\,\mu\nu} x_\nu. \end{equation}
It obeys $[P_\mu,x^\nu]=-i\delta^\nu_\mu$ (so it generates spatial translations, just as $\pi_\mu$ does) but satisfies
\begin{equation}\label{zinco} [P_\mu,P_\nu]=-2iM\TT^-_{\mu\nu}. \end{equation}
This is part of the supersymmetry algebra.  In fact, taking the nonrelativistic limit of eqn. (\ref{susy5d}), the nonrelativistic limit of
the supersymmetry algebra is
\begin{align}\label{susy5dnonrel} [P_\mu,P_\nu]& = -2iM \TT^-_{\mu\nu} \cr
											[\eJ, P_\mu] &=2i\TT^-_{\mu\nu}P^\nu \cr
											[\eJ, Q_{Ai}] &=-\frac{i}{2}\TT^-_{\mu\nu}\Gamma^{\mu\nu}_{AB}Q^B_i \cr
                                          [P_\mu,Q_{\dA i}]& =    \TT^-_{\mu\nu}\Gamma^\nu_{\dA B} Q^B_{i}, \cr
                                          \{Q_{A i},Q_{B j}\}& =2M \varepsilon_{A B}\varepsilon_{ij} \cr
                                           \{Q_{Ai},Q_{\dA j}\}& = -i\Gamma^\mu_{A\dA}\varepsilon_{ij}P_\mu\cr
                                           \{Q_{\dA i},Q_{\dB j}\}&=\varepsilon_{\dA\dB}\varepsilon_{ij}(\hH+\eJ).
                                                    \end{align}
                                                    
The nonrelativistic conserved Hamiltonian $\hH$ is not simply $P^2/2M$, which does not commute with the $P_\mu$.   $\hH$ can be conveniently written in terms of the
conserved quantities $P_\mu$ and $\TT^{-\mu\nu}L_{\mu\nu}$, where
$L_{\mu\nu}=x_\mu\pi_\nu-x_\nu\pi_\mu$:
\begin{equation}\label{heff}\hH=\frac{P^2}{2M}-\TT^{-\mu\nu}L_{\mu\nu}.\end{equation}

What are the conserved supersymmetries?  Clearly -- because we have not added the magnetic moment term (\ref{urmox}) -- the charges
$Q_{Ai}=M\sqrt 2\psi_{Ai}$ are conserved.  To define conserved supercharges $Q_{\dA j}$ of the opposite chirality, we must modify the definition  in (\ref{dorfic})
by replacing $M\dot x^\mu$ with the conserved charge $P^\mu$.  Thus the supersymmetry generators are
\begin{equation}\label{luft}Q_{Ai}=M\sqrt 2\psi_{Ai},~~~ Q_{\dA j}=-\frac{i}{\sqrt 2}P^\mu \Gamma_{\mu A \dA}\psi_j^A. \end{equation}
The nonrelativistic supersymmetry algebra (\ref{susy5dnonrel}) is satisfied.

The constant $-M$ in the Lorentz signature action contributes $+M$ to the energy.  So after compactifying the time direction on a (Euclidean signature) circle of circumference $2\pi e^\sigma$, the constant term in the action contributes to the path
integral  a factor
$\exp(-2\pi e^\sigma M)=\exp(-2\pi e^\sigma\sum_Iq_I h^I)$.  However, after compactifying, the gauge fields can have constant
components $\alpha^I$ in the $t$ direction and these give an imaginary contribution to the Euclidean action.   As explained in eqns. (\ref{pofgo}) and
(\ref{merot}), the effect of this is to replace $e^\sigma h^I$ by $-iZ^I$, and so to replace $\exp(-2\pi e^\sigma M)$ by 
$\exp(2\pi i \sum_I q_I Z^I)$.  To take account of the fermionic collective coordinates of the particle orbit,  we just have to extend the factor $\exp(2\pi i \sum_I q_I Z^I)$ to a superfield, namely
\begin{equation}\label{urtz} \exp\left(2\pi i \sum_I q_I\ZZ^I\right).  \end{equation}

This factor, of course, must be multiplied by a one-loop determinant computed using the action (\ref{hurmex}).  The product of boson and fermion
determinants is independent of $M$, since $M$ can be removed by  rescaling  $x^\mu$ and $\psi_{Ai}$ by a common factor $1/\sqrt M$.
(This scaling does not affect the path integral measure, or the zero-mode measure $\d^4x \,\d^4\psi^{(0)}$.)
The one-loop computation can therefore be performed using the action
\begin{equation}\label{murtz} I'= \frac{1}{2}\int\d t\,\left(\dot x^\mu \dot x_\mu +i\veps^{AB}\veps^{ij}\psi_{Ai}\frac{\d}{\d t}\psi_{Bj}+
2\TT^-_{\mu\nu}
x^\mu\dot x^\nu\right). \end{equation}

The particle mass $M$ has disappeared in eqn. (\ref{murtz}), so the one-loop determinant depends only on the radius $e^\sigma$ of
the circle and on $\TT^-_{\mu\nu}$.  
We can constrain this dependence using the scaling symmetry
\begin{equation}\label{wurtz}t\to \lambda t,~~x\to \lambda^{1/2}x,~~~\psi_{Ai}\to \psi_{Ai},~~~\TT^-_{\mu\nu}\to \lambda^{-1}\TT^-_{\mu\nu}. \end{equation}
The measure $\d^4x\,\d^4\psi^{(0)}$ scales as $\lambda^2$, and the circumference $2\pi e^\sigma$ scales as $\lambda$, 
so  scale-invariance implies that the one-loop
determinant has the form $\d^4x \,\d^4\psi^{(0)}\, e^{-2\sigma}f(e^\sigma \TT^-)$ for some function $f$. 
The meaning of the factor $\d^4x \,\d^4\psi^{(0)}$ is that we cannot integrate over the zero-modes or collective coordinates (the integral over the $x^\mu$
gives $\infty$ and the integral over $\psi^{(0)}$ gives 0), so we leave them unintegrated and interpret the result as a measure on the collective
coordinate moduli space rather than a number.  
Lorentz invariance implies that $f$
is really a function of $e^{2\sigma}\TT^-_{\mu\nu}\TT^{-\,\mu\nu}$, so the one-loop determinant has the form
\begin{equation}\label{murgly} \d^4x\,\d^4\psi^{(0)} \sum_{\sg=0}^\infty c_\sg e^{(2\sg-2)\sigma} (\TT^-_{\mu\nu}\TT^{-\mu\nu})^\sg\end{equation}
wth some constants $c_\sg$. 

 In this derivation, we  used the metric $\d s^2=-(\d t-V)^2+\sum_\mu(\d x^\mu)^2$, as in eqn.
(\ref{miro}), with a periodicity in imaginary time of $2\pi e^\sigma$.
  To write the effective action
  in conventional 4d variables, we use $\d^4x\d^4\psi^{(0)}=\d^4x\d^4\theta\sqrt{g^E} \frac{e^{-\sigma}}{4}$ (eqn. (\ref{tormox})).
  Also, to express (\ref{murgly}) in 4d terms, we should re-express $\TT^-_{\mu\nu}\TT^{-\mu\nu}$ in terms of the
corresponding 4d quantity $\WW^-_{\mu\nu}\WW^{-\mu\nu}=16 e^{-\sigma}\TT^-_{\mu\nu}\TT^{-\mu\nu}$ (eqn. (\ref{welz})).    Actually, here
we should replace $\WW^-$ with the chiral superfield $\W$ whose bottom component is $\WW^-$.  
Setting $\W^2=\W_{\mu\nu}\W^{\mu\nu}$, 
 eqn. (\ref{murgly}) becomes
\begin{equation}\label{urgly}\frac{1}{4}\d^4x\d^4\theta \sqrt{g^E}\, \frac{e^{(3\sg-3)\sigma}}{16^{\sg}} ({\W^2})^{\sg}.\end{equation}
Now we recall that $\X^0=-ie^{-3\sigma/2}/2$ (eqn. (\ref{x0val})).  Having generalized the fields to superfields, we can
integrate over the collective coordinates to get a contribution to the effective action:
\begin{equation}\label{likeable}-\int\d^4x \d^4\theta \sqrt{g^E} (-64)^{-\sg}\frac{(\W^2)^\sg}{(\X^0)^{2\sg-2}}.\end{equation}

This -- and its generalization with a classical factor  $\exp\left(2\pi i \sum_Iq_I\ZZ^I\right)$ included -- is a chiral interaction of the sort
described in the GV formula and discussed throughout this paper.  It is a non-trivial $F$-term in the sense explained in section
 \ref{trivial}; it cannot be written
as $\int\d^4x\d^8\theta(\dots)$. 
Suppose on the other hand that we add non-minimal terms to the action (\ref{hurmex}).  
Any translation-invariant and supersymmetric interaction that we might add that
has not already been included in eqn. (\ref{hurmex}) would scale as a negative power of $\lambda$.  Hence, a contribution to the superparticle
path integral that depends on such interactions would be similar to (\ref{murgly}) but with extra powers
of $e^{-\sigma}$.   Such interactions are trivial $F$-terms, and it is difficult to learn very much about them.

\def\O{{\mathcal O}}

\subsubsection{The Computation}\label{thecomp}

Having come this far, the actual computation of the one-loop determinant using the action (\ref{murtz}) is not difficult.

The one-loop path integral gives a zero-mode integral times $\sqrt{\det'\,\D_F/\det'\,\D_B}$, where  $\D_B$, $\D_F$ are the bosonic and fermionic kinetic operators
\begin{align}\label{mozzo}\D_B&=-\frac{\d^2}{\d t^2}\delta_{\mu\nu}+2\TT^-_{\mu\nu}\frac{\d}{\d t}\cr
\D_F&=i\frac{\d}{\d t}\veps_{AB}\veps_{ij}\end{align}
and $\det'$ is a determinant in the space orthogonal to the zero-modes.
Moreover, the real symmetric operator $\D_B$
 can be conveniently  factored as a product of two imaginary, self-adjoint (and skew-symmetric) operators
\begin{equation}\label{merz}\D_B=\D_1\D_2,~~~\D_1=i\frac{\d}{\d t}\delta_{\mu\nu},~~~\D_2=i\left(\frac{\d}{\d t}\delta_{\mu\nu}-2\TT^-_{\mu\nu}\right). \end{equation}
So $\det'\,\D_B=\det'\,\D_1\cdot \det\,\D_2$.   Since $\D_1$ is conjugate to $\D_F$, the ratio $\det'\,\D_F/\det'\,\D_B$  actually equals
$1/\det\,\D_2$.    This determinant  can be evaluated by writing down the eigenfunctions of $\D_2$ (which are simple
 exponentials) and computing the regularized product of the corresponding eigenvalues.  This is a rather standard computation.

However, we will here take a shortcut.  We use the fact that a path integral on a circle has a Hamiltonian interpretation; the path integral we want equals
$\int\d^4x\,\d^4\psi^{(0)} \,\Tr'\,\exp(-2\pi e^\sigma \hH)$, where $\Tr'$ is a trace with the zero-modes removed.  
We can pick coordinates on $\R^4$ in which $\TT^-$ is the direct sum of two $2\times 2$ blocks:
\begin{equation}\label{morzo}\TT^-=\frac{1}{2}\begin{pmatrix} 0 & \TT &&\cr
                                                                                          -\TT & 0&&\cr && 0 & -\TT\cr && \TT& 0 \end{pmatrix}.\end{equation}
(We reversed the sign in the lower right block to make $\TT^-$ anti-selfdual if the four coordinates are oriented in the standard way.)               
A factor of $1/2$ was included in eqn. (\ref{morzo}) so that 
\begin{equation}\label{worzo}\TT^2=\TT^{-\mu\nu}\TT_{-\mu\nu}.\end{equation}                                                                           
Let us compute the desired trace in the subspace corresponding to the upper block.  The corresponding Hamiltonian describes a particle moving in
two dimensions in a constant magnetic field $\TT$.  The energy eigenstates are Landau bands\footnote{In speaking of Landau bands, we assume that
$\TT$ is real, while in the graviphoton background it is imaginary.  The determinant that we are trying to compute is holomorphic in $\TT$, so it is determined
for all $\TT$ by what happens for $\TT$ real.} with energies $\left(\frac{1}{2}+m\right)\TT$, $m=0,1,2,\dots$.
The density of states per unit area in any one Landau band is $\d^2x\, \TT/2\pi$.  So   for  the bosonic variables that describe motion in this plane, the one-loop path integral equals
\begin{equation}\label{morz}\frac{\d^2 x}{2\pi}\TT \sum_{m=0}^\infty\exp\left(-\pi e^\sigma \TT\left(1+2m\right)\right)=\frac{\d^2x}{2\pi}\frac{\TT e^{-\pi e^\sigma \TT}}
{1-e^{-2\pi e^\sigma \TT}}=\frac{\d^2x}{4\pi}\frac{\TT}{\sinh (\pi e^\sigma \TT)}. \end{equation}
Including an identical factor for the lower block in eqn. (\ref{morzo}), and including the fermion zero-modes, the full one-loop path integral gives
\begin{equation}\label{orz}\frac{\d^4x \d^4\psi^{(0)}}{(4\pi)^2} \frac{\TT^2}{\sinh^2(\pi e^\sigma \TT)}.\end{equation}

Including also the  classical factor that was described in eqn. (\ref{urtz}), the contribution
of the BPS hypermultiplet to the GV formula is
\begin{equation}\label{welx} \frac{\d^4x\,\d^4\psi^{(0)}}{(4\pi)^2}  \exp\left(2\pi i\sum_I q_I \ZZ^I\right)\frac{\TT^2}{\sinh^2(\pi e^\sigma \TT)}. \end{equation}

Though we have derived this formula by a computation in the supersymmetric G\"odel background, it gives part of the effective action in a more general background.  Since the function $\TT/\sinh(\pi e^\sigma\TT)$ is
regular for real $\TT$, this contribution to the effective action is regular as long as the 5d graviphoton field is real.
However, in the supersymmetric G\"odel background, $\TT$ is imaginary and the effective action has poles if $\TT$ is large.
This does not really affect our derivation.  The GV formula governs couplings (\ref{zell}) that are each perturbative in $\TT$,
and the computation we have performed can be understood as a convenient way to evaluate 
all such perturbative contributions together.

To express the result (\ref{morz}) in four-dimensional terms, we follow the same steps that led to eqn. (\ref{likeable}).
We write $\TT=\sqrt{\TT^{-\mu\nu}\TT^-_{\mu\nu}}=\frac{e^{\sigma/2}}{4}\sqrt{(\WW^-)^2}$, and interpret $\WW^-$  as the bottom component
of a superfield $\W$.    We also use $e^{3\sigma/2}=-i/2\X^0$, and $\d^4x\d^4\psi^{(0)}=\frac{1}{4}\d^4x\d^4\theta\sqrt
{g^E}e^{-\sigma}$ (eqn. (\ref{tormox})).  The resulting contribution to the 4d effective action is
\begin{equation}\label{zelmo}-\int \frac{\d^4x\d^4\theta}{(2\pi)^4} \sqrt{g^E} \exp\left(2\pi i\sum_I q_I \ZZ^I\right)\frac{\frac{1}{64}\pi^2\W^2}{\sin^{2}\left(\frac{\pi\sqrt{\W^2}}{8\X^0}\right)}.
\end{equation}

\subsection{Multiple Winding, Bubbling, And Comparison To String Theory}\label{winding}

Following \cite{GV1,GV2}, we will now explain the interpretation of this formula in string theory.

In perturbative string theory (either physical or topological string theory), we should distinguish the string worldsheet $\Sigma^*$ from its image $\Sigma\subset Y$.
$\Sigma^*$ does not necessarily have the same genus as $\Sigma$. The map from $\Sigma^*$ to $\Sigma$ must be holomorphic if it is to contribute to the amplitudes $\F_\sg$
in  topological
string theory (or in physical string theory, given the relation between the two), but  is not necessarily an isomorphism.

In general, a non-constant holomorphic map $\varphi:\Sigma^*\to\Sigma$ may have any degree $k\geq 1$.  The $\ZZ$-dependence of a contribution from a map of degree $k$ is
a factor $\exp(2\pi ik \sum_I q_I\ZZ^I)$.  The formula (\ref{welx}) evidently corresponds to contributions with $k=1$.  This is not surprising, since in deriving the formula,
we considered an M2-brane wrapped just once on $\Sigma$ and
assumed that the superparticle trajectory winds just once around the M-theory circle. 

A string theory map $\varphi:\Sigma^*\to\Sigma$ of degree $k$ will correspond in M-theory to a configuration in which, roughly
speaking, an M2-brane worldvolume has a degree $k$ map to $S^1\times \Sigma$.   There are two distinct effects in M-theory that combine to produce this
result.  First, the M2-brane may wrap $k_1$ times over $\Sigma$.   Since multiple M2-branes cannot be treated semiclassically,
the rigorous meaning of this statement is that a BPS state in M-theory may have an M2-brane charge that is
$k_1$ times the homology class $[\Sigma]$ 
(in other words, $k_1$ times the charge of an M2-brane wrapped once on $\Sigma$).  Second, regardless of what BPS
state we consider and what its quantum numbers may be, when we use this BPS state to make an instanton in M-theory
compactified on a circle, this state may wind $k_2$ times around the circle.   The relation between the degree $k$ measured
in string theory, the charge $k_1$ of the BPS particle in units of $[\Sigma]$, 
and the number $k_2$ of times  that the  particle winds around the circle
is $k=k_1k_2$.  

Thus what in string theory is the sum over the degree of the map $\varphi$ is in the context of the GV formula a combination
of two effects: a BPS particle may be multiply-charged and it may wind any number of times around the M-theory circle.
In this section, we describe the effect of multiple winding, and in section \ref{moregenmass}, we consider
the effects of multiple charge.

It is easier to write down the formula that  governs contributions with multiple winding than to explain properly what it means.  
So we will first write down the formula.  If a point superparticle wraps $k$ times around the M-theory circle, 
the effective circumference of the circle becomes $2\pi k e^\sigma$.
The winding also multiplies the classical action $2\pi i\sum_I q_I\ZZ^I$ by $k$.
To evaluate the contribution of a particle orbit of winding number $k$, we also have to divide 
by a factor of $k$ to account for the cyclic symmetry between the $k$ branches of the particle orbit.  
So the analog of eqn. (\ref{zelmo}) with $k$-fold wrapping is obtained by multiplying $\sum_I q_I \ZZ^I$ and $e^\sigma$ by $k$, and dividing the whole formula by $k$.  
Summing over $k$ gives the contribution of the given BPS state with any winding:
\begin{equation}\label{zelmox}-\int \frac{\d^4x\d^4\theta}{(2\pi)^4} \sqrt{g^E}\sum_{k=1}^\infty\frac{1}{k} \exp\left(2\pi ik\sum_I q_I \ZZ^I\right)\frac{\frac{1}{64}\pi^2\W^2}{\sin^{2}\left(\frac{\pi k\sqrt{\W^2}}{8\X^0}\right)}.
\end{equation}
(In the denominator, the
 factor of $k$ in $\sin^{2}\left(\frac{\pi k\sqrt{\W^2}}{8\X^0}\right)$ comes from substituting $e^\sigma\TT\to ke^\sigma\TT$ in the denominator in
(\ref{welx}).)

However, a careful reader may find this formula puzzling.  A weakly coupled elementary point particle could wind $k$ times around a circle, and such a contribution
could be evaluated along the lines of the previous paragraph.  Does this make sense for a wrapped M2-brane, whose self-interactions are not small?  A multiply-wound
M2-brane is not a concept that makes  sense semiclassically, since a system of $k$ parallel M2-branes for $k>1$ is actually strongly coupled.   If parallel M2-branes
are separated in a transverse direction, their interactions
remain strong until the separation exceeds  the eleven-dimensional Planck scale $\varrho$.  (Beyond this scale, the long range forces
due to graviton and $C$-field exchange cancel for nonrelativistic BPS particles.)

Ignoring the strong interactions between BPS particles in this derivation can be justified as follows.
In nonrelativistic quantum mechanics in $D>2$ spatial dimensions, a short-range interaction, no matter how strong, is irrelevant in the renormalization
group sense and is unimportant at low energies, except for the possibility that it may generate a bound state.  In explaining this, we take the Hamiltonian
of a free particle to be $\hH_0=P^2$, where $P$ is the momentum; hence $P$ has dimension  $E^{1/2}$ (energy to the one-half power).  A short-range
interaction is equivalent to  $\hH'=c\delta^D(x)$, for some constant $c$, modulo less relevant couplings involving derivatives of a delta function.  The constant $c$ has dimension $E^{1-D/2}$ and so is an irrelevant coupling  if $D>2$.  For application to the GV formula,
we have $D=4$, so the interactions are safely irrelevant except for possibly generating bound states.  (Bound states are a short-range phenomenon that
cannot be analyzed by renormalization group scaling.)  Concretely, the irrelevance of a short-range coupling for $D>2$ means the following.   
  If a particle of mass $M$ 
propagates a Euclidean distance $L$ on a classical orbit (in our case, the classical orbit is a copy of the M-theory circle and $L=2\pi e^\sigma$ is its circumference), 
the fluctuations in its position in
the directions normal to the classical orbit
 are typically of order $\sqrt{L/M}$.  No matter how large  $M$ may be, $\sqrt{L/M}$ is much greater than
the interaction range $\varrho$  if $L$ is sufficiently large.  Thus, at any given time, two branches of an orbit that wraps $k$ times around the M-theory circle  are
unlikely to be within range of the interaction.  The condition $D>2$ ensures that this is unlikely to happen at any time along the orbit.  

What we learn from this reasoning is that we can ignore the interactions among $k$ parallel M2-branes except for the possibility that, when they are close
together, they form a bound state.  Since the M2-brane states under consideration are BPS states, 
such a bound state would  be a bound state at threshold -- a new BPS
state with larger charges.  In fact, a bound state of $r$ BPS states that each have charges $q_I$ would have charges $\tilde q_I=r q_I$.  

Because M2-branes are strongly coupled,
it is not straightforward to determine if such bound states exist (and if so for what values 
of $r$ and with what spin).   If bound states exist, they are new BPS states
that can themselves be treated  as elementary superparticles, when they wrap around a sufficiently large M-theory circle.   Their contribution can be evaluated by methods
similar to what we have already described, with modifications to account
for their spins; see section \ref{moregenmass}.  The full GV formula involves a sum over all 
M-theory BPS states, possibly including bound states.  

A further comment is called for.  In eqn. (\ref{zelmo}), we consider only $k>0$.  Exchanging $k>0$ with $k<0$ amounts to a reflection 
of the M-theory circle.  When combined
with a reflection of $\R^4$, which reverses the four-dimensional chirality, this is a symmetry of M-theory.   So orbits of $k<0$ generate anti-chiral couplings,
just as orbits of $k>0$ generate chiral couplings.    But what about $k=0$?  For $k=0$, the BPS state has no net winding, so generically it does not propagate
a macroscopic distance, even if the M-theory circle is large.  Since we do not have a 
microscopic theory of M2-branes, we cannot make sense of a configuration in which an M2-brane
propagates over a non-macroscopic distance.  So we have no way to make sense of a 
$k=0$ contribution.  But intuitively, what we would want to say about such a contribution
is as follows.  In M-theory on $\R^4\times S^1\times Y$, an M2-brane wrapped on $\Sigma\subset Y$ and propagating a small distance in $\R^4\times S^1$ 
is not, in leading order (in the inverse radius of $S^1$), affected by 
the compactification from $\R^5$ to $\R^4\times S^1$.  So whatever contribution it 
makes is part of the effective action for M-theory on $\R^5\times Y$, compactified classically
from $\R^5$ to $\R^4\times S^1$.   As we stressed in sections \ref{detprinc} and 
\ref{classred}, an important input to the GV formula is that one knows the relevant effective action in five dimensions,
before compactification.  So there is no need to study the $k=0$ contributions.

Now let us look more closely at what the formula (\ref{zelmo}) means in terms of perturbative string theory.  Since
\begin{equation}\label{morof}\frac{1}{\sin x}=\frac{1}{x}\left(1+\frac{x^2}{6}+\dots \right), \end{equation}
we can expand eqn. (\ref{zelmox}) in a power series in $\W$:
\begin{equation}\label{yelx} -\int\frac{\d^4x\,\d^4\theta}{(2\pi)^4}(\X^0)^2
\sum_{k=1}^\infty \frac{1}{k^3}\exp\left(2\pi ik\sum_I q_I \ZZ^I\right)\left(1+\frac{\pi^2 k^2\W^2}{192(\X^0)^2}+\mathcal{O}(\W^4)\right).
 \end{equation}
In perturbative string theory, the contribution proportional to $\W^{2\sg}$ comes from worldsheets of genus $\g$, as we have explained in section \ref{compper}.  Thus, 
the formula (\ref{yelx}), even though it reflects a single wrapped M2-brane of genus 0 and degree 1, is interpreted in perturbative string theory as a sum of contributions
with all values $k\geq 1$ and $\g\geq 0$.

One might expect to compute these contributions 
in topological string theory (and therefore also in physical string theory, given their relationship) by counting degree $k$ maps from   a string worldsheet $\Sigma^*$ of genus $\g$ 
to  a given holomorphic curve $\Sigma\subset Y$.  However, in general this counting is not straightforward.

Let us look at a few cases.  We can specialize to $\g=0$ by setting $\W=0$ in eqn. (\ref{yelx}).  The $k=1$ contribution is 
\begin{equation}\label{relx}-\int \frac{\d^4x\,\d^4\theta}{(2\pi)^4}(\X^0)^2\cdot\exp\left(2\pi i\sum_I q_I \ZZ^I\right)\cdot 1. \end{equation}
This contribution is not hard to understand.
 A genus 0 worldsheet $\Sigma^*$ with a holomorphic map $\Sigma^*\to\Sigma$ of degree 1 is unique up to isomorphism; it is
isomorphic to $\Sigma$, with the map being the isomorphism.  This
uniqueness means that the contribution of genus 0 worldsheets singly wrapped on $\Sigma$ to the topological
string amplitude
 is  precisely $\exp\left(2\pi i\sum_I q_I \ZZ^I\right)\cdot 1$. 
 The occurrence of this factor in (\ref{relx}) is the most
basic relation between topological string amplitudes and the physical string amplitudes that are described by the GV formula.
  The remaining factor $-\d^4x\,\d^4\theta\,(\X^0)^2/(2\pi)^4$ in (\ref{relx}) represents the embedding of the topological string amplitude in physical string theory.    

Still with $\g=0$, we see in eqn. (\ref{yelx}) that if we take $k>1$, in addition to the classical action being multiplied by $k$,
the amplitude acquires a factor of $1/k^3$.  This is not an integer, so this answer cannot come from a straightforward ``counting''
of holomorphic maps.  The factor $1/k^3$ for a $k$-fold cover of a genus 0 curve
was first discovered using mirror symmetry \cite{Cand}.  Its interpretation in topological
string theory depends on the fact that for $k>1$, there is a nontrivial moduli space of degree $k$ holomorphic maps from a genus 0 worldsheet $\Sigma^*$ to $\Sigma$; this moduli space has orbifold singularities, because of which the ``counting'' does not give
an integer.  See \cite{AM} for a derivation along these lines.  The GV formula has given this rather subtle factor of $1/k^3$ without
much fuss \cite{GV1,GV2,MoMa}.

\begin{figure}[ht]
 \begin{center}
   \includegraphics[width=3in]{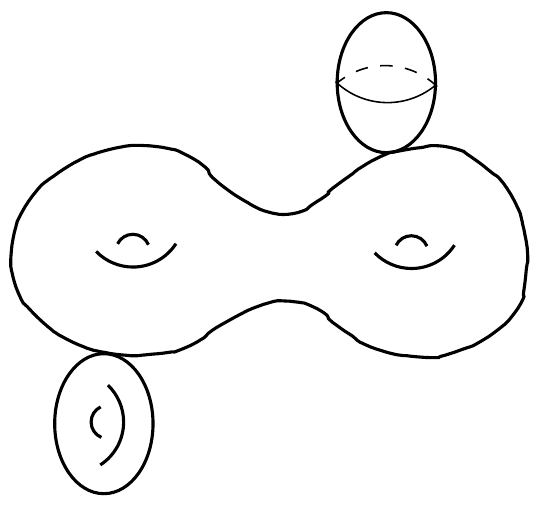}
 \end{center}
\caption{\small A Riemann surface $\Sigma^*$ splits into several components $\Sigma^*_i$.  In the context of topological string theory,
ordinarily all components except one are mapped to points in $Y$.  The splitting off from $\Sigma^*$ of one or more components that are mapped
to points in $Y$ is called ``bubbling.''
In the example shown, $\Sigma^*$ has genus 3 and from top to bottom the components have genus 0, 2, and 1.   \label{bubbling}}
\end{figure}

One need not look far to find further subtleties that are nicely resolved by the GV formula.
For example, for $\g>0$, assuming that $\Sigma^*$ is smooth, and
with $\Sigma$ of genus 0,  there does
not exist a degree 1 holomorphic map $\Sigma^*\to \Sigma$.  Thus naively  perturbative string theory does not generate 
contributions to the chiral interactions $\F_\sg$ with $k=1$ and $\g>0$.  But such contributions are clearly visible in (\ref{yelx}).  
As explained in \cite{GV1,GV2}, these contributions are interpreted 
in topological string theory in terms of contributions
in which $\Sigma^*$ is not smooth but is a union of various components $\Sigma^*_i$ that are glued together at singularities.  
For $k=1$, $\g>0$, one of these components is of genus 0 and is mapped isomorphically
onto $\Sigma$ by a degree 1 map, and the others are mapped to $Y$ by maps of degree 0 (thus, they are mapped to points in $Y$).  
Such a degeneration of $\Sigma^*$ and its
map to $Y$  is sometimes called ``bubbling'' (fig. \ref{bubbling}).   By integration over the moduli of such bubbled configurations, one can compute
topological string amplitudes with $k=1$ and $\g>0$.   More generally, such bubbled configurations contribute to a variety of topological string amplitudes.

Thus \cite{GV1,GV2}, the GV formula when interpreted in topological string theory describes a variety of multiply-wrapped
 and/or   bubbled configurations associated
to a given BPS state.

 \subsection{More General Massive BPS States}\label{moregenmass}
 
 \subsubsection{The General Answer}\label{genan}
 
It takes only a few steps to generalize the hypermultiplet computation of section \ref{hyperplet} to arbitrary massive BPS states.
We give here a general description and explain some explicit formulas in section \ref{expform}.

In general, it is inconvenient to describe a nonrelativistic action for an arbitrary BPS multiplet, but it is straightforward to describe a Hilbert space  for this
system, with an action of a Hamiltonian $\hH$,  conserved momentum and angular momentum operators $P^\mu$ and $J_{\mu\nu}$, and supercharges
$Q_{Ai}$ and $Q_{\dA{} j}$.  We already know how to describe the hypermultiplet in this language.   
The appropriate Hilbert space\footnote{Our notation will be as follows: a Hilbert space like $\widehat \H_0$ labeled with a hat describes the bosonic center
of mass motion as well as the quantization of fermion zero-modes; an unhatted Hilbert
space represents the quantized fermion zero-modes only.} $\widehat\H_0$ is an irreducible representation of the action of free bosons $x^\mu$ and their canonical
momenta $\pi_\mu$ as well as free fermions $\psi_{Ai}=Q_{Ai}/M\sqrt 2$.   The conserved momenta are $P^\mu=\pi^\mu$, the nonrelativistic Hamiltonian is
\begin{equation}\label{consh}\hH=\frac{P^2}{2M},\end{equation}
the rotation generators act in the natural way, and the  supercharges were defined in eqn. (\ref{luft}). These modes are needed to realize the spatial
translations and spacetime supersymmetries, but a general set of BPS states may have additional degrees of freedom.  So
 in general, the Hilbert space that describes  BPS states with charge $\vec q=(q_1,\dots,q_{b_2})$ and mass $M=\zeta(\vec q)$ is
 $\widehat\H_{\vec q}=\widehat\H_0\otimes {\eV}_{\vec q}$, where $\widehat\H_0$ is the Hilbert space 
 for a hypermultiplet and ${\eV}_{\vec q}$
is a vector space with an  action\footnote{This action is not necessarily irreducible, as there may be several BPS multiplets 
of charge $\vec q$.} of the rotation group $SU(2)_\ell\times SU(2)_r$.   The action on $\widehat\H_{\vec q}$  of the supercharges, the momentum,
and the Hamiltonian come entirely from their action on $\widehat\H_0$, but the rotation group acts also on ${\eV}_{\vec q}$.

Now let us turn on the graviphoton field $\TT^-_{\mu\nu}$.   
We can still define free fermions by $\psi_{Ai}=Q_{Ai}/M\sqrt 2$. The momentum generators still commute with the Hamiltonian, but instead of commuting with each
other they generate the Weyl algebra $[P_\mu,P_\nu]=-2iM\TT^-_{\mu\nu}$.  The minimal way to satisfy this is   to deform the $P^\mu$
as in eqn. (\ref{puft}):
\begin{equation}\label{zong}P^\mu=\pi^\mu-M\TT^{-\,\mu\nu} x_\nu.\end{equation}
We may use this formula,  since the representation of the canonical commutation relations $[P_\mu,P_\nu]=-2iM\TT^-_{\mu\nu}$, $[P_\mu,x^\nu]=
 -i\delta_\nu^\mu$, $[x^\mu,x^\nu]=0$ is unique up to isomorphism, implying that  any
 further $\TT^-$-dependent contributions that we might add to $P^\mu$ that preserve the commutation relations  can be removed by a unitary transformation.  So even for $\TT^-_{\mu\nu}\not=0$, there is a decomposition of the Hilbert space as
 $\widehat\H_{\vec q}=\widehat\H_0\otimes {\eV}_{\vec q}$ with the property that the supercharges $Q_{Ai}=M\psi_{Ai}$ and momentum generators $P_\mu$  act only on the first factor.
The most obvious way to satisfy the supersymmetry algebra  is to take $Q_{\dA j}$ to similarly act only on $\widehat\H_0$ and to be given by the same formula as
for the  hypermultiplet:
\begin{equation}\label{mertz} Q_{\dA j}=-\frac{i}{\sqrt 2}P^\mu \Gamma_{\mu A \dA}\psi_j^A. \end{equation}
To explain why eqn. (\ref{mertz}) gives a sufficiently good approximation to
 $Q_{\dA j}$, we use the scaling symmetry (\ref{wurtz}), which we extend to act on $\widehat\H_{\vec q}$
(and not just $\widehat\H_0$) by saying
that ${\eV}_{\vec q}$ is invariant under scaling.   To compute the desired effective action in four
dimensions, we have to evaluate the trace $\Tr'\,(-1)^F\exp(-2\pi e^\sigma \hH)$  (the symbol $\Tr'$ means that the trace is defined without an integral over collective coordinates). To do this computation after scaling the time by a large factor $\lambda$,  since $e^\sigma$ scales as $\lambda$,
 we are not interested in terms in $\hH$ that
scale as a power of $\lambda$ more negative than $\lambda^{-1}$.  Since $\hH$ can be computed from the anticommutator $\{Q_{Ai},Q_{\dot Aj}\}$, this means that we are not interested in corrections to $Q_{\dA j}$ that scale
as a power more negative than $\lambda^{-1/2}$.   In order for $Q_{\dA j}$ to be conserved, it must be possible to write it  with no explicit dependence on $x^\mu$,
just in terms of the conserved charges
$\psi_{Ai}=Q_{Ai}/M\sqrt 2$ and  $P^\mu$ as well as matrices acting on ${\eV}_{\vec q}$.  (In particular, we cannot make 
use of the conserved angular momentum
without spoiling the commutator $[Q_{\dot A i},P_\mu]$, which comes out correctly if we use (\ref{mertz}).)
 These requirements mean that no correction to (\ref{mertz}) involving $\TT^-_{\mu\nu}$ is possible: $\TT^-_{\mu\nu}$ scales as $\lambda^{-1}$ and the other possible ingredients
 in a hypothetical correction to  the right hand side of (\ref{mertz}) scale with nonpositive powers of $\lambda$ (indeed, $P^\mu$, $\psi_{Ai}$, and
 a matrix acting on $\eV_{\vec q}$ scale respectively as $\lambda^{-1/2}$, 1, and 1).
 
 Since the $Q_{\dA j}$ act only on $\widehat\H_0$ and not on ${\eV}_{\vec q}$, the same is true of 
 \begin{equation}\label{mortox} \{Q_{\dA i},Q_{\dB j}\}=\veps_{\dA \dB}\veps_{ij}(\hH+\eJ). \end{equation}
 However,  $\eJ=\TT^-_{\mu\nu}J^{\mu\nu}$ is the sum of operators $\eJ_0$ and $\eJ_{\vec q}$ that act on $\widehat\H_0$ and ${\eV}_{\vec q}$, respectively. For the hypermultiplet,
 ${\eV}_{\vec q}$ is trivial so $\eJ_{\vec q}=0$ and direct evaluation of the left hand side of (\ref{mortox}) using (\ref{mertz}) leads to the formula for $\hH$ given in eqn. (\ref{heff}).  In general, (\ref{mortox}) implies that
 \begin{equation}\label{norox} \hH=\frac{P^2}{2M}-\TT^{-\mu\nu}L_{\mu\nu}-\eJ_{\vec q}. \end{equation}
 The role of the $\eJ_{\vec q}$ term is to ensure that $\hH+\eJ_{\vec q}$ acts only on $\widehat\H_0$ and not on ${\eV}_{\vec q}$.  
 
 Now to evaluate the contribution of the BPS states of charge $\vec q$  to the GV formula, we need to evaluate $\Tr_{\widehat\H_{\vec q}}'\,(-1)^F\exp(-2\pi e^\sigma \hH)$.
 The trace factors as a trace in $\widehat\H_0$ times a trace in ${\eV}_{\vec q}$.  The trace in $\widehat\H_0$ is the one that we already evaluated in discussing the hypermultiplet.  
 In acting on ${\eV}_{\vec q}$, $\hH$ can be replaced by $-\eJ_{\vec q}$, so the  trace in ${\eV}_{\vec q}$ simply gives $\Tr_{{\eV}_{\vec q}}(-1)^F \exp(2\pi e^\sigma \eJ_{\vec q})$.   Using $\eJ_{\vec q}=\frac{e^{\sigma/2}}{4}\JJ_{\vec q}$ (eqn. (\ref{lixt})) and the usual formula $e^{3\sigma/2}=-i/2\X^0$, this trace is 
  $\Tr_{{\eV}_{\vec q}}(-1)^F \exp(-i\pi  \JJ_{\vec q}/4\X^0)$.
 
 The contribution of BPS states of  charges $\vec q$ propagating once around the circle to the
 GV formula is obtained by just including this trace in (\ref{zelmo}):
 \begin{equation}\label{zomobo}-\int \frac{\d^4x\d^4\theta}{(2\pi)^4} \sqrt{g^E}\Tr_{{\eV}_{\vec q}}\left[(-1)^F \exp(-i\pi \JJ_{\vec q}/4\X^0)\right]
  \exp\left(2\pi i\sum_I q_I \ZZ^I\right)\frac{\frac{1}{64}\pi^2\W^2}{\sin^{2}\left(\frac{\pi\sqrt{\W^2}}{8\X^0}\right)}.
\end{equation}
This can be extended as before to include multiple windings:
 \begin{equation}\label{zomoboz}-\int \frac{\d^4x\d^4\theta}{(2\pi)^4} \sqrt{g^E}\sum_{k=1}^\infty\frac{1}{k}\Tr_{{\eV}_{\vec q}}\left[(-1)^F 
 \exp(-i\pi k
 \JJ_{\vec q}/4\X^0) \right]
 \exp\left(2\pi ik\sum_I q_I \ZZ^I\right)\frac{\frac{1}{64}\pi^2\W^2}{\sin^{2}\left(\frac{\pi k\sqrt{\W^2}}{8\X^0}\right)}.    \end{equation}
 To get the complete GV formula, we need to sum this formula over all possible charges $\vec q$.  But
 states with $\zeta(\vec q)<0$ do not contribute to the GV formula since they preserve the wrong supersymmetry.  The complete GV formula is thus
 \begin{equation}\label{womombo}-\int \frac{\d^4x\d^4\theta}{(2\pi)^4} \sqrt{g^E}\sum_{q|\zeta(q)\geq 0}\sum_{k=1}^\infty\frac{1}{k}\Tr_{{\eV}_{\vec q}}\left[(-1)^F 
 \exp(-i\pi k
 \JJ_{\vec q}/4\X^0) \right]
 \exp\left(2\pi ik\sum_I q_I \ZZ^I\right)\frac{\frac{1}{64}\pi^2\W^2}{\sin^{2}\left(\frac{\pi k\sqrt{\W^2}}{8\X^0}\right)}.  \end{equation}
 Actually, our derivation has assumed that $\zeta(\vec q)>0$, not just $\zeta(\vec q)\geq 0$, because we have assumed that the BPS states under discussion
 have a strictly positive mass in five dimensions.   This means that our analysis does not apply to BPS states with $\vec q=0$, since such states are always
 massless in five dimensions.   This case requires a different
 derivation, but with a suitable interpretation of what is meant by ${\eV}_{\vec q}$ and with $\JJ_{\vec q}$ set to 0,
 the formula (\ref{womombo}) also gives correctly the contribution of BPS states with $\vec q=0$, as we will learn in section \ref{schwfields}.
Our derivation also breaks down for BPS states with $\vec q\not=0$ and $\zeta(\vec q)=0$, but this case is nongeneric in the sense that it arises
only if the Kahler moduli of $Y$ are varied to approach a boundary of the Kahler cone.

 \subsubsection{Concrete Formulas}\label{expform}
 
 To make the formula  (\ref{womombo}) explicit, we need to know how to compute the space of BPS states of charge $\vec q$.
 M-theory in general and M2-branes in particular are not sufficiently well understood for it to be possible at present, for a given Calabi-Yau manifold $Y$,
 to give a full answer to this question.  Potential complications include strong coupling
 of multiply-wrapped M2-branes, singularities in the moduli space of holomorphic curves in a Calabi-Yau manifold, 
 bubbling, and the interplay of all of these. 
 Luckily, there are favorable situations in which it is possible to explicitly determine the space of BPS states with charge $\vec q$ and show that the action 
 of the supersymmetry algebra is as described above.    Here, essentially 
 following \cite{GV2}, we will just summarize a few highlights, leaving some further details for appendix \ref{holc}.

To quantize an M2-brane with worldvolume  $\R\times \Sigma\subset \R^5\times Y$, we have to quantize the fermions
that live on the M2-brane.  These transform as spinors on $\Sigma$ with values in (positive chirality) spinors of the 
normal bundle to the M2-brane worldvolume. As is usual in Kaluza-Klein
reduction,
the modes that have to be included in the low energy description are the zero-modes along the compact manifold
 (in this case, the zero-modes of
fields propagating on $\Sigma$).  
    Half of the M2-brane  fermions transform under $SU(2)_\ell\times SU(2)_r$ as
$(1/2,0)$ and half transform as $(0,1/2)$. Zero-modes of the $(0,1/2)$ fermions are related by supersymmetry to 
infinitesimal deformations of the complex submanifold
$\Sigma\subset Y$. We say that $\Sigma$ is ``rigid'' or ``isolated'' if there are no such fermion zero-modes, and we consider this case first.

For $\Sigma$ rigid,   the efective quantum mechanics is obtained just by quantizing the fermion 
zero-modes that transform as $(1/2,0)$, along
with the center of mass coordinates $x^\mu$; there are no other bosonic or fermionic zero-modes.  
The M2-brane fermions that transform as $(1/2,0)$ can be interpreted
as differential forms on $\Sigma$.  If $\Sigma$ is of genus 0, its non-zero Betti numbers are $b_0=b_2=1$.  
The corresponding zero-modes are precisely the
fermionic collective coordinates $\psi_{A i}$ that we included in studying the hypermultiplet.   
However, for $\g>0$, one has $b_1=2\g$, leading to additional zero-modes consisting of $2\g$ copies of the $(1/2,0)$
representation of the rotation group.  In the effective quantum mechanical problem on $\R\times \Sigma$, where $\R$ parametrizes the time, these zero
modes lead to $2\g$ fields $\rho_{A\sigma}(t),\,\sigma=1,\dots,\g$ that correspond to $(1,0)$-forms on $\Sigma$ and $2\g$ more fields 
$\t\rho_{A\sigma}(t)$, $\sigma=1,\dots,
\g$ that correspond to $(0,1)$-forms.  The action for these modes is 
\begin{equation}\label{oxic}\S_\rho=\int\d t\,\sum_{\sigma=1}^\sg\left(i\t\rho_{A\sigma}\frac{\d}{\d t}\rho^A_\sigma+
\frac{i}{2}\TT^-_{AB} \t\rho^A_\sigma\rho^B_\sigma \right), \end{equation}
and the corresponding Hamiltonian is 
\begin{equation}\label{helox} \hH_\rho=-\eJ_\rho=-\sum_{\sigma=1}^\sg \frac{i}{2} \TT^-_{AB}\t\rho^A_\sigma \rho^B_\sigma.\end{equation}

The problem of quantizing the four fermions $\t\rho_{A\sigma},\,\rho_{B\sigma}$, for $A,B=1,2$ and a fixed value of $\sigma$,
is isomorphic to the problem of quantizing the four fermions $\psi_{Ai}$ that appear already in the study of the hypermultiplet.
Quantization of this system gives the familiar spin content $2(0,0)\oplus (1/2,0)$
of a massive BPS hypermultiplet, described by a four-dimensional Hilbert space $\sH$.  For this set of four states, $\Tr\,(-1)^F\exp(-i\pi\JJ/8\X^0)=\Tr\,(-1)^F\exp(2\pi e^\sigma\eJ)=-4\sin^2(\pi \sqrt{W^2}/8\X^0)$.  The full set of
fermions $\t\rho_{A\sigma},\,\rho_{B\sigma}$ consists of $\g$ copies of this spectrum, leading to  $\Tr\,(-1)^F\exp(2\pi e^\sigma\eJ)=(-1)^\sg(4\sin^2(\pi \sqrt{W^2}/8\X^0))^\sg$.
From (\ref{zomoboz}), it then follows that the contribution to the GV formula of BPS states that arise from an M2-brane wrapped on $\Sigma$ is
\begin{equation}\label{morpho}\int\frac{\d^4x\d^4\theta}{(2\pi)^4}  \sum_{k=1}^\infty\frac{1}{k} (-1)^{\sg-1}   \exp\left(2\pi ik\sum_I q_I \ZZ^I\right)\frac{\frac{1}{64}\pi^2\W^2}{\sin^{2-2\sg}(\pi k \sqrt{\W^2}/8\X^0)}. \end{equation}
We write $\sH_\sg$ for the space obtained by quantizing $\t\rho_{A\sigma},\rho_{B\sigma}$, $\sigma=1,\dots,\g$.  Thus, $\sH_\sg$ is the tensor
product of $\g$ copies of $\sH$, that is $\g$ copies of $2(0,0)\oplus (1/2,0)$. 

\def\hh{\widehat}
This description of  $\sH_\sg$ makes manifest the action of $SU(2)_\ell$ (and the trivial action of $SU(2)_r$). However, 
as preparation for the case that $\Sigma$ is not rigid, it is helpful 
to describe $\sH_\sg$
in another way.   Here, for each value of $A=1,2$, we combine together   $\rho_{A\sigma}$, which is a $(1,0)$-form on $\Sigma$,
with $\t\rho_{A\sigma}$, which is a $(0,1)$-form on $\Sigma$, to make a field $\hh\rho_{A y}$, where $y=1,\dots, 2\g$ labels the choice of a harmonic 1-form on $\Sigma$.
Let $\veps_{yz}$ be the intersection pairing on $H^1(\Sigma,\Z)$.  The canonical anticommutation relations of $\hh\rho_{Ay}$
are $\{\hh\rho_{A y},\hh\rho_{Bz}\}=\veps_{AB}\veps_{yz}$.  To make this look slightly more familiar, let us denote $\rho_{Ay}$ as
$\rho_{+y}$ or $\rho_{-y}$, depending on the value of $A$.  We also define $\rho_-^y=\veps^{yz}\rho_{-z}$. Then the canonical anticommutators are 
\begin{align}\label{cancom} \{\rho_{-}^y,\rho_{-}^z\}=\{\rho_{+y},\rho_{+z}\}&=0\cr
                                                 \{\rho_{-}^y,\rho_{+z}\}&=\delta^y_z. \end{align}
We can regard $\rho_{+z}$ as a set of fermion creation operators and $\rho_-^y$ as the corresponding annihilation operators.  
The full space of states is a fermion Fock space, made by repeatedly acting with $\rho_+$ on a ``ground state'' that is annihilated
by $\rho_-$.  Since $\rho_+$
is an element of the first de Rham cohomology group $H^1(\Sigma)$, the one-particle states are a copy of $H^1(\Sigma)$.  The space of $k$-fermion
states is then the antisymmetric tensor product of $k$ copies of $H^1(\Sigma)$ ; we denote this as
 $\wedge^k H^1(\Sigma)$. The space $\sH_\sg$ obtained by quantizing the fermions is obtained
by summing this over $k$:
\begin{equation}\label{ancom}\sH_\sg=\oplus_{k=0}^{2\sg}\wedge^k H^1(\Sigma). \end{equation}
This description of $\sH_\sg$ emphasizes its dependence on $\Sigma$, but hides the action of $SU(2)_\ell$.  As explained in \cite{GV2}, to understand the $SU(2)_\ell$ action, it helps to recognize that
$\sH_\sg$ can be interpreted as the cohomology of the Jacobian of $\Sigma$.        Recall that the Jacobian of $\Sigma$ 
is a Kahler manifold and that there is a natural
Lefschetz $SU(2)$ action on the cohomology of any Kahler manifold (and more generally on the cohomology of a Kahler manifold with values in a flat vector
bundle).  The $SU(2)_\ell$ action on $\sH_\sg$ can be understood as the natural Lefschetz $SU(2)$ action on the cohomology of the Jacobian.                                          

If $\Sigma$ is not rigid, then in general there is a moduli space $\M$ that parametrizes the bosonic deformations of $\Sigma$.  The only case in which it is straightforward
to describe the BPS states that arise from an M2-brane simply-wrapped on the curves parametrized by $\M$
 is the case that $\M$ is smooth\footnote{Technically, by saying that $\M$ is ``smooth,'' we mean that the deformation theory of $\SIgma\subset Y$ is unobstructed
 so that in particular the infinitesimal deformations of $\Sigma$ represent tangent vectors to $\M$.} and compact  and parametrizes a family
of smooth curves $\Sigma\subset Y$.  
(It is quite exceptional for all these conditions to be satisfied.)  In this case, the effective quantum mechanics problem
that describes the degrees of freedom parametrized by  $\M$ and takes into account the zero-modes of $(0,1/2)$ fermions is the theory of differential forms on $\M$. However, remembering the $(1/2,0)$ fermions, these are differential forms with values in $\sH_\sg$.  As $\Sigma$ varies, 
$H^1(\Sigma)$ varies as the fiber of a flat vector bundle over $\M$.  (We do not have to worry about $\Sigma$ developing singularities
because of our assumption that $\Sigma$ is always smooth.)  Since $\sH_\sg$ is constructed from $H^1(\Sigma)$ as in 
(\ref{ancom}), it also varies as the fiber of a flat vector bundle.
The space of supersymmetric
states in this situation is the  de Rham cohomology of $\M$ with values in $\sH_\sg$; we denote this as $H^*(\M;\sH_\sg)$.  
This is the  contribution of $\Sigma$ to  
${\eV}_{\vec q}$.  The $SU(2)_r$ action is the natural Lefschetz $SU(2)$ action on $H^*(\M;\sH_\sg)$ (now making use of the fact that $\M$ is a Kahler
manifold), and the $SU(2)_\ell$ action comes
from the action of $SU(2)_\ell$ on $\sH_\sg$. 

It is explained in \cite{GV2} that $H^*(\M;\sH_\sg)$ has a natural interpretation in terms of 4d BPS states in Type IIA superstring theory on
$Y$.   To determine the space of 4d BPS states, one has to quantize a suitable D-brane moduli space. Given
the assumption that $\Sigma$ is always smooth, one can argue
that this quantization gives again $H^*(\M;\sH_\sg)$.  (The argument involves describing the D-brane moduli space as a fiber
bundle over $\M$ and computing its cohomology by a Leray spectral sequence.)  Thus, under the hypothesis that $\Sigma$ is always smooth, one expects that 4d BPS
states always descend in a simple way from 5d BPS states (recall from section \ref{tenc} that in general one does not expect something as simple
as this).  The hypothesis that $\Sigma$ is always smooth is almost never satisfied.  However \cite{HST}, there is a well-developed mathematical
theory in which  the hypothesis
that $\Sigma$ is always smooth is replaced with the hypothesis that $\Sigma$ varies only in a Fano subvariety $W\subset Y$.  Under this hypothesis,
it is proved (with a precise set of mathematical definitions that hopefully match correctly  the physics) that $H^*(\M,\sH_\sg)$,
which is the space of 5d BPS states, agrees with what one would get in $d=4$ by quantizing the D-brane moduli space.  Thus this really does
seem to give an interesting and perhaps surprising situation in which 4d BPS states descend simply from five dimensions.

The GV formula is often written in the following way.  If one permits oneself to take formal sums and differences of vector
spaces, then any $\Z_2$-graded representation of $SU(2)_\ell$ can be expanded as $\oplus_{\sg=0}^\infty A_{\sg}\otimes \sH_\sg$,
where $\sH_\sg$ is the $SU(2)_\ell$ representation defined as the tensor product of $\g$ copies of $2(0)\oplus (1/2)$
 and $A_{\sg}$ is a $\Z_2$-graded vector space with trivial action of $SU(2)_\ell$.  In particular,
the space of BPS states of charge $\vec q$  is formally a sum $\oplus_{\sg=0}^\infty A_{\sg,
\vec q}\otimes \sH_\sg$.    Set $a_{\sg,\vec q}=\Tr_{A_{\ssg,\vec q}}(-1)^F$.  (Thus,
a rigid curve $\Sigma\subset Y$ of genus $\g$ and homology class $\vec q$ contributes 1 to $a_{ \sg,\vec q}$, and 0 to $a_{\sg',\vec q}$ for $\g'\not=\g$.)  The complete GV formula can be written
\begin{equation}\label{morphoxic}\int\frac{\d^4x\d^4\theta}{(2\pi)^4}\sum_{\vec q|\zeta(\vec q)\geq 0}\sum_{\sg=0}^\infty  \sum_{k=1}^\infty\frac{1}{k} (-1)^{\sg-1}  a_{\sg,\vec q} \exp\left(2\pi ik\sum_I q_I \ZZ^I\right)\frac{\frac{1}{64}\pi^2\W^2}{\sin^{2-2\sg}(\pi k\sqrt{\W^2}/8\X^0)}. \end{equation}

 \section{The Schwinger Calculation With Fields}\label{schwfields}
 
 \subsection{Preliminary Reduction}\label{prelim}
 
The group of rotations that preserves the momentum vector of a massless particle in five dimensions is $SO(3)$, a subgroup of the corresponding
group $SO(4)$ for a massive particle at rest.  The spin of a five-dimensional massless particle is measured by a representation
 of the double cover of this $SO(3)$.  This double cover is a diagonal subgroup  $SU(2)_\Delta\subset SU(2)_\ell\times SU(2)_r$ of the group
 that measures the spin of a massive particle.  
 
In a five-dimensional theory with minimal supersymmetry (eight supercharges), the states of a massless particle with specified momentum
are annihilated by four of the supercharges and furnish a representation of the other four supersymmetries along with $SU(2)_\Delta$.  The minimal
such representation transforms under $SU(2)_\Delta$ as $W=2(0)\oplus (1/2)$  (that is, two copies of spin 0 and one copy of spin $1/2$).  A general
irreducible representation of supersymmetry and $SU(2)_\Delta$ is simply the tensor product of $W$ with the spin $j$ representation of $SU(2)_\Delta$,
for some $j\in \frac{1}{2}\Z$.  We write this tensor product as  $R_j=(j)\otimes W$.  In particular, $R_0=(0)\otimes W\cong W$, since $(0)$ is the trivial
1-dimensional representation of $SU(2)_\Delta$.

Taking CPT into account, $R_j$ must appear in the spectrum 
an even number of times if $j$ is an integer  but may appear any integer number of times if $j$ is a half-integer.
The basic massless  hypermultiplet $H$, vector multiplet $V$, and supergravity multiplet $G$ are
\begin{align}\label{zelk} H & = 2R_0 \cr V&=R_{1/2} \cr G&=R_{3/2}. \end{align}

We expect that a combination of massless supermultiplets that could be deformed in a supersymmetric fashion to a massive non-BPS multiplet does not
contribute to the GV formula. 
(We have seen in section \ref{collective} that a massive non-BPS supermultiplet does not contribute to the GV formula.)
 For example, the  combination $H\oplus V$ can be deformed by Higgsing to a massive non-BPS vector multiplet.
  To see this, notice that such a supermultiplet must realize eight supercharges,  four of which transform as $(1/2,0)$ under
$SU(2)_\ell\times SU(2)_r$ and four as $(0,1/2)$.  The basic such representation is the massive vector multiplet $W_\ell\otimes W_r$, 
where $W_\ell $ admits the action of one set of four
supersymmetries and $W_r$ admits the action of the other set.  ($W_\ell$ consists of four states transforming as $2(0,0)\oplus (1/2,0)$ while $W_r$ consists
of four states transforming as $2(0,0)\oplus (0,1/2)$.)    When we restrict to $SU(2)_\Delta\subset SU(2)_\ell\otimes SU(2)_r$ and only four supersymmetries, 
we can identify both $SU(2)_\ell$ and $SU(2)_r$ with $SU(2)_\Delta$ and also ignore the supersymmetries that act on (say) $W_\ell$.  Then the massive non-BPS vector multiplet becomes
$W\otimes W = (2(0)\oplus (1/2))\otimes W = 2 R_0\oplus R_{1/2}=H\oplus V.$  Accordingly, we expect that the combination $H\oplus V$ of massless supermultiplets
does not contribute to the GV formula.

Since  $ W\otimes W$ can be deformed in a supersymmetric fashion to a massive non-BPS multiplet, the same is true, for any $j\in \frac{1}{2}\Z$, of
$(j)\otimes W\otimes W$.  For $j>0$, this is the
same as $(j)\otimes (2(0)\oplus (1/2))\otimes W\cong R_{j+1/2}\oplus 2 R_j\oplus R_{j-1/2}$.  So we expect that any such combination does not contribute to the GV
formula.  For $j=1$ or $j=1/2$, we get the combinations $R_{3/2}\oplus 2 R_1\oplus R_{1/2}$ and $R_1\oplus 2R_{1/2}\oplus R_0$. Taking linear combinations of
these expressions and $H\oplus V=2R_0\oplus R_{1/2}$,
we are led to expect  that $R_{3/2}\oplus 4R_0=G\oplus 2H$ does not contribute to the GV formula.

Granted this, the contribution of $n_H$ hypermultiplets, $n_V$ vector multiplets, and $n_G$ supergravity multiplets is equivalent to the contribution of
$n_H-n_V-2n_G$ hypermultiplets. This combination has an interesting interpretation.  The Betti numbers of a Calabi-Yau three-fold $Y$ obey $b_0=1$, $b_1=0$,
and $b_i=b_{6-i}$.  Accordingly, the Euler characteristic of $Y$ is $\chi(Y)=2+2b_2-b_3$.    Generically (as long as one stays away from boundaries of the Kahler
cone of $Y$), massless states in M-theory compactification on $Y$ come entirely from classical dimensional reduction on $Y$ of the eleven-dimensional supergravity
multiplet.  With this assumption, the number of vector multiplets is $n_V=b_2-1$, the number of hypermultiplets
is $n_H=b_3/2$ (in six dimensions, $b_3$ is always even), and the number of supergravity multiplets is $n_G=1$.  Therefore, $n_H-n_V-2n_G=\frac{b_3}{2}
-b_2-1=-\frac{1}{2}\chi(Y)$.  So the total contribution to the GV formula from massless states in five dimensions, away from boundaries of the
Kahler cone, is $-\frac{1}{2}\chi(Y)$ times the contribution of a single massless hypermultiplet.

In section \ref{hypercalc}, we will calculate the contribution to the GV formula of a massless hypermultiplet.  As explained in section \ref{pf}, this calculation cannot
be naturally performed in the approach via 5d particles.  But instead, since there is a natural field theory for a 5d massless
hypermultiplet, there is no problem to perform the calculation in terms of fields.  In fact, it is straightforward to generalize the field theory computation to a 5d {\it massive}
BPS hypermultiplet, and we will do so.  (As explained in section \ref{pf}, near a boundary of the Kahler cone of $Y$, there 
can be a massive BPS hypermultiplet that is light enough so that a description in 5d field theory makes sense.)   Once one formulates the computation in 5d field
theory, it is natural to make a Kaluza-Klein reduction to four dimensions and to write the answer as a sum over contributions of 4d mass eigenstates.  This
gives a representation of the answer in terms of a sum over states of definite momentum around the M-theory circle, in contrast to the particle approach of
section \ref{schwpart} that gives the answer as a sum over configurations of definite winding number.  The two representations are related by a Poisson resummation.
The winding number representation is usually  more useful.

The upshot of the computation is to show that the contribution to the GV formula of a massless hypermultiplet is the obvious zero mass limit of the contribution
of a massive hypermultiplet, which was computed in section \ref{schwpart}.  When the answer is stated this way, one may feel that one does not  need
to actually do the field theory computation for the hypermultiplet: the computation of section \ref{schwpart} is valid for a hypermultiplet of any non-zero mass, so could not we understand the zero mass case as a limit from non-zero mass? However, we find it instructive
to do the explicit computation with 4d mass eigenstates.   It is particularly illuminating to see how an answer emerges
 that is holomorphic in the 4d central charge $\cZ  $, even though naively a Schwinger-like calculation depends only on the particle mass $|\cZ  |$.   Moreover, it turns out that there is a subtlety in the zero mass case, first identified in \cite{MD},
 that is best understood by performing a computation with 4d mass eigenstates.

We should stress that we consider the argument that was used to express the contribution of the supergravity multiplet as $-2$ times the contribution
of a massless hypermultiplet to be somewhat heuristic.  In the particle treatment of section \ref{schwpart}, we had a very clear argument that a massive non-BPS
superparticle cannot contribute to the GV formula.  In general, we do not have an equally clear argument for the 
analogous statement in field theory. For the special case of $H\oplus V$, there is a clear argument, since the deformation to a massive non-BPS 
multiplet can be realized physically by Higgsing.  

To interpret in the language of eqn. (\ref{womombo})  the statements that $H\oplus V$ and $2H\oplus G$ do not contribute to the GV formula, 
we have to be slightly formal about what we mean by the contribution
of a massless vector multiplet $V$ or of the supergravity multiplet $G$ to the vector space ${\eV}_{\vec q}$ for $\vec q=0$.    
To define ${\eV}_{\vec q=0}$, we are supposed to write the space of BPS
states of given momentum as the tensor product of the space of states for a hypermultiplet with some vector space ${\eV}_{\vec q=0}$.  Because of the
fact that the hypermultiplet $H= 2R_0$ is two copies of $R_0$, while $V$ and $G$ are not divisible by 2,
 to define ${\eV}_{\vec q=0}$ we would have to divide by 2 -- an operation that does not make
sense for vector spaces, though it makes sense for the trace that we actually need in eqn. (\ref{womombo}).  Since $H=2R_0$ and $V=(1/2)\otimes R_0$,
the contribution of $V$ to ${\eV}_{\vec q}$ is formally $\frac{1}{2}(1/2)$, that is one-half a copy of the spin $1/2$ representation.  This
answer means that the contribution
of $V$ to $\Tr_{{\eV}_{\vec q}}\,(-1)^F=\frac{1}{2}\Tr_{(1/2)}(-1)^F=-1$, so that the contribution of $V$ to the GV formula is the same
as the contribution of $-H$.  Likewise, the contribution of $G$ to ${\eV}_{\vec q}$ is formally $\frac{1}{2} (3/2)$, meaning that its contribution to 
$\Tr_{{\eV}_{\vec q}}(-1)^F$ is $\frac{1}{2}\Tr_{(3/2)}\,(-1)^F=-2$, reproducing the fact that $G$ makes the same contribution to the GV formula as $-2H$.   
It is unappealing that $\eV_0$ does not exist and we must formally divide by 2.  A possibly more natural approach is
to place a factor of $1/2$ in front of the $\vec q=0$
contribution in (\ref{womombo}).  The intuition in doing this would be that since BPS states with $\zeta(\vec q)>0$ contribute with weight 1
 in (\ref{womombo}) and those
with $\zeta(\vec q)<0$ contribute with weight 0, it is fairly natural to say that BPS states with $\vec q=0$ and hence  $\zeta(\vec q)=0$ contribute with weight 1/2.
If we do this, we would say that the contributions of $H$, $V$, and $G$ to $\eV_{\vec q=0}$ are respectively $2(0)$, $(1/2)$ and $(3/2)$.

\subsection{Calculation For The Hypermultiplet}\label{hypercalc}

\subsubsection{Preliminaries}\label{afew}
We aim here to compute the one-loop effective action for a 4d BPS hypermultiplet in the graviphoton background.
Actually, in a field theory calculation, it is not difficult to be more general, as long as the fields are slowly varying on
a length scale set by the hypermultiplet mass (or by the graviphoton field strength if the hypermultiplet mass vanishes).
There is actually a good reason to be more general.  In contrast to the particle computation of section  \ref{thecomp}, it
is difficult to do this computation in a manifestly supersymmetric fashion, because there is no convenient and simple
superfield description of a hypermultiplet.  Being able to perturb slightly around the graviphoton background will
help in expressing the answer we get in a supersymmetric form.  A simple perturbation will suffice for our purposes:
rather than taking the four-manifold on which the hypermultiplet propagates to be flat (as in the reduction to
four dimensions of the supersymmetric G\"{o}del solution), we take this metric to be hyper-Kahler, with anti-selfdual
Weyl tensor.  

As remarked in footnote \ref{tumox}, the particle computation for a hypermultiplet, as opposed to a more general 
BPS multiplet, makes use only of the negative chirality supersymmetries $Q_{Ai}$.  The same is true in
the field theory computation: the negative chirality supersymmetries are enough to determine the action
we will use.  An anti-selfdual graviphoton preserves the $Q_{Ai}$ even if
its field strength is not constant,\footnote{This is clear from eqn. (\ref{omfo}). 
By contrast, a gauge field in a vector multiplet must
be selfdual, not anti-selfdual, to preserve the $Q_{Ai}$.  See the remarks following eqn. (\ref{zerg}).}  but these supersymmetries are broken by anti-selfdual Riemannian curvature.

In our calculation, we will treat the scalars in vector multiplets as constants (so that the 5d and 4d central charges $\zeta$ and $\cZ$ 
are constants).  This means our calculation will not determine a contribution to the effective action that does not depend on $\W_{AB}$, that is a prepotential
term $-i\int\d^4x\d^4\theta
\F_0(\X^\Lambda)$.  We will see that our calculation is also not powerful enough to fully understand $\F_1$.  

We will assume that the minimal
hypermultiplet action is sufficient to determine the quantum effective action modulo $D$-terms.  For the particle
computation, we proved the analogous statement in section \ref{graviback}   by a scaling argument.

\subsubsection{The Action}\label{moreaction}

We start with a 5d metric on $M_5=S^1\times M_4$ of the form
\begin{equation}\label{toff}\d s^2=e^{2\sigma}(\d y + B)^2+e^{-\sigma}g_{\mu\nu}\d x^\mu\d x^\nu.\end{equation}
Here $g_{\mu\nu}$ is a hyper-Kahler metric with anti-selfdual Weyl curvature on the four-manifold $M_4$, and $B_\mu$ is a Kaluza-Klein gauge field. We require that
$B_\mu=-i\frac{e^{-3\sigma/2}}{4}U_\mu$, where $U_\mu$ is  the four-dimensional gauge field whose curvature
is the 4d anti-selfdual graviphoton $\WW^-_{\mu\nu}$. It is related to the 5d graviphoton $\TT^-_{\mu\nu}$ by the usual equation (\ref{plorox}). We also adjust curvatures of the 5d gauge fields to be $\d V^I = h^I\TT^-$, just like in the graviphoton background. We will calculate the effective action in a region of $M_4$ in which
the metric is very nearly flat and $\WW^{-}_{\mu\nu}$ is very nearly constant -- so that the background is very close
to the standard graviphoton background. And this slightly curved background exactly preserves one of the useful features of the graviphoton background -- 
the only nonzero gauge field is the graviphoton.

In general,  in supergravity, hypermultiplets parametrize a quaternionic manifold $\mathcal X$.  
However, for a 1-loop computation, we can approximate $\mathcal X$ by a flat manifold, which in the 
case of a single hypermultiplet is just $\R^4$.    We consider the general
case that the hypermultiplet has charges $q_I$ and hence a bare mass $M=\sum_I q_Ih^I$ in five dimensions.  
The action is not just the obvious minimal
coupling of bosons and fermions to a gravitational background, because in five dimensions, the fermions in a hypermultiplet have a non-minimal magnetic moment coupling to the graviphoton field.

We denote the scalars in the hypermultiplet as $q^X$, $X=1,\dots,4$.  The fermions are a pair of spinors
 $\xi^{\alpha p}$ where
$\alpha=1,\dots,4$ is a spinor index of $SO(1,4)$ and $p=1,2$.  The Lagrangian density is 
\begin{equation}\label{hyperaction}\L=-\frac{1}{2}D_M q^X D^Mq^X -\frac{1}{2}M^2 (q^X)^2-
2M\bar\xi^1\xi^2
+ \bar\xi_p\slashed{\mathcal{D}}\xi^p +\frac{3i}{8}h_I (\d V^I)_{MN} \bar\xi_p \Gamma^{MN}\xi^p.
\end{equation}  For $M=0$, this action can be found in \cite{Cucu} (along with its generalization to an arbitrary
system of hypermultiplets).  The mass terms can be generated by a coupling
to a $U(1)$ vector multiplet and giving an expectation value to the scalar in the vector multiplet.
Notice that the fermion mass term explicitly
breaks  an $SU(2)$ symmetry of the massless action (acting on the $p$ index) down to $U(1)$.

The explicit magnetic moment term in (\ref{hyperaction}) does not actually mean that the fermions have a magnetic
moment.  We have to recall that the minimal Dirac Lagrangian for a charged fermion
describes a particle with a magnetic moment; one usually says
 that the particle has a  $g$-factor of 2.  
 Also, in reduction to $d=4$, a contribution to the effective magnetic moment comes from the coupling of fermions to the 5d spin-connection in the action (\ref{hyperaction}), which upon dimensional reduction with the metric (\ref{toff}) generates a magnetic moment coupling to the Kaluza-Klein gauge field $B_\mu$
and hence to the graviphoton.   (This shifts the coefficient of the magnetic moment term in
 eqn. (\ref{act4d}).)
 The net effect 
 for $M\not=0$ (the $M=0$ case has
some subtleties that will appear later) is that the effective magnetic moment vanishes in four
dimensions. That must be so, at least for $M\not=0$, since the particle description of section \ref{graviback} 
made it clear that the fermions in a BPS hypermultiplet
have no magnetic moment.   Vanishing of the effective magnetic moment will be clear in eqn. (\ref{sureg}).

The effective action generated by the hypermultiplet is simply the difference of the logarithms of  boson
and fermion determinants.  A convenient way to calculate this difference is to reduce to four dimensions, expressing
the answer as a sum of contributions of Kaluza-Klein modes of definite mass.   This contrasts with the particle computation, where it is easier to consider orbits of definite winding number around the Kaluza-Klein circle.  A Poisson resummation will be needed to convert the answer obtained as a sum over mass eigenstates
to the answer expressed as a sum over orbits of definite winding.

The  Kaluza-Klein mode with $-n$ units of momentum around the circle is a 4d hypermultiplet with a 4d central charge
\begin{equation}\label{central}
\cZ   = e^{-\sigma/2}(M - i e^{-\sigma}n - ie^{-\sigma}q_I\alpha^I),
\end{equation}
where $\alpha^I$ are the constants that determine the holonomy around the circle of the gauge fields $V^I$.
The mass of the hypermultiplet is $|\cZ  |$.  It will be convenient to simply think of the hypermultiplet as a pair of
complex scalars $\phi^i$, $i=1,2$, and a Dirac fermion $\psi$.    We write $\psi_L$ and $\psi_R$ (or $\psi_A$ and $\psi_{\dot A}$) for the components
of $\psi$ transforming with spin $(1/2,0)$ or $(0,1/2)$ under $SU(2)_\ell\times SU(2)_r$.   After Kaluza-Klein reduction,
the action density  for the
$n^{th}$ mode is
\begin{align}
\label{act4d}
\frac{\mathcal{L}_n}{2\pi} =& -\sum_{i=1}^2 (|\nabla_\mu\phi^i|^2 + |\cZ  \phi^i|^2) + 2\bar\psi_L^c\slashed{D}\psi_R+2\bar\psi_R^c\slashed{D}\psi_L \cr&- 2\cZ  \bar\psi^c_L\psi_L - 2\bar{\cZ  }\,\bar\psi^c_R\psi_R + \frac{i}{4}\WW^-_{\mu\nu}\bar\psi_L^c\gamma^{\mu\nu}\psi_L,
\end{align}
where 
\begin{align}
\nabla_\mu\phi^i&=\partial_\mu\phi^i - i\frac{\bar{\cZ  }}{4}U_\mu\phi^i,\cr
D_\mu\psi &= \partial_\mu\psi + \frac{1}{4}\omega_{\mu}^{ab}\gamma_{ab}\psi - i\frac{\bar{\cZ  }}{4}U_\mu\psi.
\end{align}
Here $\omega_\mu^{ab}$ is the Levi-Civita connection on $M_4$, and $U_\mu$ is the gauge field whose curvature
is $\WW^-_{\mu\nu}$.
Modulo possible terms of higher dimension, this action is actually determined by the 4d supersymmetry algebra (\ref{morfic}),
even if we consider only the $\Qb_{A i}$ supersymmetries and not $\Qb_{\dB j}$.   

\subsubsection{The Computation}\label{morecomp}

As long as $\cZ  \not=0$, the problem of evaluating the bosons and fermion determinants in this problem can be simplified
by integrating out $\psi_R$.  (The case $\cZ  =0$, which means that $M=q_I=0$ and $n=0$, needs special care and
will be treated separately.)  If we eliminate $\psi_R$ classically by solving its equation of motion, the action 
density for $\psi_L$ becomes 
\begin{equation}\label{suref}\frac{\L_n^{\psi_L}}{2\pi}=\frac{2}{\bar \cZ  }\bar\psi^c_L\biggl(\slashed{D}^2 +\frac{i}{8}\bar \cZ  \WW^-_{\mu\nu}
\gamma^{\mu\nu} -\bar \cZ   \cZ  \biggr)\psi_L. \end{equation}
Standard Dirac algebra gives $\slashed{D}^2 +\frac{i}{8}\bar \cZ  \WW^-_{\mu\nu}
\gamma^{\mu\nu} =D_\mu D^\mu$, showing the disappearance of the magnetic moment.  Finally, by absorbing
a factor of $\sqrt{\bar \cZ  }$ in $\psi_L$ (and the same factor in $\bar\psi^c_L$), 
we eliminate the ugly factor of $1/\bar \cZ  $ in front of the kinetic enegy of $\psi_L$.
After these manipulations the action becomes
\begin{equation}\label{sureg}\frac{\L_n^{\psi_L}}{2\pi}={2}\bar\psi^c_L\bigl(D_\mu D^\mu -\bar \cZ   \cZ  \bigr)\psi_L. \end{equation}
We have to be careful to include some constant factors generated by these manipulations.  The Gaussian integral over $\psi_R$ that is used to eliminate $\psi_R$ generates a factor of $\bar \cZ  $ for every mode of $\psi_R$.  The rescaling
of $\psi_L$ multiplies the path integral measure by a factor of $1/\bar \cZ  $ for every mode of $\psi_L$.  
Including these factors and also the determinants coming from functional integrals over $\phi^i$  and $\psi_L$,
the path integral for the $n^{th}$ Kaluza-Klein mode gives
\begin{equation} \label{longo} (\bar \cZ  )^{n_R-n_L}\frac{\det_L(-D^2+|\cZ  |^2)}{\det^2(-\nabla^2+|\cZ  |^2) },\end{equation}
where $\det_L$ is the determinant in the space of left-handed fermions.  Also,  $n_R-n_L$ is formally the difference
between the number of right- and left-handed fermion modes; we interpret this difference as the index of the Dirac
operator, which we denote as $\I$. 

In what follows, we write $\Tr_L$ for a trace in the space of left-handed fermions, and $\Tr$ for a trace in the space
of scalar fields.  Also, we drop the distinction between $D$ and $\nabla$ and write simply 
$-D^2=-g^{\mu\nu}D_\mu D_\nu$ for the Laplacian acting on a field of any spin.  The desired
contribution to the effective action is minus the logarithm of (\ref{longo}) or
\begin{equation}\label{minusac}
-\I\ln(\bar{\cZ}) - \textrm{Tr}_L\ln(|\cZ|^2-D^2) + 2\textrm{Tr} \ln(|\cZ|^2-D^2).
\end{equation}
With the help of
\begin{align}\label{diff}
\ln A = \int_0^{\infty}\frac{\d s}{s}(e^{-s} - e^{-sA}),
\end{align}
we can rewrite (\ref{minusac}) in the form
\begin{equation}\label{wingo}
-\I\ln \bar{\cZ}-
\int_0^{\infty}\frac{\d s}{s}\left(2\Tr-\Tr_L\right)\left(e^{-s(|\cZ|^2-D^2)}-e^{-s}\right).
\end{equation}
This formula is obtained by using the representation (\ref{diff}) of the logarithm for every mode.   When we sum over all modes,
the coefficient of $e^{-s}$ in eqn. (\ref{wingo}) is formally what we might call $n_L-2n_0$, where $n_0$ is the total number of modes of spin 0.
On a hyper-Kahler manifold $M_4$ with anti-selfdual Weyl curvature, the positive chirality spin bundle is simply a trivial bundle of rank 2,
so $n_R$ is the same as $2n_0$ and hence $n_L-2n_0=-\I$.
So an equivalent formula is
\begin{equation}\label{wingorm}
-\I \,\ln\bar{\cZ} -
\int_0^{\infty}\frac{\d s}{s}\left[\left(2\Tr-\Tr_L\right)\left(e^{-s(|\cZ|^2-D^2)}\right)-\I e^{-s}\right],
\end{equation}
Using (\ref{diff}) one more time, this is
\begin{equation}\label{wingor}-
\int_0^{\infty}\frac{\d s}{s}\left[\left(2\Tr-\Tr_L\right)\left(e^{-s(|\cZ|^2-D^2)}\right)-\I e^{-s\bar\cZ}\right].\end{equation}

A Hilbert space $\sH$ consisting of two states transforming under $SU(2)_\ell\times SU(2)_r$ as $(0,0)$ and two
transforming as $(1/2,0)$ was encountered in section \ref{collective}.  It arises upon quantizing a Clifford algebra
generated by four fermions $\psi_{Ai}$ with the familiar anticommutation relations
$\{\psi_{Ai},\psi_{ Bj}\}=\veps_{A B}\veps_{ij}$.   Now we regard $\sH$ as the fiber of a vector bundle over $M_4$
and write $\hsH$ for the space of sections of this bundle.
Clearly, we can rewrite (\ref{wingor}) in the form
\begin{equation}\label{wingox}
 -
\int_0^{\infty}\frac{\d s}{s}\left(\Tr_\hsH(-1)^Fe^{-s(|\cZ|^2-D^2)}-\I e^{-s\bar\cZ}\right).
\end{equation}
We can interpret the operator $\exp(-s(|\cZ|^2-D^2))$ acting on the space $\hsH$ as $\exp(-sH)$ where $H$
is the Hamiltonian derived by quantizing the following superparticle action:
\begin{equation}
\label{supart}
\eurm S=\int \d t\left(-|\cZ|^2 + {\dot{x}^2\over 4}+{\bar{\cZ}\over 4}U_{\mu}\dot{x}^{\mu} +\frac{i}{2}\varepsilon^{ij}\varepsilon^{A B}\psi_{Ai}\nabla_{t}\psi_{Bj}\right).
\end{equation}
Here $x^\mu$ are local coordinates for a point in $M_4$, so that $x^\mu(t)$ describes a particle orbit\footnote{We have normalized
the kinetic energy of $x^\mu$ so that the bosonic Hamiltonian is $P^2$; if the kinetic energy were $\frac{1}{2}\dot x^2$, the Hamiltonian
would be $P^2/2$.} in $M_4$; the $\psi_{Ai}$ are fermi
fields defined along the particle orbit;
and  $\nabla_t=\partial_t + {1\over 4}\dot{x}^{\mu}\omega_{\mu}^{ab}\gamma_{ab}$ is the pullback of the Levi-Civita connection of $M_4$
to the orbit.   To compute $\Tr\,(-1)^F\exp(-sH)$,
we perform a path integral on a circle of circumference $s$, with periodic boundary conditions for fermions, and using
the Euclidean version of the above action:
\begin{equation}\label{onno}
\eurm S_E=\int_0^\beta \d\tau \left(|\cZ|^2+{\dot{x}^2\over 4}-i{\bar{\cZ}\over 4}U_{\mu}\dot{x}^{\mu} +\frac{1}{2}\varepsilon^{ij}\varepsilon^{AB}\psi_{Ai}\nabla_{\tau}\psi_{Bj}\right).
\end{equation}

Clearly, we have arrived at something very similar to what we had in the particle-based calculation.  However,
there are a few key differences.  In section \ref{thecomp}, we had to compute a Euclidean path integral on a circle
of definite radius; this circle was effectively just the M-theory circle.  Now the radius of the circle is an integration
variable, the proper time $s$.  Related to this, in section \ref{thecomp}, we were computing the contribution
of an orbit of definite winding number.  Now we are computing the contribution of a particle of definite Kaluza-Klein
momentum.  

Also, the  computation in \ref{thecomp} was performed in a manifestly supersymmetric framework.  In our present
computation, the starting point was not manifestly supersymmetric (because we lacked a convenient and manifestly supersymmetric description of the hypermultiplet).  It is easy to guess  from eqn. (\ref{onno})
how to express our present computation
in a supersymmetric form.  But to be sure, we will perturb slightly around the supersymmetric G\"odel solution, allowing anti-selfdual Weyl curvature, and verify that the result can be expressed in terms of superfields in the expected way.

In doing this computation, we can assume that the radius of curvature is very large (on a scale set by the particle
mass or the graviphoton field), and that the graviphoton field is nearly constant. 
This being so, the problem can be analyzed in a standard way, using the fact that if $M_4$ were flat and the graviphoton field
exactly constant, the action would be quadratic and the path integral would be simple.  The $F$-terms that
are described by the GV formula have contributions that, when expressed in terms of ordinary fields (and taking the fermions to vanish and
the scalars to be constants),
take the form of $R^2$ times a function of $\WW^-$ only, where $R_{\mu\nu}^{ab}$ is the Riemann tensor.  So
in evaluating the path integral, it suffices to work to quadratic order in $R$, and to ignore covariant derivatives
of $R$ or $\WW^-$.

We set $x^\mu(\tau)=x^\mu+z^\mu(\tau)$, where $x^\mu$ labels a  point in $M_4$, and $z^\mu(0)=z^\mu(s)=0$.
The path integral over $x^\mu(\tau)$ splits as an integral over a field $z^\mu(\tau)$ that vanishes at $\tau=0$ and
an ordinary integral  over $x^\mu$.   Near $x^\mu$, we use Riemann normal coordinates, which
are Euclidean up to second order in $z^\mu$.  In these coordinates, the spin connection is 
\begin{equation}
\omega_{\mu}^{ab}(z)=\frac{1}{2}z^{\nu}R_{\nu\mu}^{\ \ ab} + \O(z^2),
\end{equation}
where the $\O(z^2)$ terms can be ignored as they are proportional to the covariant derivative of the Riemann tensor.
Up to terms of order $z^3$, the part of the action that involves fermions is 
\begin{equation}
\label{kinexp}
\frac{1}{2}\varepsilon^{ij}\varepsilon^{AB}\psi_{Ai}\dot\psi_{Bj} - \frac{1}{16}\dot{z}^\mu z^\nu R_{\nu\mu}^{\ \ ab}\varepsilon^{ij}\psi_{Ai}\gamma^{AB}_{ab}\psi_{Bj}.
\end{equation}
The fermions $\psi_{Ai}(\tau)$ have four zero-modes $\psi_{Ai}^{(0)}$ -- the modes that are independent of $\tau$.  
The action (\ref{kinexp}) contains a coupling $R\psi^{(0)}\psi^{(0)}$, which is the only coupling that can saturate
the fermion zero-modes.  Using this coupling to saturate the zero-modes gives an explicit factor of $R^2$ in the 
path integral, and as we do not wish to compute terms of higher order in $R$, we can drop the coupling of $R$ to
other fermion modes.  The action then reduces to
\begin{equation}\label{hobo}
S_E = \int_0^\beta \d\tau \left[ \frac{\dot{z}^2}{4} - i\frac{\bar{\cZ}}{8}\left(\WW^-_{\nu\mu} - 
\frac{i}{2\bar \cZ}R_{\nu\mu}^{-\,ab}\varepsilon^{ij}\psi^{(0)}_{Ai}
\gamma^{AB}_{ab}\psi^{(0)}_{Bj}\right)z^{\nu}\dot{z}^{\mu} + 
\frac{1}{2}\varepsilon^{ij}\varepsilon^{AB}\psi_{Ai}\dot\psi_{Bj}+|\cZ|^2 \right].
\end{equation}
Now we observe that replacing $iR\psi^{(0)}\psi^{(0)}/\bar \cZ$ by $R\psi^{(0)}\psi^{(0)}$ has the effect of just
multiplying the path integral by $-\bar \cZ^2$.   If we make this replacement, and also set $\psi_{Ai}^{(0)}={\sqrt 2}
\theta_{Ai}$, and finally set $z^\mu=\sqrt 2 y^\mu$, then the action becomes 
\begin{equation}\label{hobot}
S_E = \int_0^s \d\tau \left[ \frac{\dot{y}^2}{2} -
 i\frac{\bar{\cZ}}{4}\W^-_{\mu\nu}y^{\nu}\dot{y}^{\mu} + 
 \frac{1}{2}\varepsilon^{ij}\varepsilon^{AB}\psi_{Ai}\dot\psi_{Bj}+|\cZ|^2 \right],
\end{equation}
where 
\begin{equation}\label{doofol}\W_{\mu\nu}(x,\theta)=
\WW^-_{\mu\nu}(x) +\dots - R^-_{\mu\nu\lambda\rho}(x)
\varepsilon_{ij}\bar\theta^i\sigma^{\lambda\rho}\theta^j +\dots\end{equation}
is the superfield whose bottom component is $\WW^-_{\mu\nu}$.

The constant term $|\cZ|^2$ in the Lagrangian density just multiplies the path integral by $\exp(-s|\cZ|^2)$.
So
\begin{align}\label{ofin}\Tr_{\hsH}(-1)^F& \exp(-sH)=-\frac{e^{-s|\cZ|^2}}{\bar \cZ^2}\int\d^4y\d^4\theta\sqrt{g}\cr & \int\mathcal D' y\,\mathcal D'\psi
\exp\left( - \int_0^s \d\tau\left(  \frac{\dot{y}^2}{2} -
 i\frac{\bar{\cZ}}{4}\W^-_{\mu\nu}y^{\nu}\dot{y}^{\mu} + 
 \frac{1}{2}\varepsilon^{ij}\varepsilon^{AB}\psi_{Ai}\dot\psi_{Bj}\right) \right),\end{align}
 where $\mathcal D'$ represents a path integral over non-zero modes only.   
 Apart from the decoupled fermions $\psi_{Ai}$,  the remaining path integral 
 describes a particle in a constant magnetic field 
 $\bar \cZ\,\W$.  This is a very standard path integral, 
 and one way to evaluate it was described in section \ref{thecomp}.  
  We finally learn that
  \begin{equation}
\label{transwer}
\Tr_{\hsH}(-1)^F\exp\left(-s(|\cZ|^2- D^2)\right)=-{e^{-s|\cZ|^2}\over (2\pi)^4}\int \d^4x \d^4\theta\sqrt{g} \frac{\pi^2\mathcal{W}^2/64}{\sinh^2{s\bar{\cZ}\sqrt{\mathcal{W}^2}\over 8}}.
\end{equation}
(The measure $\d^4\theta$ was defined in eqn. (\ref{wormox}), and the same derivation applies here.)
 
When this is inserted in (\ref{wingor}), we get
\begin{equation}\label{tormo}\int_0^\infty\frac{\d s}{s}\left( \frac{e^{-s|\cZ|^2}}{(2\pi)^4}
\int \d^4x \d^4\theta\sqrt{g} \frac{\pi^2{\W}^2/64}{\sinh^2{s\bar{\cZ}
\sqrt{{\W}^2}\over 8}}+\I e^{-s\bar\cZ}\right). \end{equation}
To see that the integral converges near $s=0$, we observe first that expanding the integrand gives a term
proportional to $1/s^3$, but this term is independent of $\W$ and is annihilated by the $\d^4\theta$ integral.
The next term in the expansion is proportional to $1/s$, but the index theorem for the Dirac operator ensures
that this contribution cancels, so the integral converges for small $s$.   In fact, with the help of the index theorem, (\ref{tormo}) is equivalent to  
\begin{equation}\label{tormon}\int_0^\infty\frac{\d s}{s}\left( \frac{1}{(2\pi)^4}
\int \d^4x \d^4\theta\sqrt{g} \left(e^{-s|\cZ|^2}\frac{\pi^2{\W}^2/64}{\sinh^2{s\bar{\cZ}
\sqrt{{\W}^2}\over 8}}+\frac{\pi^2\W^2}{3\cdot 64} e^{-s\bar\cZ}\right)\right). \end{equation}
Since $\cZ$ has non-negative real part, the integral also converges at large $s$.

To establish holomorphy in $\cZ$, we simply rescale
$s\to s/\bar \cZ\cZ$, to get:
\begin{equation}\label{lormo}\int_0^\infty\frac{\d s}{s}\left( \frac{1}{(2\pi)^4}
\int \d^4x \d^4\theta\sqrt{g} \left(e^{-s}\frac{\pi^2{\W}^2/64}{\sinh^2{s
\sqrt{{\W}^2}\over 8\cZ}}+\frac{\pi^2\W^2}{3\cdot 64}e^{-s/\cZ}\right)\right). \end{equation}
It is very satisfying to see holomorphy emerging even though the particle mass is certainly not holomorphic in $\cZ$.

In this calculation, we have taken $\cZ$ to be a complex constant, rather than a field.  This means that we have not taken into
account fluctuations in the scalar fields in vector multiplets.  When such fluctuations are included, $\cZ$ becomes
a chiral superfield, and the effective action may have an additional contribution,\footnote{Since
the action (\ref{act4d}) depends on the 4d superfields $\ZZ^I$ only in the combination $\sum_I q_I \ZZ^I$
which appears in the central charge, any $\W$-independent function
that we have not computed is a function of $\cZ$ only.  Even without any computation of terms that are independent of $\W$, eqn. (\ref{lormo}) clearly
needs some modification when $\cZ$ is not
constant, if only to ensure that it converges for $s\to 0$.  } not determined in our computation, that depends only on $\cZ$.

\subsubsection{The Case $\cZ=0$}\label{funnycase}

We recall that in this derivation, we assumed at the beginning that $\cZ\not=0$.  Let us separately consider
the case that $\cZ=0$.  This case only arises if $M=q_I=0$ and in addition the Kaluza-Klein momentum $n$ vanishes.
Looking back to the 4d action (\ref{act4d}) with which we started, we see that for $\cZ=n=0$, the scalars $\phi^i$
do not couple to $\WW^-$, but the fermions have a magnetic moment coupling.  (Thus, the case $\cZ=0$ is the only
case in which the fermions have a magnetic moment.)  
This case is simple enough that we can get a very general answer, for arbitrary $M_4$.  

We recall that a Dirac fermion is equivalent to a pair of Weyl (or Majorana) fermions.    So in a notation slightly
different from that in (\ref{act4d}), we have two right-handed fermions $\psi_R^1,$ $\psi_R^2$, and two left-handed
fermions $\bar\psi_L^1$, $\bar\psi_L^2$.  The fermion kinetic energy is
\begin{equation}\label{mixof}\int\d^4x\sqrt g \left(\bar\psi^1_L\slashed{D}\psi_R^1+\bar\psi_L^2\slashed{D}\psi_R^2\right).
\end{equation}
Classically, there is a $U(1)$ symmetry under which $\psi_R^1$ has charge 1,  $ \psi_L^1$ has charge $-1$,
and $\psi_R^2$, $\psi_L^2$ are neutral. ($\bar\psi_L^1$ is just the transpose of $\psi_L^1$, with no complex conjugation involved,
so it has charge $-1$ just like $\psi_L^1$, ensuring the invariance of the fermion kinetic energy.)  However, this $U(1)$ symmetry is violated in a gravitational field.  The net
violation of the symmetry is given by the index $\I$ of the Dirac operator.  Hence, on a four-manifold 
$M_4$ on which $\I\not=0$, the fermion path integral vanishes when the graviphoton field vanishes.
The graviphoton field $\WW^-$ couples to a pair of left fermions
\begin{equation}\label{nixof}\int\d^4x\sqrt g \WW^{-\mu\nu}\bar\psi_L\negthinspace{}^2\gamma_{\mu\nu}\psi_{L}^1. \end{equation}
Thus $\WW^-$ effectively has charge 1 under the symmetry.   If $\I<0$, so that generically a left fermion has
$|\I|$ zero-modes and a right fermion has none, then the insertion in the path integral 
of $|\I|=-\I$ copies of this interaction can give a nonzero result.  The path integral is then proportional
to $(\WW^-)^{-\I}$.  For $\I>0$, the path integral vanishes, if the graviphoton is anti-selfdual as assumed in the above
formulas for the action.
 (It would be inconvenient to restrict our discussion to the case that $M_4$ is hyper-Kahler
with  anti-selfdual
Weyl curvature, as this forces $\I>0$, while we have just seen that the more interesting case is $\I<0$.)

For some purposes, one can
 describe this result by saying that the effective action contains a term $-\frac{\chi}{2}\I\log\WW^-$,
which in supersymmetric language could be derived from an $F$-term
\begin{equation}\label{refro}\frac{\chi}{2}\int\d^4x\d^4\theta\W^2\log\sqrt{\W^2}. \end{equation}
(We recall from section \ref{prelim} that the effective number of massless hypermultiplets is $-\chi/2$.)
For the original calculation leading to a result along these lines, see \cite{MD}.  

However,  this interpretation has some limitations. First, technically, the $\WW^{-\I}$ behavior of the path integral
arises only for $\I<0$, not for $\I>0$.  For $\I>0$, it is not possible to get a nonzero path integral by making
a negative number of $\WW^-$ insertions.  Moreover, the path integral of the fermions under discussion
on a four-manifold of $\I>0$ does not blow up as $\WW^{-\I}$ for $\WW^-\to 0$. Rather, it vanishes identically for 
all $\WW^-$.

Furthermore, if one carries out the $\d^4\theta$ integral in (\ref{refro}),  one gets, in addition to an $\I\log\WW$ ``coupling,''
 a variety of interactions that are singular for $\WW\to 0$ and look  difficult to interpret.

Most fundamentally, the problem with trying to describe this effect by a term in the effective
action  such as (\ref{refro})  is that the effect is fundamentally non-local.  Our derivation has shown that the effect comes entirely
from integrating out particles that are massless in four dimensions, so one should not try to incorporate it into
a 4d Wilsonian effective action.

It was observed in \cite{MD} that an $F$-term of the form $\W^2\log \sqrt{\W^2}$  is not part of  the relation between the
$\F_\sg$'s and the topological string.  Indeed, such an effect is certainly not seen in the perturbative string theory
calculation of \cite{AGNT}.  The reason is clear from a low-energy point of view:  perturbation theory with the interaction
(\ref{nixof}) will never generate a coupling of any number of gravitons to any (positive) number of graviphotons,
since this is prevented by the $U(1)$ symmetry.

\subsubsection{Comparison With The Particle-Based Calculation}\label{compar}

Since we have taken $\cZ$ to be constant, we cannot compute the hypermultiplet contribution to $\F_0$.  But
we can compute its contribution to $\F_\sg$, $\g\geq 1$.  

Using the expansion
\begin{equation}\label{bern}\frac{(x/2)^2}{\sinh^2 (x/2)}=\sum_{n=0}^\infty (1-2n)B_{2n} \frac{x^{2n}}{(2n)!}, \end{equation}
where the $B_{2n}$ are Bernoulli numbers (of alternating sign), and integrating over $s$ term by term, we get
\begin{align}\label{tomox}\F_1&=-i\frac{\log \cZ}{3(32\pi)^2} \cr
\F_\sg & =-i\frac{1}{(16\pi)^2}\frac{B_{2\sg}}{2\g(2\g-2)}(4\cZ)^{2-2\sg}. \end{align}
Since $\mathrm{Re}\,\cZ\geq 0$, the $s$ integral converges and determines in $\F_1$ a definite branch of $\log\cZ$,
namely the one with $|\mathrm{Im}\log\cZ|\leq \pi/2$.  

Eqn. (\ref{tomox}) determines the contribution to $\F_\sg$ of a 4d hypermultiplet of given $\cZ$.  To get the contribution
of a 5d hypermultiplet of mass $M$, we have to sum over the Kaluza-Klein momentum $n$.  This is particularly simple for
$M=0$, which only occurs for $q_I=0$, in which case eqn. (\ref{central}) for the central charge reduces to $\cZ=-in e^{3\sigma/2}=2n\X^0$.    The sum over $n$ can also be performed for $M\not=0$ (see \cite{GV1}), but this does not affect the qualitative point that we
wish to make.

As discussed in  section \ref{funnycase}, for $M=0$, we sum only over $n\not=0$.     The contribution
of a massless 5d hypermultipet to $\F_\sg$ for $\g\geq 2$ is 
\begin{equation}\label{funsum}\F_\sg^{M=0}=-i\sum_{n\not=0}\frac{1}{(16\pi)^2}\frac{B_{2\sg}}{2\g(2\g-2)}(8n\X^0)^{2-2\sg}=-
\frac{i}{(16\pi)^2}\frac{B_{2\sg}}{\g(2\g-2)} (8\X^0)^{2-2\sg}\zeta(2\g-2).\end{equation}

For $\g=1$, the sum over $n$ is divergent.  This should be interpreted as follows.  The one-loop effective action
in five dimensions is potentially ultraviolet divergent, but any such divergence is the integral of a gauge-invariant local expression.  Such an
integral cannot contribute to $\F_1$, as we explained in section \ref{classred}.  Similarly, although our knowledge
of M-theory does not give us much insight about how to regularize the one-loop computation in five dimensions, 
any two regularizations that preserve
5d covariance will differ only by the integral of a gauge-invariant local expression, and will therefore give the same result for
$\F_1$.  Consequently, a computation that preserves 5d symmetry will give a finite and unambiguous answer  for $\F_1$.
The computation that we have performed was based on an expansion in Kaluza-Klein harmonics and did not
preserve the 5d symmetry.  This is why it does not give a satisfactory understanding of  the contribution of a 5d field to $\F_1$.  
   
We would like to compare the hypermultiplet contribution to the effective action as computed in the field-based approach
to the earlier particle-based result (\ref{zelmox}).  The field-based calculation involved a sum over states of definite momentum
around the Kaluza-Klein circle, and the particle-based calculation involved a sum over orbits of definite winding number.
As usual (and essentially as in \cite{GV2}), to convert one to the other, one should perform a Poisson resummation.    In doing this resummation, we should remember
two facts, which turn out to be related.  We cannot really compare the two computations for $\F_1$, because our field-based computation
was not powerful enough to determine the sum over Kaluza-Klein momenta in $\F_1$.  And in the particle-based computation, we do not
want to include a contribution with winding number zero, because this contribution is not meaningful in the context of the particle-based computation.

Since we will not try to make a comparison for $\F_1$, we will ignore the $\W^2 e^{-s/\cZ}$ term in eqn. (\ref{lormo}),
which only contributes to $\F_1$.  To avoid having to worry about the potential divergence of the $s$ integral for $s\to 0$,
 we simply remember that the result of the Poisson resummation should be expanded
in powers of $\W$, keeping only terms of order $\geq 4$.  Also, in performing the Poisson resummation, we will
discard by hand the contribution of winding number $k=0$; it will be clear that this term only contributes to $\F_1$.   

As in eqn. (\ref{pofgo}), we  define $S(\vec q)=2\pi(e^\sigma M-i q_I\alpha^I)=-2\pi iq_IZ^I$, so that the central charge of a particle of Kaluza-Klein
momentum $n$ is $\cZ=e^{-3\sigma/2}(-in+S(\vec q)/2\pi)$.  We also rescale the Schwinger parameter by $s\to s\cZ$ (this can be accompanied by a rotation
of the integration contour in the complex plane, so that we still integrate over the positive $s$ axis).  The  sum and integral to be performed are
then
\begin{equation}\int\frac{\d^4x\d^4\theta}{(2\pi)^4}\sqrt g\sum_{n\in\Z}\int_0^\infty\frac{\d s}{s}\exp\left(-se^{-3\sigma/2}S(\vec q)/2\pi\right)\exp(inse^{-3\sigma/2})\frac{\pi^2\W^2/64}
{\sinh^2(s\sqrt{\W^2}/8)}. \end{equation}
Upon using $\sum_{n\in\Z}e^{in\theta}=2\pi\sum_{k\in\Z}\delta(\theta-2\pi k)$, we get 
\begin{equation}\label{weldo}\int\frac{\d^4x\d^4\theta}{(2\pi)^4}\sqrt g\sum_{k\in\Z}\int_0^\infty \frac{\d s}{s}\exp\left(-se^{-3\sigma/2}S(\vec q)/2\pi\right)
2\pi \delta(se^{-3\sigma/2}-2\pi k)\frac{\pi^2\W^2/64}
{\sinh^2(s\sqrt{\W^2}/8)}. \end{equation}
We see that, as expected from the particle computation, there is no contribution from $k<0$, while $k=0$ formally makes only a contribution to $\F_1$,
which we discard.   Integrating over $s$ with the help of the delta functions,  promoting $S(\vec q)=-2\pi iq_IZ^I$ to a superfield $\S(\vec q)=-2\pi iq_I\ZZ^I$ (to get a formula that is valid even when the $\ZZ^I$ are not taken to be constants), and introducing again
$\X_0=-ie^{-3\sigma/2}/2$, we recover the familiar result 
\begin{equation}\label{zelmor}-\sum_{k=1}^\infty\frac{1}{k}\int \frac{\d^4x\d^4\theta}{(2\pi)^4} \sqrt{g} \exp\left(2\pi ik\sum_I q_I \ZZ^I\right)\frac{\frac{1}{64}\pi^2\W^2}{\sin^{2}\left(\frac{\pi k\sqrt{\W^2}}{8\X^0}\right)}
\end{equation}
for the hypermultiplet contribution.

\section{D4-Branes And the OV Formula}\label{fourov}
 
\subsection{Overview}\label{ovover}
 
As explained in section \ref{oov}, the Ooguri-Vafa (OV) formula is an open-string analog of the GV formula.  It arises if in
Type IIA superstring theory on $\R^4\times Y$,
one introduces a D4-brane supported on $\R^2\times L$, with $\R^2\subset \R^4$ and $L$ a special Lagrangian submanifold of $Y$.  This breaks at least
half of the supersymmetry, leaving at most four supercharges. In the notation of eqn. (\ref{mofo}), the surviving supersymmetries are half of the
negative chirality supersymmetries $\Qb_{Ai}$ and half of the positive chirality $\Qb_{\dot A j}$.  

To get an analog of the GV formula, we want  to turn on an anti-selfdual graviphoton field $\WW^-_{\mu\nu}$ while preserving
supersymmetries of both chiralities.  Precisely as in the derivation of the GV formula, the full set of  supersymmetries -- in this case the
four supersymmetries that survive when a D4-brane is introduced -- is needed to determine the couplings of BPS states in a graviphoton background that
affect the OV formula.   The anticommutator of two positive chirality supersymmetries $\Qb_{\dot A j}$ contains the anti-selfdual rotation generator
$\JJ=\WW^-_{\mu\nu}J^{\mu\nu}$, so to preserve all four supersymmetries, the D4-brane must 
be invariant under a rotation generated by $\WW^-_{\mu\nu}$.  

This means that only one component of $\WW^-_{\mu\nu}$ can be turned on, depending on the choice of which subspace $\R^2\subset\R^4$ supports the D4-brane.
For example, let $x^1,\dots, x^4$ be Euclidean coordinates for $\R^4$ and let $\R^2\subset\R^4$ be defined by the conditions $x^3=x^4=0$.  
For $\WW^-$ to be anti-selfdual and $\JJ$ to be a symmetry of this particular $\R^2$, $\WW^-$ must be a multiple of $\d x^1\wedge \d x^2-\d x^3\wedge \d x^4$.
Thus, $\WW^-$ is completely determined by its component $\WW^-_{12}$, which we will call $\WW_\parz/2$.  
(The factor of $1/2$ is included to  match to formulas such as (\ref{worzo}) that were used in discussing the GV formula.)  When
restricted to the brane world-volume $\R^2\subset\R^4$,
  $\WW_\parz$ is the bottom component of a 2d superfield that we will call $\W_\parz$.  
  
  Similarly, the ``perpendicular'' components $\WW_\perp$ of $\WW^-$, when restricted to the brane worldvolume, 
  are the bottom components of chiral
  superfields $\W_\perp$.  It appears that, as in eqn. (4.2) of \cite{OV}, one can define chiral couplings supported on the brane
  that depend on both $\W_\parz$ and $\W_\perp^2$.  However, the OV formula determines only the special case of these
  couplings with $\W_\perp=0$, because turning on $\W_\perp$ breaks some of the supersymmetry that is used in the derivation
  of the OV formula.

In addition to $\W_\parz$,
the $F$-terms that enter the OV formula depend on two other sets of chiral superfields.  
The most obvious ones are the familiar chiral superfields $\X^\Lambda(x^\mu|\theta^{Ai})$ 
that enter the GV formula,
but now  restricted to depend on only $x^1,x^2$ and one-half of the $\theta^{Ai}$.
Thus the $\X^\Lambda$ are now chiral superfields of $(2,2)$ supersymmetry in two dimensions.  In general, there are also additional 
chiral superfields $\U^\sigma$ that describe the moduli of $L$. For the case of a single D4-brane wrapped on a smooth Lagrangian submanifold $L\subset Y$,
the number of these superfields is $b_1(L)$, the first Betti number of $L$. The microscopic origin of the $\U^\sigma$ in this case will be reviewed in 
sections \ref{centch} and \ref{comptop}. As explained in section \ref{oov}, the OV formula determines chiral couplings
 \begin{equation}\label{mellop}J_\sn=\int_{\R^2}\d^2x \,\d^2\theta \,\RR_\sn(\X ^\Lambda;\U^\sigma) \W_\parz^\sn. ~~\n\geq 0. \end{equation}
One of the main ideas in the derivation is to consider not the individual $\J_\sn$ but the sum
\begin{equation}\label{elxop}\J=\sum_{\sn=0}^\infty J_\sn=\sum_{n=0}^\infty\int_{\R^2}\d^2x \,\d^2\theta \, 
\RR_\sn(\X ^\Lambda;\U^\sigma)\W_\parz^\sn\end{equation}
as a contribution to the effective action in a background with $\W_\parz$ turned on.

We have formulated this discussion as if we are interested in just one D4-brane on $\R^2\times L$, but there 
is no problem generalizing  to a case involving an arbitrary collection of Lagrangian submanifolds $L_i$ $i=1,\dots,s$, 
with $N_i$ D4-branes wrapped on $L_i$.  Until one turns on
the graviphoton field, one could take any set of parallel two-planes $\R^2_i\subset \R^4$ and then a 
configuration with $N_i$ D4-branes wrapped on $\R^2_i\times L_i$
preserves $(2,2)$ supersymmetry. (The $\R^2_i$ must be parallel to preserve the 2d translation invariance that is part of the
two-dimensional $(2,2)$ supersymmetry algebra.)  However, to get something like the OV formula, we have to turn on an 
anti-selfdual graviphoton field.  Moreover, to preserve the full
supersymmetry, we  need to maintain the anti-selfdual rotation symmetry $\JJ$, and this is only possible 
if the $\R^2_i$ all coincide so that the same anti-selfdual rotation generator can leave them all fixed. Thus the general
form of the OV formula involves the case of $N_i$ D4-branes wrapped on $\R^2\times L_i$, with a common factor $\R^2\subset \R^4$, but
arbitrary special Lagrangian submanifolds $L_i\subset Y$.  (Moreover, each stack of D4-branes can be endowed with  its own Chan-Paton
vector bundle.) 

In the rest of this section, we describe some further details about this construction.

\subsubsection{Infrared Problems}\label{dvortex}

From a macroscopic point of view, if $Y$ is compact, a D4-brane wrapped on $\R^2\times L\subset \R^4\times Y$ or an  
M5-brane wrapped on $\R^3\times L\subset \R^5\times Y$
is effectively supported on $\R^2\subset \R^4$ or $\R^3\subset \R^5$.  This is of real codimension 2, and the 
brane behaves as a vortex, producing a monodromy for a certain scalar field.
This will produce long range effects, meaning that for some purposes it will not be a sufficient approximation to view
the D4-brane or M5-brane as living in a pre-existing spacetime. We will not grapple with any such 
issues in this paper, though they may be important for some questions concerning the OV formula.

Even if $Y$ is not compact, one should worry about infrared effects if $L$ is compact.  Indeed, if $L$ is compact, then $\R^3\times L$
is macroscopically three-dimensional.  If there are massless gauge fields propagating on $\R^3$ (as there will be if $L$ has a positive first Betti
number), then particles that are charged under these gauge fields are confined.  Thus, a wrapped M2-brane that is charged under
a gauge field that propagates along $\R^3$ does not give rise to a BPS state; the would-be BPS state is actually confined.

This is a special case of a more general fact (pointed out to us by N. Seiberg).  In four-dimensional gauge theory, 
there can be Coulomb branches in which electric
and magnetic charges are not confined.  In a 4d gauge theory with $\N\geq 2$ supersymmetry, 
the central charges in the supersymmetry
algebra can receive contributions from electric and magnetic charges as well as from global conserved 
charges.  By contrast, in three dimensions, electric charge is confined even in a Coulomb phase, 
there is no magnetic charge, and supersymmetric central charges
are linear combinations of global conserved charges only.  

In field theory, in a certain sense, chiral couplings can be generated by loops of would-be BPS particles that actually
are confined.  A simple example is given by a 3d $U(1)$ gauge theory with $\N=2$ supersymmetry, coupled to massive 
charged chiral multiplets.  Compactification from $\R^3$ to $\R^2\times S^1$ generates an  effective 2d theory with
$\N=(2,2)$ supersymmetry.  The field strength of the vector multiplet is  a twisted chiral superfield $\varSigma$.  A certain chiral coupling -- a twisted chiral superpotential  $\RR_0(\varSigma)$, analogous to one of the couplings described by the OV formula  -- is generated
by a one-loop diagram with charged chiral multiplets running around the loop.  The particles in these multiplets would be BPS
particles if the gauge coupling $e$ were set to 0.  For $e\not=0$, they are confined but they still make sense in loop diagrams.  (If $e$ is small enough, they are not confined on the length scale of $S^1$.)

For the OV formula, the analog of $e\to 0$ is a limit in which the volumes of $Y$ and $L$ are taken to infinity.  In such a limit,
the gauge symmetries associated to gauge fields that propagate on $\R^3\times L$ become global symmetries, and there
can exist BPS particles that are charged under these symmetries.  

It may be that for compact $Y$ and $L$, there is a version
of the OV formula in which the chiral couplings $J_\sn$ are expressed as a sum over BPS particles that exist in the infinite
volume limit.
We do not know if such a version of the OV formula exists.  However, clearly it is safer to assume from the beginning that
$Y$ and $L$ are not compact and that the relevant symmetries are global symmetries.  This is the case considered
in \cite{OV} and the subsequent literature.  

Another basic problem arises whenever some of the moduli ($\ZZ^\Lambda=\X^\Lambda/\X^0$ and $\U^\sigma$) entering in the $J_\sn$'s are dynamical.\footnote{The $\ZZ^\Lambda$ are non-dynamical if  $Y$ is not compact  and  the harmonic two-forms associated to its Kahler moduli are
not square-integrable.}  As was essentially just pointed out in a mirror version, one of the interactions
determined by the OV formula, namely $J_0$,  is independent of $\W_\parallel$ and is simply a superpotential of $(2,2)$ supersymmetry.
In the  presence of a non-trivial superpotential,\footnote{This issue does not have a close analog in the GV case, because the analog of $J_0$
is there the prepotential $\F_0$, which determines the Kahler metric on the vector multiplet moduli space but does not contribute to the vacuum energy.}
 supersymmetry is broken for  generic values of $\ZZ^\Lambda$ and $\U^\sigma$.
It seems that the instanton computation of the $J_\sn$'s makes sense anyway, since when the radius of the M-theory circle is large,
the instanton-generated $J_\sn$'s are small and it makes sense to compute them while ignoring the dynamics they induce. (To be more confident about this
argument, one would want to ensure that contributions to $J_0$ from massless 3d BPS particles, 
which possibly cannot be considered small, do not trigger supersymmetry breaking.)  These questions did not arise in the original work \cite{OV} on the OV formula, 
since the geometry considered was such that none of the moduli
were dynamical.  

\def\ad{{\mathrm{ad}}}
What we have described is just a sampling of infrared questions concerning the OV formula.  We will
explain more in section \ref{parfields}.  
  We will not try to resolve any of these questions in this paper.  The classic applications of the OV formula involve 
  cases in which $Y$ and $L$ are not compact and most of these  questions do not arise. 
  
\subsubsection{Central Charges}\label{centch}

M-theory on a Calabi-Yau manifold $Y$ has Kahler moduli $v^1,\dots,v^{b_2}$ where $b_2=b_2(Y)$, the second Betti number of $Y$,
is the rank of the cohomology group $H^2(Y;\Z)$ or equivalently of the homology group $H_2(Y;\Z)$.
In M-theory on $\R^5\times Y$, we can wrap an M2-brane on a
 two-manifold $\Sigma\subset Y$ only if $\Sigma$ is closed and oriented.  Such a
 $\Sigma$ has a homology class in 
$H_2(Y;\Z)$.  The components of this class (in a suitable basis) are the M2-brane charges $q_1,\dots,q_{b_2}$ that appear in the GV formula.

A Lagrangian submanifold $L\subset Y$ may itself have moduli.  The number of moduli is $b_1=b_1(L)$, the first Betti number of $L$.
The OV formula involves these moduli as well as the $v$'s. 
Once we include an M5-brane supported on $L\subset Y$, 
a BPS state can arise from an M2-brane wrapped on $\Sigma\subset Y$
where $\Sigma$ still must be oriented, but may have a boundary on $L$.  In fact, for the OV formula, we are only interested in the case that the
boundary $\partial \Sigma$ of $\Sigma$ is non-empty.  This condition ensures that the M2-brane state in question is bound to the M5-brane and
therefore (in the Type IIA language) will contribute to the effective action a term supported along $\R^2\subset\R^4$.

An oriented two-dimensional surface $\Sigma\subset Y$  that may have a boundary  on $L$ has a homology class in the relative homology group
 $H_2(Y,L;\Z)$.    This group is related to $H_2(Y;\Z)$ by an exact sequence that reads in part
\begin{equation}\label{zox}\dots H_2(L;\Z)\xrightarrow{\alpha} H_2(Y;\Z)\xrightarrow{\beta} H_2(Y,L;\Z) \xrightarrow{\gamma} H_1(L;\Z)
\xrightarrow{\alpha'} H_1(Y;\Z) \dots \end{equation}
The maps $\alpha$ and $\alpha'$ take a cycle in $L$ and map it to $Y$ using the embedding $L\subset Y$.  The map $\beta$
is defined using the fact that a cycle $\Sigma\subset Y$ that has no boundary is a special case of a cycle whose boundary is on $L$.
And the map $\gamma$ maps a cycle $\Sigma\subset Y$ whose boundary is in $L$ to its boundary $\partial\Sigma\subset L$.

The physical meaning of the map $\alpha$ is as follows.  Once we introduce an M5-brane, since an M2-brane can end on an M5-brane,
some M2-brane charges might not be conserved any more.  If  $\Sigma\subset Y$ is homologous to a cycle in $L$, then an M2-brane
wrapped on $\Sigma$ can annihilate and disappear.  Hence for the purpose of the OV formula, $H_2(Y;\Z)$ should be replaced
by the quotient $H_2(Y;\Z)/\alpha(H_2(L;\Z))$, which parametrizes charges that are carried by M2-branes without boundary and are
conserved in the presence of an M5-brane wrapped on $\R^3\times L$.  To keep our terminology familiar, we will assume in what follows
that $\alpha=0$.  Otherwise, in all statements
one replaces $H_2(Y;\Z)$ by $H_2(Y;\Z)/\alpha(H_2(L;\Z))$ and replaces $b_2(Y)$ by the rank of that group, which we might call $b_2'=b_2'(Y,L)$.

The interpretation of the map $\alpha'$ is as follows.  If a 1-cycle in $L$ is the boundary $\partial \Sigma$ of some $\Sigma\subset Y$
that represents a class in $H_2(Y,L;\Z)$, this means by definition that $\partial \Sigma$, when embedded in $Y$, is a boundary (of $\SIgma$) and so
vanishes in $H_1(Y;\Z)$.
So the image of $H_2(Y,L;\Z)$ under $\gamma$ is not all of $H_1(L;\Z)$, but only the kernel of $\alpha'$.

In any event, for a Calabi-Yau manifold $Y$, $H_1(Y;\Z)$ is always a finite group.  This means that we can set $\alpha'=0$ if there is no torsion
or we reduce modulo torsion.  Until section \ref{discpars}, we ignore torsion and consider only $\Z$-valued charges.  With also $\alpha$ assumed
to vanish, the long exact sequence (\ref{zox}) reduces to a short exact sequence
\begin{equation}\label{turono} 0\to H_2(Y;\Z) \xrightarrow{\beta} H_2(Y,L;\Z)\xrightarrow{\gamma} H_1(L;\Z)\to 0. \end{equation}
This implies that the rank of $H_2(Y,L;\Z)$ is the sum $b_2(Y)+b_1(L)$.  That  number (or $b_2'+b_1(L)$ if $\alpha\not=0$) 
is the total number of $\Z$-valued charges of a BPS state in this situation.  

By mapping the   $H_2(Y,L;\Z)$-valued charge of an M2-brane with boundary on $L$ to $H_1(L;\Z)$ via $\gamma$, we learn that such an M2-brane
has a charge valued in $H_1(L;\Z)$, or in other words that it carries $\Z$-valued charges $r_1,\dots, r_{b_1}$ that are determined
by its boundary.   Concretely, these charges are dual to oriented circles $\ell^\rho\subset L$ that provide a basis
  of $H_1(L,\Z)$ (modulo possible torsion).   An M2-brane wrapped on $\Sigma$ has charges $r_\rho$, $\rho=1,\dots, b_1(L)$ if its boundary $\partial \Sigma$
  is homologous in $H_1(L;\Z)$ to $\sum_\rho r_\rho \ell^\rho$.   
  
  Since there is no natural map from $H_2(Y,L;\Z)$ to $H_2(Y;\Z)$, there is no
  equally natural definition of the ``bulk'' charges of an M2-brane that is allowed to end on $L$.  However, modulo torsion, we can always pick
  a splitting of the exact sequence (\ref{turono}), and this enables us to define the bulk charges $q_1,\dots , q_{b_2}$.  
  We will pick a fixed splitting in what follows, though one
  could proceed more intrinsically.    If we use a different
  splitting, the $q_I$ are shifted by integer linear combinations of the $r_\rho$.

In M-theory, if $L$ is compact, then just like the charges $q_I$ that entered the GV formula, the new charges $r_\rho$ 
 also couple to abelian gauge fields.  These are abelian gauge fields that only propagate along the support $\R^3\times L$
 of the M5-brane, so that macroscopically, they propagate along $\R^3\subset
 \R^5$.   These abelian gauge fields have a simple microscopic origin.  Along the world-volume of an M5-brane, there propagates a two-form field (whose
 curvature is constrained to be selfdual).  When we compactify the M5-brane on $\R^3\times L$, the Kaluza-Klein expansion of the two-form field gives $b_1(L)$
 abelian gauge fields on $\R^3$.  As we have discussed in section \ref{dvortex}, states that are charged with respect
 to these gauge fields are actually confined.  The derivation and interpretation of the OV formula are more straightforward
 if $L$ does not admit any square-integrable harmonic 1-forms, either because $L$ is compact with $b_1(L)=0$ or because $L$ is not compact
 and its geometry and topology ensure that harmonic 1-forms on $L$ are not square-integrable.  In this case, the symmetries
 associated to the moduli of $L$ behave as global symmetries and the $r_\rho$ are global charges that can contribute to the central
 charge of a BPS state.
 
The area of a holomorphic curve $\Sigma\subset Y$ whose boundary is on $L$ is determined by its homology class or in other words by the
charges $q_1,\dots, q_{b_2}$ and $r_1,\dots,r_{b_1}$.  This area in M-theory units
is \begin{equation}\label{murox}A=\sum_I q_I v^I+\sum_\rho r_\rho w^\rho,\end{equation} where $v^I$ are the Kahler moduli
of $Y$ and $w^\rho$, $\rho=1,\dots, b_1(L)$, are the moduli of $L$.  The $v^I$ are familiar; one can think of the formula (\ref{murox})
as the definition
of a convenient set of coordinates $w^\rho$ that parametrize the moduli space of $L$.  (If one changes the splitting that was used to define the $q_I$,
then the $w^\rho$ are shifted by integer linear combinations of the $v^I$.)  To find the mass of an M2-brane wrapped on $\Sigma$ measured in the 5d Einstein
frame, we make a Weyl transformation to 5d variables \begin{equation}\label{survi}h^I=\frac{v^I}{v},~~~k^\rho=\frac{w^\rho}{v},\end{equation}
 as in eqn. (\ref{buyt}).  Generalizing eqn. (\ref{merf}), the mass of
a BPS particle with charges $\vec q,\vec r$ (in units in which the M2-brane tension is 1) is then
\begin{equation}\label{merfo}m(\vec q,\vec r)=\sum_I q_Ih^I+\sum_\rho r_\rho k^\rho. \end{equation} 
Assuming that $\Sigma$ has a non-empty boundary on $L$,  and that $L$ is suitably noncompact,
this particle propagates on $\R^3$ and 
is a BPS particle in a 3d theory that has $\N=2$ supersymmetry (four superchanges).
The 3d $\N=2$ supersymmetry algebra has a real central charge $\zeta$ that equals the mass of a BPS particle, so it
is given by the formula (\ref{merfo}):
\begin{equation}\label{werfo}\zeta(\vec q,\vec r)=\sum_I q_Ih^I+\sum_\rho r_\rho k^\rho. \end{equation}

Now let us compactify from $\R^5\times Y$ to $\R^4\times S^1\times Y$, so that the M5-brane worldvolume becomes $\R^2\times S^1\times L$.
As usual, we suppose that the $S^1$ has circumference $2\pi e^\sigma$.  The real part of the action of a BPS particle of mass $m(\vec q,\vec r)$ propagating around the $S^1$ is 
then $2\pi e^\sigma m(\vec q,\vec r)=2\pi e^\sigma(\sum_I q_Ih^I+\sum_\rho r_\rho k^\rho)$.  However, just as in the derivation of the GV formula,
the action also has an imaginary part that arises from the fact that when we compactify on a circle, the abelian gauge fields may have holonomies
around the circle.    As in our study of the GV formula, we write $\exp(2\pi i \alpha^I)$, $I=1,\dots ,b_2(Y)$, for the holonomies of the gauge fields
that arise from M-theory compactification on $Y$, and we similarly write $\exp(2\pi i\beta^\rho)$, $\rho=1,\dots,b_1(L)$ for the holonomies of the gauge
fields that live on the M5-brane.  (Again the definition of the $\beta^\rho$ depends on a choice of splitting;
in a change of splitting, they are shifted by integer linear combinations of the $\alpha^I$.) Then a particle of charges $\vec q,\vec r$ propagating around the circle acquires a phase $\exp\left(2\pi i (\sum_Iq_I\alpha^I+\sum_\rho r_\rho \beta^\rho)\right)$.
As in eqn (\ref{pofgo}), we can interpret this to mean that the action for such a particle is
\begin{equation}\label{mofgo}S(\vec q,\vec r)= 2\pi\left(\sum_Iq_I(e^\sigma h^I-i\alpha^I)+\sum_\rho r_\rho (e^\sigma k^\rho-i\beta^\rho)\right)=-
2\pi i\left(\sum_I q_I Z^I +\sum_\rho r_\rho U^\rho\right),  \end{equation} where $Z^I$ was defined in eqn. (\ref{merot}) and
similarly  \begin{equation}\label{perot} U^\rho=\beta^\rho+ie^\sigma k^\rho.\end{equation}

Actually, $Z^I$ and $U^\rho$ are the bottom components of 2d chiral superfields $\ZZ^I$ and $\U^\rho$.  As in the derivation of the GV formula,
a BPS particle propagating around the circle in $\R^2\times S^1$
has bosonic collective coordinates (its position along $\R^2$) 
and also fermionic collective coordinates.  To take account of the fermionic collective
coordinates, it is better to write the action as a superfield
\begin{equation}\label{tofgo}\S(\vec q,\vec r)=-2\pi i\left(\sum_I q_I \ZZ^I +\sum_\rho r_\rho \U^\rho\right).\end{equation}
As in the derivation of the GV formula, 
the contribution of a BPS particle to the OV formula is given by $\exp(-\S)$ multiplied by a product of one-loop determinants.

\subsubsection{String Theory Interpretation}\label{comptop}

In Type IIA superstring theory, rather than an M5-brane wrapped on $\R^3\times L$, we consider a D4-brane wrapped on $\R^2\times L$.
While the M5-brane supports a two-form field with self-dual curvature, the D4-brane 
supports\footnote{In what follows, we ignore various refinements
involving the K-theory interpretation of D-branes.} a $U(1)$ gauge field.  To define a supersymmetric 
D4-brane state, we must endow $L$ with a flat Chan-Paton
bundle $\L$.  This means we must specify the holonomy $\exp(2\pi i \beta^\rho)$ of $\L$ around 
each circle $\ell^\rho\subset L$.   From a Type IIA
point of view, the $\beta^\rho$ are moduli of a D4-brane wrapped on $L$.  The D4-brane moduli are complex parameters $U^\rho$ whose
real parts are the $\beta^\rho$ and whose imaginary parts are geometric moduli of $L$.  In the 
5d units used in section \ref{centch}, those imaginary
parts are $e^\sigma k^\rho$.  The complex fields $U^\rho$ are the bottom components of 
chiral superfields $\U^\rho$ that appear in the action (\ref{tofgo}) and
therefore will enter the OV formula.   
The angles $\beta^\rho$  will play a similar role to the angles $\alpha^I$ that arise as periods of the Neveu-Schwarz sector $B$-field
of Type IIA superstring theory.

Consider the contribution to Type IIA superstring perturbation theory of a string wrapped on $\Sigma\subset Y$, with $\partial\Sigma\subset L$.
Suppose that $\partial\Sigma$ is homologous in $L$ to $\sum_\rho r_\rho \ell^\rho$ and more generally that its class in $H_2(Y,L;\Z)$
is determined by the charges $q_I,r_\rho$.  The contribution of $\Sigma$ in Type IIA superstring perturbation theory then
contains a factor $\prod_\rho \exp(2\pi i r_\rho \beta^\rho+2\pi i q_I\alpha^I)$ coming from the holonomy of the Chan-Paton
bundle $\L$ around the boundary components
of $\Sigma$ as well as the coupling to the $B$-field.

In M-theory, there is no Chan-Paton line bundle or $B$-field
and instead of a D4-brane on $\R^2\times L$, there is an M5-brane on $\R^3\times L$.  This M5-brane
has only the real  moduli that we have called $k^\rho$, rather than complex moduli $U^\rho$ whose real parts are the holonomies $\beta^\rho$.  
 In M-theory,
the choice of M5-brane configuration that is used to deduce the OV formula does not depend on the $\beta^\rho$; 
likewise, the relevant space of BPS
states does not depend on the $\beta^\rho$. (As we discuss momentarily, the same is true for the $\alpha^I$.)  
The $\beta^\rho$ only enter when one compactifies from $\R^3\times L$ to $\R^2\times S^1\times L$
and introduces a holonomy around the circle.  Microscopically, in  M-theory terms, this ``holonomy'' means that the two-form that lives on the M5-brane 
is non-zero; it is flat, but
 has a period $\beta^\rho$ when integrated on $S^1\times \ell^\rho$.  More macroscopically, in terms of 3d BPS states on $\R^2\times S^1$,
the meaning of this holonomy is that in representing the path integral as a trace in a space of physical states, one includes a factor 
$\exp\left(2\pi i\sum_\rho 
\beta^\rho Q_\rho\right)$  where
here $Q_\rho $ is the charge that measures the winding of an M2-brane around $\ell^\rho$.  

What we have said here has a perfect analog in the derivation of the GV formula.  Type IIA compactification on $\R^4\times Y$ involves $B$-field periods $\alpha^I$
which in the M-theory lift are not parameters of the compactification, and therefore also do not influence the space of BPS states that contribute to the GV
formula.   The B-field periods appear in M-theory when one compactifies on a circle and turns on appropriate $C$-field periods, leading to a factor 
\begin{equation}\label{zondo}\exp\left(2\pi i \sum_I \alpha^IQ_I\right)\end{equation} when one represents the path integral as a trace.
This combines  with the analogous $\beta^\rho$-dependent phase mentioned in the last paragraph to give a phase factor
\begin{equation}\label{tornoz}\exp\left(2\pi i\left(\sum_\rho 
\beta^\rho Q_\rho+\sum_I\alpha_IQ^I\right)\right)\end{equation}
weighting the contribution of an M2-brane state that winds once around the M-theory circle.

All this becomes slightly less trivial when one considers discrete parameters in the compactification, which we have so far ignored.

\subsubsection{Discrete Parameters}\label{discpars}

Until now, we have ignored torsion in our discussion of the charges carried by a BPS state.  In general, 
there may be discrete parameters either in the moduli of Type IIA
compactification on $Y$, or in the choice of a flat line bundle on $L$.    Such discrete parameters lead to torsion -- 
that is to discrete symmetries of the space of
BPS states.

Let us first consider discrete parameters in the GV formula.  When we compactify Type IIA on a Calabi-Yau manifold 
$Y$, we pick a flat $B$-field,\footnote{Supersymmetry
is preserved if we also introduce a flat Ramond-Ramond $p$-form field on $Y$ for $p=1,3,5$.  But a flat RR field does 
not affect superstring perturbation
theory and so is not relevant to the GV or OV formulas, which govern interactions that can be computed in perturbation theory.}  an element
of $H^2(Y;U(1))$.  The connected component of this group is parametrized by the angles $\alpha^I$ that we have already 
incorporated.  In general, however,
$H^2(Y;U(1))$ is not connected.  Because of the exact sequence
\begin{equation}\label{zomalt} 0\to \Z\to \R \to U(1)\to 0,\end{equation}
there is an exact sequence of cohomology groups
\begin{equation}\label{omalt}\dots \to H^2(Y;\Z)\to H^2(Y;\R) \to H^2(Y;U(1))\to H^3(Y;\Z)\to H^3(Y;\R)\to \dots. \end{equation}
The torsion subgroup of $H^3(Y;\Z)$, which we denote $H_{\mathrm {tors}}^3(Y;\Z)$, is defined as the kernel of the map 
$H^3(Y;\Z)\to H^3(Y;\R)$.  It is always a finite group.
From (\ref{omalt}), we deduce the exact sequence
\begin{equation}\label{pomalt} 0 \to H^2(Y;\R)/H^2(Y;\Z)\to H^2(Y;U(1))\to H^3_{\mathrm{tors}}(Y;\Z)\to 0. \end{equation}
Here $H^2(Y;\R)/H^2(Y;\Z)$ is the torus parametrized by the angles $\alpha^I$.  The exact sequence shows that this 
torus is the connected subgroup of
$H^2(Y;U(1))$, and that in general the group of components of $H^2(Y;U(1))$ is $H^3_{\mathrm{tors}}(Y;\Z)$.   
This result can be interpreted as follows.
In general, a $B$-field on a manifold $Y$ has a characteristic class  $x\in H^3(Y;\Z)$, but if $B$ is flat, then $x$ is valued
in the torsion subgroup of $H^3(Y;\Z)$.  This is the discrete data labeling the compactification.

In M-theory on $\R^5\times Y$, one sees  the Kahler moduli of $Y$, but not the additional moduli associated to the choice of a $B$-field.   Consequently,
the space of BPS states that is used to compute the GV formula does not depend on either the angles $\alpha^I$ that parametrize the continuous moduli of a flat
$B$-field, or the discrete choice of $x\in H^3_{\mathrm{tors}}(Y;\Z)$.  

Rather, as we have said already in the case of the continuous moduli,
the torsion part of the $B$-field
 enters the M-theory description as part of the symmetry group.  In M-theory, there is a three-form $C$-field, not a two-form $B$-field.
In general, compactification of the $C$-field on any manifold $Y$ generates a group $H^2(Y;U(1))$ of unbroken gauge symmetries -- the same group that
parametrizes flat $B$-fields in the Type IIA interpretation.  A shortcut to understanding this statement is that symmetries in compactification on $Y$
appear, if one compactifies on another circle, as monodromies around the circle.  A ``monodromy'' in this context is a flat $C$-field on $S^1\times Y$
that is trivial when restricted to $Y$ (that is to $p\times Y$ for $p$ a point in $S^1$).  The group of flat $C$-fields on $S^1\times Y$ is $H^3(S^1\times Y;U(1))$
and the subgroup consisting of flat $C$-fields on $S^1\times Y$ that are trivial when restricted to $Y$ is isomorphic to $H^2(Y;U(1))$.  So this  group is interpreted in M-theory as the group 
of unbroken gauge symmetries coming from the $C$-field.

If we turn on the continuous part of the $B$-field in the Type IIA description, then in the M-theory description we have to include the holonomy factor
(\ref{zondo}) that we have already discussed.  Including this factor  amounts to including in the trace that gives the GV formula 
the  element $\exp(2\pi i\sum_I \alpha^I Q_I)$ of the connected part (the identity component) of the global symmetry group $H^2(Y;U(1))$.  (This factor
is part of the familiar factor $\exp(2\pi i \sum_I q_I \ZZ^I)$  in the GV formula.)
If we want to turn on a discrete $B$-field in the Type IIA description, then in the M-theory description we have to 
proceed in the same way, including in the trace an element  of $H^2(Y;U(1))$ that might not be connected to the identity.
If we pick a splitting of the exact sequence  (\ref{pomalt}) (such a splitting always exists but is not unique), this means
that we multiply $\exp(2\pi i\sum_I\alpha^I Q_I)$ by an element 
 $x\in H^3_{\mathrm{tors}}(Y;\Z)$.  (We write both $H^2(Y;U(1))$ and its group of components $H^3_{\mathrm{tors}}(Y;\Z)$ multiplicatively;
 in particular, if $x\in H^3_{\mathrm{tors}}(Y;\Z)$ is of order $n$, we write $x^n=1$.)
If a BPS state winds $k$ times around the circle, its contribution is weighted by $x^k$ as well as by a factor $\exp(2\pi i k\sum_I \alpha^I Q_I)$ that is
already present in eqn. (\ref{womombo}).  Thus, 
the generalization of eqn. (\ref{womombo}) to include discrete $B$-fields is obtained simply by including an additional  factor of $x^k$ inside
the trace:
 \begin{align}\label{womombox}-\int \frac{\d^4x\d^4\theta}{(2\pi)^4} \sqrt{g^E}\sum_{q|\zeta(q)\geq 0}\sum_{k=1}^\infty\frac{1}{k}\Tr_{{\eV}_{\vec q}}\left[(-1)^F x^k
 \exp(-i\pi k
 \JJ_{\vec q}/4\X^0) \right]
 \exp\left(2\pi i\sum_I q_I \ZZ^I\right)\frac{\frac{1}{64}\pi^2\W^2}{\sin^{2}\left(\frac{\pi k\sqrt{\W^2}}{8\X^0}\right)}.  \end{align}
 This is the GV formula with discrete $B$-fields.   The formula (\ref{womombox}) could be written more intrinsically, without making use
 of a splitting of the exact sequence (\ref{pomalt}).  (For this, basically one would just interpret the product $x \exp(2\pi i\alpha_I Q_I)$ as
 an arbitrary element of $H^2(Y;U(1))$, without trying to write it as a product of two factors.   This would require rephrasing many of our
 statements in a slightly more abstract way.)

Discrete data in the flat Chan-Paton line bundle $\L\to L$ can be treated similarly.  Flat unitary line bundles over $L$ are classified by $H^1(L;U(1))$.  This
group appears in an exact sequence 
\begin{equation}\label{pollyo}0\to H^1(L;\R)/H^1(L;\Z)\to H^1(L;U(1))\to H^2_{\mathrm{tors}}(L;\Z)\to 0 \end{equation}
that is just analogous to eqn. (\ref{pomalt}). Here the element of 
$H^2_{\mathrm{tors}}(L;\Z)$ corresponding to a flat line bundle $\L$ is the
first Chern class $y=c_1(\L)$.   So a flat line bundle $\L$ has continuous moduli 
$\beta^\rho$ that parametrize the torus $H^1(L;\R)/H^1(L;\Z)$, and
also a discrete modulus $y\in H^2_{\mathrm{tors}}(L;\Z)$.  (Here again we 
implicitly choose a splitting of the exact sequence (\ref{pollyo}), and we write the group $H^2_{\mathrm{tors}}(L;\Z)$ multiplicatively.)

Though a D4-brane wrapped on $\R^2\times L$ has moduli valued in $H^1(L;U(1))$, upon lifting to
 an M5-brane on $\R^3\times L$, these moduli disappear.  The space of BPS states bound to the M5-brane
does not depend on the choice of an element of $H^1(L;U(1))$.  Rather, $H^1(L;U(1))$ is a 
group of gauge symmetries that act on the space of BPS states.  
The continuous symmetries are the angles $\beta^\rho$; they will enter the OV formula 
because they appear in the action $\S(\vec q,\vec r)$.  The discrete
symmetry $y$ should be included in the trace that defines the OV formula (eqn. (\ref{zeloxy}) below) just as we have 
included the symmetry element $x$ in the analogous trace in
eqn. (\ref{womombox}).

\subsubsection{The Nonabelian Case}\label{nonab}

To reach the limits of present knowledge, all we have to do is to consider in a similar spirit 
the case of $N\geq 2$ D4-branes wrapped on $\R^2\times L$.
In this case, instead of a flat unitary line bundle $\L\to L$, the definition of a supersymmetric D4-brane state involves the
choice of a flat $U(N)$ bundle  over $ L$.
For $N=1$, we used the fact that flat unitary line bundles $\L\to L$ are parametrized 
by the group $H^1(L;U(1))$.  The M-theory lift involves a space of BPS
states that does not depend on $\L$, but has  $H^1(L;U(1))$ as a group of symmetries.

By analogy with what happened for $N=1$, one might now hope that the M-theory description for 
$N\geq 2$ would involve a space of BPS states that does not
depend on the choice of a flat $U(N)$ bundle $V\to L$, but possesses some symmetry group $G$.   
Now in general we might expect $G$ to be nonabelian; moreover, 
by analogy with what happened for $N=1$, we may expect
$G$ to be a gauge group in an effective description on $\R^3$ of a system of M5-branes 
on $\R^3\times L$.  In this case, if we compactify 
from $\R^3\times L$ to $\R^2\times S^1\times L$,
then the possible holonomies around the circle, up to gauge-equivalence, are 
labeled by conjugacy classes in $G$.  So to imitate what happened for $N=1$,
we would want the flat $U(N)$ bundles $V\to L$ to be parametrized by the conjugacy classes in some group $G$.  
However, for a general $L$, this is too much to hope for.  

A case in which this actually does work, and which was important in the original work of Ooguri and Vafa \cite{OV}, as well
as many later papers, is that $L$ (or  
one or more of the $L_i$ in a more general problem involving multiple Lagrangian submanifolds  $L_i\subset Y$) 
is topologically $\R^2\times S^1$.  This example works
because for any group $G$, the moduli space of flat $G$-bundles on $\R^2\times S^1$ is the space 
of conjugacy classes in a  group (namely $G$).
So in the case of $N$ D4-branes wrapped on $\R^2\times L$, where  $L$ is topologically 
$\R^2\times S^1$, the M-theory description can plausibly involve 
a fixed space of BPS states, independent of the choice of a flat $U(N)$ bundle over 
$L$, with a $U(N)$ symmetry acting on this fixed space of states.  In fact, we would expect this, since if $L=\R^2\times S^1$, then
$\R^3\times L=\R^5\times S^1$, and a system of $N$ M5-branes on $\R^5\times S^1$ is believed to be described at long distances
by $U(N)$ gauge theory on $\R^5$.

For a more generic $L$, one cannot expect such a nice answer.
A proper understanding presumably depends in part on the following. 
A system of $N\geq 2$ M5-branes wrapped on $\R^3\times L$, with compact $L$, can be described at long
distances by a quantum field theory on $\R^3$.  But in general this quantum field theory is not infrared-free\footnote{For a simple
example, take $L=S^3$.  The relevant 3d theory is then a $U(N)$ gauge theory with a Chern-Simons coupling of level 1 coupled
to massless chiral multiplets in the adjoint representation.  This theory is not infrared-free.}  
and cannot be described 
even at long distances by classical fields and classical
concepts such as gauge symmetries.  Analogous problems do not arise in the GV formula because 
in that case, one is studying the propagation of particles on
$\R^5\times Y$, and (unless one adjusts the moduli of $Y$ to reach a very special singularity) the physics 
on $\R^5\times Y$ is infrared-free.  Similarly, the
observation in the last paragraph depends among other things on the fact that if $L=\R^2\times S^1$, then 
$\R^3\times L=\R^5\times S^1$, with enough non-compact dimensions
so that classical concepts are valid.  

The infrared questions mentioned in the last paragraph arise even if $b_1(L)=0$, so they are independent of the
questions concerning confinement that were discussed in section \ref{dvortex}. 
 
\subsubsection{What We Can Calculate And How}\label{parfields}

All this is pertinent to   the OV formula, because if the relevant 3d
physics is not infrared-free, one will face new issues, lacking analogs in the GV case, even in 
understanding the contributions of massive BPS states. In fact, this may happen even if the 3d physics 
is infrared-free.  The 3d physics may be governed at long distances by a non-trivial
topological field theory, such as a pure Chern-Simons theory.  In this case, the BPS states may be anyonic; they would then have
a long range interaction of a statistical nature, which would certainly affect the contribution of a BPS particle winding multiple times
around the M-theory circle.  In this paper, we attempt to consider only the simple case in which none of this happens 
(we will see that restricting to this case is not as straightforward as it may sound).

Just as in the GV case, in deriving the OV formula, we must decide whether to treat the BPS objects as particles or as fields.
The arguments of section \ref{pf} are applicable.  Massive BPS states generically have too great a mass and too large a spin to
be usefully treated as fields. On the other hand, a particle-based computation, in which a massive BPS particle propagating around the
M-theory circle is treated as an instanton, is quite straightforward.  This computation is a simple variant of what we have explained
in section \ref{schwpart} and is presented in section \ref{morch}.   A field theory computation is useful instead for BPS states that are
massless in three dimensions.  (We will not do such a calculation, which would be an analog of the calculation described in section \ref{hypercalc}
in the GV case.)

 \subsection{The Computation}\label{morch}
 
The computation that we have to perform is not essentially new; in a sense, it is just the square root of the very simple computation
that was already described in section  \ref{thecomp}.  It is the interpretation that involves some difficulty.

As with the GV formula, we consider first a BPS superparticle that has only the two bosonic and two fermionic zero-modes that follow from supersymmetry.
The basic example is an M2-brane wrapped on a holomorphic disc $\Sigma\subset Y$ whose boundary is on $L$.
If $\Sigma$ has no infinitesimal deformations, then the 3d BPS superparticle obtained by wrapping an M2-brane on $\Sigma$ has
only the minimal set of zero-modes.  Ooguri and Vafa \cite{OV} give a useful example\footnote{\label{lofty} In their example, $Y$ is the small resolution of the conifold, which can be described via a linear $\sigma$-model with gauge group $U(1)$ and chiral superfields $u_1,u_2$ of charge 1 and $v_1,v_2$ of charge $-1$. Thus $Y$ is the quotient by $U(1)$ of the space $|u_1|^2+|u_2|^2-|v_1|^2-|v_2|^2$ =1.  An embedding $\Bbb{CP}^1\subset Y$
 is defined by $v_1=v_2=0$.  $L$ is defined by taking all $u_i$ and $v_j$ to be real.  The
symmetry  $\eS$ that is introduced below acts on $Y$ by $v_2\to -v_2$, leaving fixed $u_1,u_2,$ and $v_1$.}
in which $L$ is topologically $S^1\times \R^2$.
The $S^1$ is the ``equator'' in a holomorphically embedded $\Bbb{CP}^1\subset Y$, and taking $\Sigma$ to be the upper or lower
hemisphere of this $\Bbb{CP}^1$, one gets an example with only those bosonic or fermionic zero-modes that follow from the symmetries. 

In our problem, this gives a superparticle that propagates on a two-plane $\R^2\subset\R^4$.  We take this to be the two-plane $x^3=x^4=0$,
parametrized by $x^1$ and $x^2$.
The action that describes such a superparticle in the nonrelativistic limit
 is a simple truncation to $x^3=x^4=0$  of the one that we used in deriving the GV formula.   A way to make this truncation is to introduce 
the reflection $\eR$ that acts by $x^3\to -x^3,$ $x^4\to -x^4$ while leaving $x^1,x^2$ fixed.  This reflection is a symmetry of the bosonic
part of the action if the graviphoton field is $\eR$-invariant, which is actually the case for the choice that was already made in eqn. (\ref{morzo}).
We will extend $\eR$ to a symmetry of the full action including the fermions, and then the $\eR$-invariant part of eqn. (\ref{hurmex})
will serve as our superparticle action.  The extension of $\eR$ to the fermions  is not completely trivial since 
 on spinors $\eR$ acts as $\sigma_{34}=\gamma_3\gamma_4$,
and its square is $-1$, not $+1$.  To get an operation that squares to $+1$, we have to combine the matrix $\sigma_{34}$ acting on the
spinor index $A$ of a fermion field $\psi_{Ai}$  with a matrix that acts on the additional index\footnote{Microscopically,
the index $i=1,2$ distinguishes supersymmetries that originate from left-movers on the string worldsheet to those
that originate from right-movers.  Interaction with a D4-brane preserves a linear combination of the two types of
symmetries.  Which linear combination it is depends on the eigenvalue of $\sigma_{34}$, and that is why a reflection 
must be defined to act on the $i$ index.}  $i$ and also squares
to $-1$.    Let us call the combined operation $\eR'$.  Then the $\eR'$-invariant part of the action
(\ref{hurmex}) is the basic superparticle action relevant to the OV formula.  It possesses the $\eR'$-invariant part of the supersymmetry
algebra of the action (\ref{hurmex}), and this is the appropriate symmetry for our problem.

To perform the path-integral for this problem, we simply proceed as follows.
We have projected out half of the collective coordinates from (\ref{hurmex}), so the zero-mode measure will be $\d^2x \d^2\psi^{(0)}$
rather than $\d^4x\d^4\psi^{(0)}$.  Also, we have to compute a bosonic determinant in just one of the $2\times 2$ blocks in eqn. (\ref{morzo}). This means that the one-loop path integral just gives, in a fairly obvious 
sense, the square root of the result in eqn. (\ref{orz}).   Finally, in the classical action, we have to include the charges and moduli
associated to the D4-brane and so replace $\sum_I q_I\ZZ^I$ with $\sum_I q_I\ZZ^I+\sum_\rho r_\rho \U^\rho$.  Putting these
statements together, the result for a BPS
superparticle winding once around the circle is
\begin{equation}\label{melixy} \frac{\d^2x\d^2\psi^{(0)}}{2\pi} \exp\left(2\pi i\left(\sum_I q_I\ZZ^I+\sum_\rho r_\rho\U^\rho\right)
\right) \frac{\TT}{\sinh(\pi e^\sigma \TT)}. \end{equation}
To go from this formula to a Type IIA effective action, we follow much the same steps that were used to go from eqn. (\ref{welx}) to eqn. (\ref{zelmo}).
We write $\TT=\frac{e^{\sigma/2}}{4}\WW_{\parallel}$, and interpret $\WW_{\parallel}$ as the bottom component
of a chiral superfield $\Wp$.  We also write $\d^2x\d^2\psi^{(0)}=\frac{1}{2}\d^2x\d^2\theta \sqrt{g^E}e^{-\sigma/2}$ 
(where now $g^E$ is the Einstein
metric restricted to the brane) and use the usual formula $e^{3\sigma/2}=-i/2\X^0$.   The resulting contribution to the effective action is
\begin{equation}\label{elixy}i\int\frac{\d^2x\d^2\theta}{(2\pi)^2}\sqrt{g^E}\exp\left(2\pi i\left(\sum_I q_I\ZZ^I+\sum_\rho r_\rho\U^\rho\right)\right)
\frac{\pi\Wp/8}{\sin(\pi\Wp/8\X^0)}. \end{equation}

Before discussing the interpretation of this formula, we write its obvious extension, analogous to (\ref{zelmox}),
 to the case of a superparticle that wraps any
number of times around the circle:
\begin{equation}\label{meloxy}i\int \frac{\d^2x\d^2\theta}{(2\pi)^2}\sqrt{g^E}\sum_{k=1}^\infty\frac{1}{k} 
\exp\left(2\pi ik\left(\sum_I q_I\ZZ^I+\sum_\rho r_\rho\U^\rho\right)
\right) \frac{\pi\Wp/8}{\sin(\pi k\Wp/8\X^0)}. \end{equation}
In deriving this formula, we have ignored superparticle interactions. It is not immediately obvious that this is right,
since in  general, as explained in section \ref{winding},
short-range interactions are only irrelevant above $D=2$ (short-range interactions in $D=2$ have been studied, for example,
in \cite{HS}).  We justify ignoring the interactions, under some assumptions, in appendix \ref{supercon}.

The formulas (\ref{melixy}) and (\ref{meloxy}) have been easy to write down, but it is a little more vexing to interpret them.
First let us note that eqn. (\ref{melixy}) is  invariant under $\TT\to -\TT$, and hence the following formulas are invariant under $\W\to -\W$.  
This reflects the fact that in our model, $\TT\to -\TT$ is  a symmetry
if combined with $x^2\to -x^2$.  To extend this to a symmetry of the full M-theory construction that the model comes from, 
one combines $\TT\to -\TT$
with an operation  that {\it (i)} acts on the noncompact spatial coordinates
by $x^2\to -x^2$, $x^4\to -x^4$ (this combined operation is an orientation-preserving symmetry
of the supersymmetric G\"{o}del solution (\ref{miro}), and maps the $\R^3$
 subspace $x^3=x^4=0$ to itself, reversing its orientation), {\it (ii)}  leaves
$\Sigma$ fixed (so it maps a particular BPS superparticle to itself), and {\it (iii)} acts  holomorphically on $Y$ (in particular preserving
the orientation of $Y$) mapping $L$ to itself
but reversing the orientation of $L$ (so that it preserves the orientation of $\R^3\times L$ and 
 preserves the sign of the M5-brane charge).  
There is no difficulty finding
a suitable holomorphic symmetry of $Y$ in the example from \cite{OV} that was mentioned above; see footnote \ref{lofty}.
  We call this transformation $\eS$.  It is a symmetry of the model if
$\TT=0$ and in general it reverses the sign of $\TT$.

Now let us look more closely at the Hilbert space $\widehat\H_0$ of the above model.  Of course, this Hilbert space represents the
two bosonic coordinates $x^1,x^2$ and their canonical momenta, and also the two $\eR'$-invariant components of $\psi_{Ai}$.
  To describe those components more precisely,  first recall that the choice of the embedding $\R^2\subset \R^4$ 
 breaks the $SO(4)$ rotation symmetry to $SO(2)\times SO(2)$, generated by
\begin{equation}\label{murz}J_{12}=x_1\frac{\partial}{\partial x^2}-x_2\frac{\partial}{\partial x^1},
~~~~~J_{34}=x_3\frac{\partial}{\partial x^4}-x_4\frac{\partial}{\partial x^3}. \end{equation}
$J_{12}$ rotates the $\R^2$ on which the BPS state propagates and measures the spin of the particle.
$J_{34}$ rotates the normal plane to $\R^2$ and generates an $R$-symmetry of the superparticle.  
We note that the symmetry $\eS$ of the last paragraph
reverses the signs of both $J_{12}$ and $J_{34}$.

The spinor index $A$ of the fermion $\psi_{Ai}$
that appears in the superparticle action (\ref{hurmex}) has  components with eigenvalues $ \frac{1}{2},-\frac{1}{2}$  or 
$- \frac{1}{2},\frac{1}{2}$ for
$J_{12}$ and $ J_{34}$. The field $\psi_{Ai}$, taking into account its additional index $i=1,2$, has two components with either
set of eigenvalues.  One linear combination of the two states with eigenvalues $\frac{1}{2},-\frac{1}{2}$ and likewise
one linear combination with eigenvalues $-\frac{1}{2},\frac{1}{2}$ is  $\eR'$-invariant.  
We denote these linear combinations as $\psi_+$ and $\psi_-$, respectively.

To construct the Hilbert space $\widehat\H_0$, we have to quantize the two fermions $\psi_{\pm}$.  Since they obey $\psi_+^2=\psi_-^2=0$,
$\{\psi_+,\psi_-\}\not=0$, quantizing them will give two states that are exchanged by $\psi_+$ and $\psi_-$.  The two states, 
since they are exchanged
by fermion operators,  will have different statistics. Acting by $\psi_+$ will shift 
$J_{12}, \,J_{34}$ by $\frac{1}{2},-\frac{1}{2}$, and acting by $\psi_-$ will
shift these eigenvalues by the same amount in the opposite direction.  If the spectrum is $\eS$-invariant, then the set of 
$J_{12}$ and $J_{34}$ eigenvalues must be invariant under reversal of sign and so the eigenvalues must
be $\frac{1}{4},-\frac{1}{4}$ for one state and $-\frac{1}{4},\frac{1}{4}$ for the other state.  (This possibility was suggested in \cite{LMV}.)

This is puzzling.  The eigenvalues of $J_{12}$ appear to show that these particles are ``semions,'' of spin
$\pm 1/4$.  This suggests the existence of a long range interaction of statistical nature that, among other things, would surely interfere
with the formula (\ref{meloxy}) for multiple winding.  Though we doubt that there is such a long range interaction, we are not
quite sure why not.  We may note that in certain supersymmetric gauge theories in three dimensions with $\N=2$ supersymmetry,
it has been argued on similar grounds that certain BPS vortices that possess only the minimum spectrum of fermion zero modes
are semionic (for example, see \cite{MT}), but a careful
treatment \cite{IS} including the effects of anomalous Chern-Simons couplings associated to cancellation of discrete anomalies
shows that in these particular models, the vortices actually
have half-integer spin.  It would seem that the spin cannot be shifted in a $\eS$-invariant model, but perhaps some form of anomaly
inflow from the noncompact directions in $L$ or from the bulk of $\R^5\times Y$ is relevant in reconciling the value of the
spin with the absence of a statistical interaction.   Fractional spin in somewhat similar M-theory geometries has been discussed recently
in \cite{Ganor}.

We will have to leave this as a puzzle and move on to discuss the general case of BPS particles with arbitrary quantum numbers.  
Luckily, we find no further trouble.
  In the general case, we write the Hilbert space describing
BPS particles of charges $\vec q,\vec r$ as $\widehat\H_{\vec q,\vec r}=\widehat\H_0\otimes \eV_{\vec q,\vec r}$, where $\widehat\H_0$ is the Hilbert space described
above that realizes the supersymmetry algebra in a minimal way and  $\eV_{\vec q,\vec r}$ is
some vector space.
By arguments similar to those that we gave in the GV case, we can assume that the supersymmetry and translation generators
act only in $\widehat\H_0$, but the rotation generators $J_{12}$ and $J_{34}$ and the Hamiltonian $\hH$ will also act in $\eV_{\vec q,\vec r}$.  
In fact, it is convenient to introduce the anti-selfdual and selfdual combinations of $J_{12}$ and $J_{34}$:  $J_-=J_{12}-J_{34}$
and $J_+=J_{12}+J_{34}$. The rotation generator $\eJ=\frac{1}{2}\TT^{-\mu\nu}J_{\mu\nu}$ that enters in the supersymmetry algebra
is simply $\eJ=\TT J_-$.   We write $\eJ_{\vec q,\vec r}$ for the matrix by which $\eJ$ acts on $\eV_{\vec q,\vec r}$.  The same argument
as in eqn. (\ref{mortox}) shows that in acting on $\eV_{\vec q,\vec r}$, $\hH$ is equal to $-\eJ_{\vec q,\vec r}$.   Now we can repeat the
reasoning that led to eqn. (\ref{zomobo}).  Replacing $\widehat\H_0$ by $\widehat\H_0\otimes \eV_{\vec q,\vec r}$ has the effect of multiplying the
contribution of states of charges $\vec q,\vec r$ propagating once around the circle by $\Tr_{\eV_{\vec q,\vec r}}\,(-1)^F \exp\left(2\pi e^\sigma
\eJ_{\vec q,\vec r}\right)$.  Reasoning as in the derivation of eqn. (\ref{zomobo}), we can write this trace in two-dimensional terms as
$\Tr_{\eV_{\vec q,\vec r}}\,(-1)^F \exp\left(-i\pi
\JJ_{\vec q,\vec r}/4\X^0\right)$, where $\JJ=\Wp J_-$ acts as $\JJ_{\vec q,\vec r}$ on $\eV_{\vec q,\vec r}$.

For $k$-fold winding, we have to multiply the exponent in this trace by $k$.  The generalization of eqn. (\ref{meloxy})
is thus simply
\begin{align}\label{zeloxy}i\int \frac{\d^2x\d^2\theta}{(2\pi)^2}\sqrt{g^E}&\cdot \sum_{k=1}^\infty\frac{1}{k} 
\exp\left(2\pi ik\left(\sum_I q_I\ZZ^I+\sum_\rho r_\rho\U^\rho\right)
\right)\cr & \Tr_{\eV_{\vec q,\vec r}}\,\left[(-1)^F \exp\left(-i\pi k
\JJ_{\vec q,\vec r}/4\X^0\right)\right]\frac{\pi\Wp/8}{\sin(\pi k\Wp/8\X^0)}. \end{align}
 $J_-$ and $J_+$ are generators of the two factors of $SU(2)_\ell\times SU(2)_r\sim SO(4)$, and the way they enter the OV
 formula is similar to the way $SU(2)_\ell$ and $SU(2)_r$ enter the GV formula.  The OV and GV formulas depend respectively on the detailed
 $J_-$ and $SU(2)_\ell$ quantum numbers of the BPS states, but $J_+$ and $SU(2)_r$ only enter to the extent that they affect
 the statistics of the states.
 
 \appendix

\section{Supergravity Conventions}
Here we describe in detail our supergravity conventions in dimensions 11, 5, and 4, and also the dimensional reduction relating them.  Our conventions
are generally those of \cite{EE}.

\subsection{Gamma-Matrices and Spinors}
Euclidean gamma-matrices will always satisfy a Clifford algebra with a plus sign, e.g. $\{\Gamma_I,\Gamma_J\}=2\delta_{IJ}$. For a fermion $\psi$, sometimes we write
\begin{equation}\label{monz}
\bar\psi \equiv \psi^T C,
\end{equation}
where $C$ is usually called the charge-conjugation matrix.

\subsubsection{Four Dimensions With Euclidean Signature}
The flat space 4d gamma-matrices are denoted $\gamma_a$, while the curved-space matrices are $\gamma_\mu = e_\mu^a \gamma_a$, where $e_\mu^a$ is the 4d vielbein.   Negative chirality (or left-handed) spinor indices
are  denoted $A, B, C,\dots$, while positive chirality (or right-handed) ones are denoted  $\dA, \dB, \dot{C},\dots$.

Indices $A,B,C,\dots$  and $\dA,\dB,\dot C,\dots$ are lowered or raised by antisymmetric tensors
$\varepsilon_{AB}$ and  $\varepsilon_{\dA\dB}$, where we choose as usual $\varepsilon_{12}=\varepsilon^{12}=1$. In lowering/raising indices, we adhere to the so-called NW-SE (``Northwest-Southeast'') convention, when indices are always summed in the NW-SE direction: $\psi_A = \psi^B \varepsilon_{BA}$, $\psi^A = \varepsilon^{AB}\psi_B$.

We choose the following representation for the 4d Euclidean gamma-matrices:
\begin{align}
\gamma_i = \left(\begin{matrix}
0 & \sigma_i\cr
 \sigma_i & 0\cr
\end{matrix} \right), \,i=1,\dots,3,~~\gamma_4=\left(\begin{matrix}
0 & -i\mathbbm{1}\cr
i\mathbbm{1} & 0\cr
\end{matrix}\right),
\end{align}
where $\sigma_i$ are the usual Pauli matrices. The chirality matrix is:
\begin{align}
\gamma_5 = \gamma_1\gamma_2\gamma_3\gamma_4=\left(\begin{matrix}
-\mathbbm{1} & 0\cr
0 & \mathbbm{1}\cr
\end{matrix}\right).
\end{align}
 With this choice, the ``upper'' or ``lower'' components of a 5d spinor $\psi^\alpha$ are 4d spinors
  $\psi^A$ and $\psi^{\dA}$ of negative or positive chirality, respectively.
 Moreover, gamma-matrices with both spinor indices lowered behave under complex conjugation as follows:
\begin{equation}
\label{conj}
(\gamma^\mu_{A\dA})^*=-\varepsilon^{AB}\varepsilon^{\dA\dB}\gamma^\mu_{B\dB}.
\end{equation}
As usual in even dimensions, there are two possible charge conjugation matrices, which we will denote as $C$ and $\tilde{C}=-C\gamma_5$, satisfying $\gamma_
\mu^T=C\gamma_\mu C^{-1}$ and $\gamma_\mu^T=-\tilde{C}\gamma_\mu\tilde{C}^{-1}$ (note that $\gamma_\mu^T=\gamma_\mu^*$ in Euclidean signature):
\begin{align}
\label{Cmatrix}
C=\gamma_2\gamma_4=\left(\begin{matrix}
\varepsilon & 0\cr
0 & -\varepsilon\cr
\end{matrix} \right),\quad \tilde{C}=-C\gamma_5=\left(\begin{matrix}
\varepsilon & 0\cr
0 & \varepsilon\cr
\end{matrix} \right),\quad \varepsilon=\left(\begin{matrix}
0 & 1\cr
-1 & 0\cr
\end{matrix}\right).
\end{align}
By saying that we lower/raise both left and right 4d spinor indices by $\varepsilon$, we have automatically picked $\tilde{C}$ in $d=4$.

In Lorentz signature, fermions always carry a real structure.  This is typically not the case in Euclidean signature
(for example, if the Standard Model of particle physics is formulated in Euclidean signature, the fermions carry no real structure).  
For our purposes in this paper, spinors in 4d Euclidean
space can be considered to come by dimensional reduction from 5d Minkowski spacetime, and therefore they carry a real structure.
 Since the spinor representation of $\Spin(4)$ (or of $\Spin(4,1)$) is  pseudo-real rather than real, to define a reality condition, one has to add an extra index $i=1,2$ (which can also be lowered/raised by an antisymmetric tensor $\varepsilon_{ij}$). Then the reality conditions for left-handed and right-handed spinors $\psi^{Ai}$ and $\psi^{\dA i}$ respectively are:
\begin{align}
\label{fourdreal}
(\psi^{Ai})^* &= \psi_{Ai},\cr
(\psi^{\dA i})^* &=\psi_{\dA i}.
\end{align}

\subsubsection{5d Gamma-Matrices and Spinors}
We denote 5d gamma-matrices as $\Gamma_{\bA}$ with flat index $\bA$ (or $\Gamma_M$ with the curved index $M$). In Lorentz signature, we choose
 the following relation between  5d and  4d gamma-matrices:
\begin{align}
\label{4to5ext}
\Gamma_{\bA=a} &= \gamma_a, a=1\dots4,\cr
\Gamma_0 &= i\gamma_5.
\end{align}
In 5d Euclidean signature, we take the fifth gamma-matrix to be  $\Gamma_5=\gamma_5$.

We denote 5d spinor indices by $\alpha,\beta,\gamma,\dots$. They are lowered/raised by  the matrix $C_{\alpha\beta}$ that was defined in
eqn. (\ref{Cmatrix}) (in $d=5$, Lorentz invariance leaves no choice in this matrix)
and again a NW-SE rule is applied. We sometimes write a 5d spinor $\Psi^\alpha$ in terms of the 4d chiral basis and think of it as a pair of Weyl spinors $\Psi^A$ and $\Psi^{\dA}$, but with indices raised or lowered by the 5d matrix $C_{AB}=\varepsilon_{AB}, C_{\dA\dB}=-\varepsilon_{\dA\dB}$. In particular, that is how we usually treat the 5d supersymmetry algebra, writing it in terms of the chiral components $Q_{Ai}$ and $Q_{\dA i}$. Of course, such a splitting explicitly breaks part of the ${\rm Spin}(4,1)$ symmetry, but this part is broken by the Kaluza-Klein reduction anyway.

To define the reality condition satisfied by 5d spinors in Lorentz signature, we first introduce
\begin{align}
\label{Bmatrix}
B = -i\Gamma_0 C = \left(\begin{matrix}
-\varepsilon & 0\cr
0 & -\varepsilon\cr
\end{matrix} \right),
\end{align}
and define
\begin{equation}
\Psi^c = B^{-1}\Psi^*.
\end{equation}
To satisfy a reality condition, a spinor also needs an additional index $i=1,2$, since the spinor representation of $\Spin(4,1)$ is
pseudoreal. Finally, the
 reality condition on $\Psi^i$ is $\Psi^i = \varepsilon^{ij}(\Psi^j)^c$. In terms of the chiral components $\Psi^{Ai}$ and $\Psi^{\dA i}$, this condition is:
\begin{align}
\label{fivedreal}
(\Psi^{Ai})^* &= \Psi^{Bj}\varepsilon_{BA}\varepsilon_{ji}\equiv\Psi_{Ai},\cr
(\Psi^{\dA i})^* &= \Psi^{\dB j}\varepsilon_{\dB\dA}\varepsilon_{ji}\equiv -\Psi_{\dA i}.
\end{align}

The 5d spinor $\Psi^{i\alpha}$ in (4+1)d reduces to a pair of 4d  spinors $\psi^{Ai}$ and $\psi^{\dA i}$.  We make the identification with  the indices raised: $\psi^{Ai}=\Psi^{Ai}$, $\psi^{\dA i}=\Psi^{\dA i}$.  It is necessary to specify this because we have introduced a slightly different convention in raising and lowering 4d spinor
indices. 

\subsubsection{6d and 11d Gamma-Matrices}

We denote the  6d gamma-matrices along the Calabi-Yau manifold $Y$ as
  $\tilde\gamma_n, n=6\dots11$ (here we do not specify whether the index is ``curved'' or ``flat''). The 6d chirality matrix is $\tilde\gamma_*=i\tilde\gamma_6\cdots
  \tilde\gamma_{11}$. We think about spinors on $Y$ as $(0,p)$-forms for $p=0\dots3$. If $z^i$ are local coordinates on $Y$ and $\cQ_{i\bar{j}}$ is a metric on $Y$, the gamma-matrices act as: 
\begin{align}
\tilde\gamma_{z^i}&=\sqrt{2}\cQ_{i\bar{j}}\d\bar{z}^{\bar{j}}\wedge\cr
\tilde\gamma_{\bar{z}^{\bar i}}&=\sqrt{2}\mathlarger{\iota}_{\frac{\partial}{\partial\bar{z}^{\bar i}}}.
\end{align}  
  We choose chirality in such a way that a covariantly constant spinor $\lambda_-$ of negative chirality corresponds to an antiholomorphic $(0,3)$-form $\bar\Omega$, while a covariantly constant spinor $\lambda_+$ of positive chirality corresponds to a constant function $1$.  We choose the 6d charge conjugation matrix $C_6$ satisfying
\begin{equation}
  \tilde\gamma_n^T=-C_6\tilde\gamma_n C_6^{-1}.
\end{equation}
The choice of  $C_6$ lets us define a bilinear pairing $(~,~)$ on fermions, and we require that $(\lambda_+,\lambda_-)=(\lambda_-,\lambda_+)=1$.

Let us use calligraphic letters for the 11d indices and denote 11d gamma-matrices by slanted capital gamma.  So we write  $\varGamma_{\mathcal{A}}$ for 11d
gamma-matrices referred to a flat basis and $\varGamma_{\mathcal{M}}$ for the ones referred to a curved basis. We choose the 11d gamma-matrices to be related as follows to the 5d and 6d gamma-matrices:
\begin{align}
\varGamma_{\mathcal{A}=\bA} &= \Gamma_\bA\otimes \tilde\gamma_*,\, \bA=1\dots 5\cr
\varGamma_{\mathcal{A}=n} &= \mathbbm{1}_4 \otimes \tilde\gamma_n,\, n=6\dots 11,
\end{align}
where $\mathbbm{1}_4$ is the unit $4\times 4$ matrix. In Lorentz signature (where $\varGamma_5$ is replaced by $\varGamma_0$), we require:
\begin{equation}
\varGamma_0\varGamma_1\dots\varGamma_4\varGamma_6\varGamma_7\dots\varGamma_{11}=1.
\end{equation}
We will use large lower-case Greek letters to denote 11d spinorial indices: $\lalpha, \lbeta, \dots$ With the above choice of the 6d charge conjugation matrix, the 11d charge conjugation matrix $C_{11}$ is related to the 5d and 6d matrices in an obvious way:
\begin{equation}
C_{11}=C_5\otimes C_6.
\end{equation}

In Lorentz signature, the supersymmetry generators are an 11d  Majorana fermion $\eta$.   In compactification on $Y$, the unbroken supersymmetries are those
for which $\eta$ is the tensor product of  $\lambda_+$ or $\lambda_-$ with a 5d  spinor $\epsilon^1$ or $\epsilon^2$:
\begin{equation}
\eta=\epsilon^2\otimes \lambda_+ + \epsilon^1\otimes\lambda_-.
\end{equation}

\subsection{5d SUSY Algebra}
From (\ref{conj}), (\ref{4to5ext}) and (\ref{fivedreal}), one can find, in  5d Minkowski space, the SUSY algebra compatible with the 5d reality conditions:
\begin{equation}
\{Q_{\alpha i}, Q_{\beta j}\} = -i\varepsilon_{ij}\Gamma_{\alpha\beta}^M P_M + \varepsilon_{ij}C_{\alpha\beta}\zeta,
\end{equation}
where $\zeta$ is a real central charge. In a chiral basis, the algebra is
\begin{align}
\{Q_{Ai},Q_{Bj}\}&=\varepsilon_{AB}\varepsilon_{ij}(H+\zeta)\cr
\{Q_{Ai},Q_{\dB j}\}&=-i\varepsilon_{ij}\Gamma^\mu_{A\dB}P_\mu\cr
\{Q_{\dA i}, Q_{\dB j}\}&= \varepsilon_{\dA \dB}\varepsilon_{ij}(H-\zeta),
\end{align}
where $H=P^0$ is the 5d Hamiltonian.

\subsection{11d Supergravity}
Though not explicitly used in the main part of the paper, the following form of the 11d supergravity action is implicitly assumed:
\begin{align}
\mathcal{L} = \frac{1}{2\kappa_{11}^2}\Bigg(E R - \frac{E}{48}G^2 + \frac{1}{12^4}\epsilon^{\mathcal{MNLP}_1...\mathcal{P}_4\mathcal{Q}_1...\mathcal{Q}_4}C_{\mathcal{MNL}}G_{\mathcal{P}_1...\mathcal{P}_4}G_{\mathcal{Q}_1...\mathcal{Q}_4}\cr
-E\bar\psi_{\mathcal{M}}\varGamma^{\mathcal{MNP}}D_{\mathcal{N}}\left[{1\over 2}(\omega+\hat\omega)\right]\psi_{\mathcal{P}}\cr
-\frac{E}{192}(\bar\psi_{\mathcal{Q}}\varGamma^{\mathcal{MNLPQR}}\psi_{\mathcal{R}} + 12\bar\psi^{\mathcal{M}}\varGamma^{\mathcal{NL}}\psi^{\mathcal{P}})(G+\hat{G})_{\mathcal{MNLP}}\Big\}\Bigg),
\end{align}
where $E$ is the determinant of the 11d vielbein, $G$ is a curvature of the $C$-field, $\psi_\mathcal{M}$ is a gravitino field (a Majorana vector-spinor), and hatted quantities include some extra corrections quadratic in fermions (the exact expressions  are not important to us). The supersymmetry transformations are:
\begin{align}
\delta E_{\mathcal M}^{\mathcal A} &= {1\over 2}\bar\eta \varGamma^{\mathcal A} \psi_{\mathcal M},\cr
\delta\psi_{\mathcal M} &= D_{\mathcal M}(\hat\omega)\eta + T_{\mathcal M}^{\ \mathcal{NPQR}}G_{\mathcal{NPQR}}\eta,\cr
\delta C_{\mathcal{MNP}} &= -{3\over 2}\bar\eta\varGamma_{[\mathcal{MN}}\psi_{\mathcal P]},
\end{align}
where
\begin{equation}
T_{\mathcal M}^{\ \mathcal{NPQR}}={1\over 288}\left( \varGamma_{\mathcal M}^{\ \mathcal{NPQR}} - 8\delta_{\mathcal M}^{[\mathcal N}\varGamma^{\mathcal{PQR}]} \right).
\end{equation}

In the action above, $\kappa_{11}$ is the 11-dimensional gravitational constant. It is actually related to the M2-brane tension $T_2$ (see for example \cite{deAlwis:1996ez}), the relation being:
\begin{equation}
2\kappa_{11}^2(T_2)^3=(2\pi)^2.
\end{equation}
We will work in units with $T_2=1$ and thus:
\begin{equation}
2\kappa_{11}^2=(2\pi)^2.
\end{equation}

\subsection{Reduction from 11d to 5d}
We review the dimensional reduction of 11d supergravity on a Calabi-Yau $Y$ (an original reference is \cite{cadavid}). We denote the 5-dimensional metric as ${\rm G}$ (this will hopefully  not be confused with the field strength $G$ of the 11d $C$-field). We denote the  Calabi-Yau Ricci  metric of $Y$ as $\cQ$, and the compatible complex structure as $I$. The Kahler form of $Y$  is $\omega=\cQ (I\cdot,\cdot)$. The volume form is:
\begin{equation}
\textrm{Vol}={1\over 6}\omega\wedge \omega\wedge \omega.
\end{equation}
For arbitrary $(1,1)$-forms $\alpha$ and $\beta$, we have identities:
\begin{align}
\label{CY_formulas}
\beta\wedge \omega\wedge \omega &= (\beta,\omega)\textrm{Vol},\cr
\alpha\wedge\beta\wedge \omega &= {1\over 4}\left[ (\alpha,\omega)(\beta,\omega) - 2(\alpha,\beta) \right]\textrm{Vol},\cr
*\alpha &= -2\alpha\wedge \omega + {1\over 2}(\alpha,\omega)\omega\wedge \omega.
\end{align}
Here the inner product on 2-forms is defined by $(\alpha,\beta)=\alpha_{nr}\beta_{ms} \cQ ^{nm}\cQ ^{rs}$. 

Because $Y$ is Ricci-flat, its Kahler form is harmonic and thus can be expanded in a basis of harmonic $(1,1)$-forms $(\omega_I)$:
\begin{equation}
\omega = \sum_I v^I \omega_I,
\end{equation}
where $v^I$ are Kahler moduli. Define also:
\begin{align}
\CC_{IJK} = {1\over 6}\int_Y \omega_I\wedge \omega_J\wedge\omega_K.
\end{align}

Now let us reduce the bosonic part of the 11d action. We will be interested in the Kahler moduli of  only. (The complex structure moduli of $Y$ give rise to hypermultiplets, which decouple at low energies from the vector multiplet couplings that are described by the GV formula.)

One can find the following formula for the 11d Ricci scalar in terms of the 5d Ricci scalar and the Calabi-Yau metric:
\begin{equation}
\sqrt{\cQ }R^{(11)} = \sqrt{\cQ }R^{(5)} - \nabla_M(\sqrt{\cQ }\cQ ^{mn}\partial^M \cQ _{mn}) - \sqrt{\cQ }\left({1\over 4}(\partial_M \cQ ,\partial^M \cQ ) -{1\over 4}(\cQ ,\partial_M \cQ )(\cQ ,\partial^M \cQ )\right).
\end{equation}
Here $M$ is a 5d index, and covariant derivatives are with respect to the 5d metric. The total derivative part clearly drops out of the action.

Denote the volume of $Y$ by $V$. Introduce also $v=V^{1/3}$ and
\begin{equation}
h^I = {v^I\over v}.
\end{equation}
The volume is part of a hypermultiplet, so we are not interested in the action for it right now. Using (\ref{CY_formulas}), we can find:
\begin{align}
\int_Y R^{(11)}\textrm{Vol} &= V(x)R^{(5)} + \int_Y \partial_M \omega\wedge \partial^M \omega \wedge \omega + {1\over 4}\int_Y \partial_M \omega \wedge *(\partial^M \omega)\cr
&= V(x)\left(R^{(5)} + 3C_{IJK}h^I \partial_M h^J \partial^M h^K + {\rm hypermultiplet\ part}\right).
\end{align}

Now take a look at the 3-form field. At low energies, we expand
\begin{align}
\label{Cfieldred}
C &= \sum_I V^I\wedge \omega_I,\cr
G &= \sum_I \d V^I \wedge \omega_I,
\end{align}
where the $V^I$ are abelian gauge fields in five dimensions.
 Then \begin{equation}\label{zonk} (G,G) = 6 (\d V)^I_{MN}(\d V)^J_{PQ}{\rm G}^{MP}{\rm G}^{NQ}(\omega_I,\omega_J).\end{equation}
 The kinetic term  for $C$ in 11 dimensions reduces in $d=5$ to  
\begin{equation}
-{v\over 4}a_{IJ} (\d V^I\cdot \d V^J),
\end{equation}
where, using (\ref{CY_formulas}):
\begin{equation}
a_{IJ} = {1\over 4}\int_Y \omega_I \wedge *\omega_J = -3\CC_{IJK}h^K  + {9\over 2}(\CC hh)_I (\CC hh)_J.
\end{equation}
The 11d Chern-Simons term reduces to
\begin{equation}
-{1\over 2}\CC_{IJK}V^I\wedge \d V^J \wedge \d V^K.
\end{equation}
So, ignoring hypermultiplets, the bosonic part of the action  is (remember that $\kappa_{11}^2=2\pi^2$):
\begin{align}
2\pi^2\mathcal{L}_5 \textrm{Vol} = \Big[V(x)\left({1\over 2}R^{(5)} + {3\over 2}\CC_{IJK}h^I \partial_M h^J \partial^M h^K\right) -{v\over 4}a_{IJ} (\d V^I\cdot \d V^J)\Big]\textrm{Vol} \cr 
-{1\over 2}\CC_{IJK}V^I\wedge \d V^J \wedge \d V^K.
\end{align}
Now make a Weyl rescaling of the 5d metric ${\rm G}_{MN} \to {1\over v^2} {\rm G}_{MN}$, to  bring the 5d action to the Einstein frame:
\begin{align}
\label{5daction}
2\pi^2\mathcal{L}_5 \textrm{Vol} = \Big[{1\over 2}R^{(5)} + {3\over 2}\CC_{IJK}h^I \partial_M h^J \partial^M h^K -{1\over 4}a_{IJ} (dV^I\cdot dV^J)\Big]\textrm{Vol} \cr
-{1\over 2}\CC_{IJK}V^I\wedge dV^J \wedge dV^K+{\rm hypermultiplets}.
\end{align}
Our conventions in this action are slightly different from those often found in the literature. 
To get the action in the conventions of \cite{Cucu}, one has to rescale by $h^I \to \sqrt{3\over 2}h^I$ and $\CC_{IJK} \to {2\sqrt{2}\over 3\sqrt{3}}\CC_{IJK}$ (and also do appropriate rescalings to get rid of the factor $2\pi^2$ coming from the gravitational constant). However, the action normalized
as in  (\ref{5daction})  is more convenient for us.

Some quantities that appeared in section \ref{backsup} are 
\begin{align}
h_I &= \CC_{IJK}h^Jh^K,\cr
a_{IJ} &= -3\CC_{IJK}h^K + {9\over 2}h_I h_J,\cr
h_I &= {2\over 3}a_{IJ}h^J.
\end{align}
The constraint $\CC_{IJK}h^Ih^Jh^K=1$ (eqn. (\ref{norf}), which implies that $h_Ih^I=1$, was used in the last line.

The scalar kinetic energy in  $2\pi^2\mathcal{L}_5$ can be rewritten as:
\begin{equation}
-{1\over 2}a_{IJ}\partial_Mh^I\partial^Mh^J.
\end{equation}

\subsection{Reduction from 5d to 4d}
We reduce the $N=1, d=5$ supergravity on a circle and make the field redefinitions required to relate it to the standard $N=2, d=4$ supergravity in the Einstein frame metric (a similar procedure was performed in \cite{Gunaydin:1983bi}).

Assume that the fifth direction is a circle parametrized by an angular variable $y$ ($0\leq y\leq 2\pi$). After integrating over $y$, the overall factor of $1/(2\pi^2)$ in front of the 5d action (\ref{5daction}) will be replaced by an overall $1/\pi$ in front of the 4d action.  This factor is sometimes removed by rescalings, but we will find
it more convenient not to do so.

Take the following ansatz for the funfbein $e_M^\bA$:
\begin{align}
\label{e_red}
e_M^\bA = \left(\begin{matrix} e^{-\sigma/2} e_{\mu}^{a} & e^{\sigma}B_{\mu}\cr 0 & e^{\sigma} \end{matrix} \right),\quad e_{\bA}^M = \left( \begin{matrix} e^{\sigma/2}e_{a}^{\mu} & -e^{\sigma/2}e^{\mu}_a B_{\mu}\cr 0 & e^{-\sigma}\cr \end{matrix}\right).
\end{align}
The 5-dimensional Ricci scalar $R^{(5)}$ takes the following form in terms of the 4-dimensional Ricci scalar $R^{(4)}$ and other fields present in the funfbein:
\begin{equation}
R^{(5)} = e^{\sigma}R^{(4)} + e^{\sigma}\Box\sigma - {3\over 2}e^{\sigma}(\partial\sigma)^2 - {1\over 4}e^{4\sigma}(dB)^2.
\end{equation}

Set $\alpha^I=V^I_y$. Define the 4-dimensional gauge fields as $A^{\Lambda}, \Lambda=0\dots b_2(Y)$, where $\Lambda=I=1,\dots, b_2(Y)$ come from
reduction of the 5-dimensional vectors, while $\Lambda=0$ corresponds to the Kaluza-Klein (KK) vector:
\begin{align}
A^I_{\mu} &= V_{\mu}^I -\alpha^I B_{\mu}\cr
A^0_{\mu} &= -B_{\mu},\cr
\end{align}

The scalar kinetic term in 4d originates from the curvature term in 5d, the scalar kinetic term in 5d and the vector kinetic term in 5d. It takes the form:
\begin{equation}
-\frac{1}{\pi}\left({1\over 2} e^{-2\sigma}a_{IJ}\partial_{\mu}(e^{\sigma}h^I)\partial^{\mu}(e^{\sigma}h^J) + {1\over 2} e^{-2\sigma}a_{IJ}\partial_{\mu}\alpha^I \partial^{\mu}\alpha^J\right).
\end{equation}
If we define a complex scalar
\begin{equation}
Z^I = \alpha^I + i e^{\sigma}h^I,
\end{equation}
then the kinetic term becomes
\begin{equation}
-\frac{1}{\pi}g_{I\bar{J}}\partial_{\mu}Z^I\partial^{\mu}\bar{Z}^J,
\end{equation}
where
\begin{equation}
g_{L\bar{M}}=\frac{1}{2} e^{-2\sigma}a_{LM}=\frac{\partial}{\partial Z^L}\frac{\partial}{\partial \overline{Z^M}}\log\left[ \CC_{IJK}(Z^I-\overline{Z^I})(Z^J-\overline{Z^J})(Z^K-\overline{Z^K})\right].
\end{equation}

The vector kinetic term takes the standard form:
\begin{equation}
-\frac{i}{4\pi}\mathcal{N}_{\Lambda \Sigma}F^{\Lambda+}_{\mu\nu}F^{\Sigma+\mu\nu} + {\rm c.c.}
\end{equation}
with coefficients
\begin{align}
\N_{IJ}&=- i(e^{\sigma} a_{IJ} - 3i \CC_{IJK} \alpha^K),\cr
\N_{I0}&= i(e^{\sigma}a_{IJ}\alpha^J - \frac{3i}{2}\CC_{IJK}\alpha^J \alpha^K),\cr
\N_{00}&=- i(e^{\sigma} a_{IJ}\alpha^I \alpha^J - i \CC_{IJK}\alpha^I \alpha^J \alpha^K+\frac{1}{2}e^{3\sigma}).
\end{align}
One can check (see (\ref{yrf})) that this corresponds to the prepotential:
\begin{equation}
\mathcal{F}_0^{\rm cl}=-\frac{1}{2}{\CC_{IJK}X^I X^J X^K\over X^0}.
\end{equation}

Another useful relation in KK reduction from $d=5$ to $d=4$ is the expression for the 5d Dirac operator 
\begin{equation}
\slashed{\mathcal{D}}=\Gamma^M(\partial_M +\frac{1}{4}\omega_M^{\bA\bB}\Gamma_{\bA\bB}-iq_IV^I_M)
\end{equation}
in terms of the 4d fields:
\begin{align}
\slashed{\mathcal{D}}=e^{\sigma/2}\gamma^\mu(\partial_\mu+\frac{1}{4}\omega_\mu^{ab}\gamma_{ab}-iq_I V^I_\mu - B_\mu\partial_y + iq_I\alpha^I B_\mu) + e^{-\sigma}\gamma_5\partial_y - iq_I\alpha^Ie^{-\sigma}\gamma_5\cr
 + \frac{1}{8}e^{2\sigma}(\d B)_{\mu\nu}\gamma^{\mu\nu}\gamma_5-\frac{1}{4}e^{\sigma/2}\slashed{\partial}\sigma.
\end{align}
Taking $\sigma$ to be constant, taking $V^I_\mu=h^I V_\mu=\frac{h^I}{4}e^{-\sigma/2}U_\mu$, and acting on a field with the KK mode number $-n$, this reduces to:
\begin{align}
\slashed{\mathcal{D}}&=e^{\sigma/2}\gamma^\mu D_\mu -ie^{-\sigma}(n+q_I\alpha^I)\gamma_5 - \frac{i}{32}e^{\sigma/2}\WW^-_{\mu\nu}\gamma^{\mu\nu}\gamma_5,\cr
D_\mu&=\partial_\mu + \frac{1}{4}\omega_\mu^{ab}\gamma_{ab}-i\frac{\bar{\cZ}}{4}U_\mu.
\end{align}
One can see that the first term in the expression for $\slashed{\mathcal{D}}$ is just a 4d Dirac operator, the second term corrects the 5d mass term (replacing $M$ by $\cZ$ or $\bar{\cZ}$ depending on the 4d chirality), and the third term shifts the 5d magnetic moment coupling. This expression is important in the dimensional reduction of the 5d hypermultiplet action performed in section \ref{hypercalc}.

The fields that we have described can be organized in 4d supermultiplets $\X^\Lambda$ and $\W_{AB}$ as described in the main text.  General
references on these superfields are \cite{AA,BB,CC}.

\section{M2-brane on a Holomorphic Curve}\label{holc}

\def\MM{{\mathfrak M}}
\def\red{{\mathrm{red}}}
Here we derive the 5d BPS superparticle action describing an M2-brane wrapped on a smooth isolated holomorphic curve $\Sigma$ in a Calabi-Yau manifold $Y$.
\subsection{Membrane in  Superspace}

An M-theory membrane can be described as a submanifold $\Omega$ of dimension $3|0$ living in a superspace $\mathfrak{M}$ of dimension $11|32$.  If the background fields
are purely bosonic (as we can assume for our purposes), then $\MM$ is split, with reduced space some 11-dimensional spin-manifold $\mathfrak{M}_\red$ and odd directions
that parametrize the spin bundle $S(\MM_\red)$.  
We consider $\Omega$ as an abstract three-manifold with an embedding in $\MM$:
\begin{equation}
\hat X: \Omega \to \mathfrak{M}.
\end{equation}
Since $\MM$ projects to its reduced space $\MM_\red$, $\hat X$ can be projected to an embedding $X:\Omega\to\MM_\red$.
The additional information in $\hat X$ is 
 a fermionic section of the pull-back of the spinor bundle:
\begin{equation}
\Theta \in \Pi\Gamma\left(\Omega,X^* S(\mathfrak{M}_\red)\right).
\end{equation}
Here $\Gamma(\Omega,\cdot)$ represents the space of sections, and the symbol $\Pi$ tells us that $\Theta$ has odd statistics.
 $X$ and $\Theta$ are the fields that are governed by the membrane world-volume theory. The $\kappa$-symmetric action for these fields
 was constructed in \cite{Bergshoeff:1987cm, Bergshoeff:1987qx}. Our main reference for expanding the component action is \cite{deWit:1998tk}. Their conventions for  11d supergravity are slightly different from ours and can be translated by $\omega_\mu^{ab} \to -\omega_{\mathcal M}^{\mathcal{AB}}$, $R \to -R$, $\psi_\mu \to \frac{1}{2}\psi_{\mathcal M}$,  and $\eta \to \frac{1}{2}\eta$ 
 (here $\eta$
 is the 11d supersymmetry generator), 
 while also reversing the orientations of $\MM_\red$ and $\Omega$ and multiplying the action by an overall constant.

We parametrize $\Omega$ by local coordinates $\zeta^0, \zeta^1, \zeta^2$. We denote the fields on the membrane as  $Z^{\sM}(\zeta)=(X^{\mathcal M}(\zeta), \Theta^{\lalpha}(\zeta))$ (an index like
$\sM$ with a hat denotes a superspace index). Let $E_{\sM}^{\sA}$ be the supervielbein, where $\sM$ is a curved and $\sA$ is a flat superspace index. Let $B_{\sM\sN\sP}$ be the superspace three-form gauge superfield. $E_{\sM}^{\sA}$ and $B_{\sM\sN\sP}$ encode the target space geometry. The pull-back of the supervielbein to the membrane is $\Pi_i^{\sA}=E_{\sM}^{\sA}\partial Z^{\sM}/\partial\zeta^i$. The induced metric is $g_{ij}=\Pi_i^{\mathcal A}\Pi_j^{\mathcal{B}}\eta_{\mathcal{A}\mathcal{B}}$, where $\eta_{\mathcal{A}\mathcal{B}}$ is the 11-dimensional Minkowski metric. Here $\mathcal{A}, \mathcal{B}$ are ordinary flat 11-dimensional indices. Then the membrane action is:
\begin{equation}
S = \int d^3\zeta \left[ -\sqrt{-g} + {1\over 6}\varepsilon^{ijk}\Pi_i^{\sA} \Pi_j^{\sB} \Pi_k^{\sC} B_{\sC\sB\sA}\right].
\end{equation}

Define the matrix:
\begin{equation}
\varGamma = -{\varepsilon^{ijk}\over 6\sqrt{-g}}\Pi_i^{\mathcal{A}}\Pi_j^{\mathcal{B}}\Pi_k^{\mathcal{C}}\varGamma_{\mathcal{ABC}}.
\end{equation}
It satisfies $\varGamma^2=1$ and enters in defining the  $\kappa$-symmetry of the membrane action:
\begin{equation}
\delta Z^{\sM} E_{\sM}^{\mathcal{A}}=0,\quad \delta Z^{\sM} E_{\sM}^{\lalpha}=(1-\varGamma)^{\lalpha}_{\ \lbeta}\kappa^{\lbeta},
\end{equation}
where $\kappa(\zeta)$ is a local fermionic parameter. The $\kappa$-symmetry allows one to gauge away half of the fermionic degrees of freedom on the membrane.  (Instead of saying
that the membrane has a worldvolume of dimension $3|0$ and is governed by a $\kappa$-symmetric action, an equivalent point of view that has some advantages is to say that the membrane
worldvolume has dimension $3|8$.  The $3|8$-dimensional membrane worldvolume in the second point of view is obtained by applying all possible $\kappa$ transformations to the 
$3|0$-dimensional membrane
worldvolume in the first point of view.  This refinement will not be important for us.)

\subsection{Wrapped BPS Membrane}
We focus on the case  $\mathfrak{M}_\red = M \times Y$, where $Y$ is a Calabi-Yau manifold  and $M$ is a five-manifold with a large radius of curvature.
In our application, $M$ will eventually be 
  either Minkowski spacetime or the  supersymmetric G\"odel universe (also called the graviphoton background in this paper). Let $\Sigma \subset Y$ be a $2$-cycle inside of $Y$. Consider an M2-brane wrapping $\Sigma$. It propagates as a 5d particle on $M$, given that the radius of curvature of $M$ is large enough. More precisely, a propagating M2-brane wrapped on $\SIgma$
  generates a whole infinite set of 5d particles corresponding to its different  internal excitations. These excitations may or may not preserve some supersymmetry, and correspondingly the particles propagating on $M$ form short or long multiplets of SUSY. We are interested in those particles that preserve as much of the 5d supersymmetry as possible, namely half of it.  These arise
  from the supersymmetric ground states of the internal motion.  So those are the states that we must understand.

A supersymmetry of  the ambient superpace  $\mathfrak{M}$ remains unbroken in the presence of an M2-brane if in the M2-brane theory
the supersymmetry transformation can be compensated by a $\kappa$-transformation  \cite{bbs}.  For this to be possible,  $\Sigma$ must be\footnote{In \cite{bbs}, this is shown for a string worldsheet in a superstring theory compactified on $Y$. Our case can be reduced to this by considering an M2-brane wrapping $\Sigma$ and winding the M-theory circle once.}  a holomorphic curve in $Y$ \cite{bbs}.  In this appendix, we will consider only the case that $\Sigma$ is isolated; in other words, we 
assume  that it has no deformations (even infinitesimal ones)
as a holomorphic curve in $Y$.  Otherwise, the moduli of $\Sigma$, along with fermionic zero-modes that will be related to them by supersymmetry, must be
quantized in order to determine the supersymmetric states of the M2-brane.

So now we consider a membrane with worldvolume $\Sigma\times\gamma$, where $\gamma\subset M$ is a 5d worldline. We parametrize $\gamma$ by a coordinate
$t$ and $\Sigma$ by a local holomorphic coordinate $z$.  The membrane worldvolume is parametrized as:
\begin{align}
\label{bosmod}
X^M(t,z,\bar{z})&=x^M(t),\,M=0\dots 4,\cr
X^m(t,z,\bar{z})&=X^m(z,\bar{z}),\,m=6\dots 11,
\end{align}
where $x^M(t)$ parametrizes $\gamma\subset M$, and $X^m(z,\bar{z})$ parametrizes $\Sigma\subset Y$. 
If $M$ is taken to have Euclidean signature,
we replace here $X^0$ by  $X^5=iX^0$. We pick  local holomorphic coordinates $(z, w^1, w^2)$ on $Y$ so that $\Sigma$ is locally defined by $w^1=w^2=0$.

To describe the fermionic fields of the M2-brane, we first note that $S(M\times Y)=S(M)\otimes S(Y)$. Then we recall that on a Calabi-Yau manifold, one has
an isomorphism $S(Y) \cong \Omega^{0,\bullet}Y$, where $\Omega^{0,\bullet}$ is the space of $(0,q)$-forms, $q=0,\dots,3$.  In this isomorphism,
 the Dirac operator on $Y$ is simply $\sqrt{2}(\bar\partial + \bar\partial^*)$ (where $\bar\partial^*$ is the adjoint of $\bar\partial$  with respect to the natural $L^2$-scalar product). Thus the field $\Theta^\alpha$ has a 5d spinor index $\alpha$ and takes values in the $(0,p)$-forms on $Y$, restricted to $\Sigma$. The fact that $Y$ is Calabi-Yau implies various isomorphisms between bundles. Let $\Omega$ be a holomorphic 3-form on $Y$, normalized so that the volume form of $Y$ is
 $i\Omega\wedge\bar\Omega$.  Let $\mathcal{G}_{z\bar{z}}$ be the restriction to $\Sigma$ of the Kahler metric of $Y$ and let $G^N_{w^i\bar{w}^j}$ be the induced metric on the normal bundle $N\Sigma$  to $\Sigma$ in $Y$. We write for various components of $\Theta^\alpha$:
\begin{align}
\theta^\alpha_{\bar{w}^i} &= G^N_{\bar{w}^i w^j}\chi^{\alpha w^j}\cr
\theta^\alpha_{\bar{z}\,\bar{w}^i} &= \bar\Omega_{\bar{z}\,\bar{w}^i\bar{w}^j}\tilde\chi^{\alpha\bar{w}^j}\cr
\theta^\alpha_{\bar{w}^1\bar{w}^2}&=\bar\Omega_{\bar{w}^1\bar{w}^2\bar{z}}\mathcal{G}^{z\bar{z}}\theta^\alpha_z\cr
\theta^\alpha_{\bar{z}\,\bar{w}^1\bar{w}^2}&=\bar\Omega_{\bar{z}\,\bar{w}^1\bar{w}^2}\tilde\theta^\alpha.
\end{align}
The fields $\chi$ and $\tilde\chi$ are sections of the normal bundle $N\Sigma$ (tensored with the 5d spin bundle). They are related by supersymmetry to the normal deformations of $\Sigma$ inside of $Y$. Since we are considering the case of an isolated holomorphic curve, neither normal deformations nor fermions $\chi$, $\tilde\chi$ have any zero modes. Thus we can discard them. $\Theta^\alpha$ then reduces to
\begin{equation}
\label{fermf}
\Theta^\alpha = \theta^\alpha + \tilde\theta^\alpha_{\bar z} \d\bar{z} + \bar\Omega_{\bar{w}^1\bar{w}^2\bar{z}}\mathcal{G}^{z\bar{z}}\theta^\alpha_z\d\bar{w}^1\wedge\d\bar{w}^2 + \bar\Omega_{\bar{z}\bar{w}^1\bar{w}^2}\tilde\theta^\alpha\d\bar{z}\wedge\d\bar{w}^1\wedge\d\bar{w}^2.
\end{equation}

Because of our assumption that $\SIgma$ is rigid, in the quantization of an M2-brane wrapping $\Sigma$, the only bosonic zero-modes
are the ones associated to the center of mass motion along the five-manifold $M$.  
Hence those are the only bosonic modes in the effective action that describes such a superparticle; they  parametrize the particle orbit $\gamma\subset M$.    The  fermionic modes in this action arise as the zero-modes of the internal motion,
that is, the zero-modes of the fermionic variables on $\Sigma$.
We find these modes this by studying the part of the M2-brane action that is of order $\Theta^2$, using formulas in \cite{deWit:1998tk}.  (Terms in the action of higher order in $\Theta$ give only irrelevant contributions.)  

First of all, with bosonic fields taken as in (\ref{bosmod}), one finds
\begin{equation}
\varGamma = -\frac{i}{\sqrt{-\dot{x}^2}}\mathcal{G}^{z\bar{z}}\dot{x}^M\varGamma_{Mz\bar{z}} + \O(\Theta^2).
\end{equation}
This implies that the linearized $\kappa$-symmetry is:
\begin{align}
\delta\Theta&=(1-\varGamma)\kappa + \O(\Theta^2)\cr
\delta X^{\mathcal{M}}&=\bar{(1-\varGamma)\kappa}\varGamma^{\mathcal{M}}\Theta + \O(\Theta^2).
\end{align}
This can be used to gauge-away the $\varGamma=-1$ part of $\Theta$ (up to higher orders in $\Theta$). So we may assume that $\varGamma=1+ \O(\Theta^2)$ when acting on $\Theta$, that is $(\varGamma-1)\Theta = \O(\Theta^3)$. This implies:
\begin{equation}
\label{kap_cons}
(\dot{x}^M\varGamma_M -i\mathcal{G}^{z\bar{z}}\varGamma_{z\bar{z}}\sqrt{-\dot{x}^2})\Theta=\O(\Theta^3).
\end{equation}
Using this and taking the ansatz (\ref{Cfieldred}) for the C-field, we find the action (using results of \cite{deWit:1998tk}):
\begin{align}
\label{act1}
S=\int \d t\,\d^2z &\Big[ -2\mathcal{G}_{z\bar{z}}\sqrt{-\dot{x}^2}-2i\dot{x}^M V_M^I\omega_{Iz\bar{z}} + 4i\bar\Theta\varGamma_{z\bar{z}}\nabla_t \Theta - 4\sqrt{-\dot{x}^2}\bar\Theta(\varGamma_z\nabla_{\bar{z}} + \varGamma_{\bar{z}}\nabla_z)\Theta\cr &
-\frac{1}{2}\sqrt{-\dot{x}^2}\mathcal{G}^{z\bar{z}}\bar\Theta\varGamma_{z\bar{z}}\varGamma^{MN}\Theta (\d V^I)_{MN}\omega_{Iz\bar{z}}\cr &-\frac{1}{2}\sqrt{-\dot{x}^2}\mathcal{G}_{z\bar{z}}\bar\Theta\varGamma^{w^i\bar{w}^j}\varGamma^{MN}\Theta (\d V^I)_{MN}\omega_{Iw^i\bar{w}^j} + \O(\Theta^4)\Big],
\end{align}
where $\d^2 z = \frac{i}{2}\d z\wedge\d\bar{z}$.   Here the covariant derivative $\nabla_t$ is defined using the pullback to the membrane worldvolume of the Levi-Civita connection of $M$.

\subsection{Fermionic Zero-Modes}

Now we can find the fermionic zero-modes.
Expanding around a membrane that wraps $\Sigma$ and at is rest in $M=\R^5$, so that $\dot x^2=-1$, the fermionic part of the action becomes
\begin{align}
\int \d t\d^2z &\Big[4i\bar\Theta\varGamma_{z\bar{z}}\nabla_t\Theta - 4\bar\Theta(\varGamma_z\nabla_{\bar{z}} + \varGamma_{\bar{z}}\nabla_z)\Theta-\frac{1}{2}\mathcal{G}^{z\bar{z}}\bar\Theta \varGamma_{z\bar{z}}\varGamma^{MN}\Theta(\d V^I)_{MN}\omega_{Iz\bar{z}}\cr &
-\frac{1}{2}\mathcal{G}_{z\bar{z}}\bar\Theta\varGamma^{w^i\bar{w}^j}\varGamma^{MN}\Theta (\d V^I)_{MN}\omega_{Iw^i\bar{w}^j} \Big].
\end{align}
If the $U(1)$ background fields vanish, i.e. at $V^I=0$, then only the first two terms in the action survive, the Hamiltonian becomes simply $H=4\bar\Theta(\varGamma_z\nabla_{\bar{z}} + \varGamma_{\bar{z}}\nabla_z)\Theta$, and thus the fermion zero-modes are characterized by
\begin{equation}
\label{fermmod}
(\varGamma_z\nabla_{\bar{z}} + \varGamma_{\bar{z}}\nabla_z)\Theta=0.
\end{equation}
Once we find the solutions in this idealized case, we can turn on the curvature of $M$ and a graviphoton background as small perturbations.

To solve eqn. (\ref{fermmod}), we first note that 
\begin{align}
(\varGamma_z\nabla_{\bar{z}} + \varGamma_{\bar{z}}\nabla_z)=\mathbbm{1}_4\otimes \mathcal{G}_{z\bar{z}}\mathcal{D},\quad
\mathcal{D}=\mathcal{G}^{z\bar{z}} (\tilde\gamma_z\nabla_{\bar{z}} + \tilde\gamma_{\bar{z}}\nabla_z),
\end{align}
where $\mathbbm{1}_4$ is the identity operator acting on $S(\R^{4,1})$ and $\mathcal{D}$ is simply the natural  Dirac operator 
on $\Sigma$ acting on spinors with values in the pullback to $\Sigma$ of $S(Y)$, the spinors on $Y$.
 If we expand $\Theta$ as in (\ref{fermf}), then
the components all obey the most obvious equations:
\begin{align}
\bar\partial\theta &= 0\cr
\partial(\tilde\theta_{\bar{z}}\d\bar{z}) &=0\cr
\bar\partial(\theta_{z}\d z) &=0\cr
\partial\tilde\theta &=0.
\end{align}
Because $\Sigma$ is compact, these equations  imply that $\theta$ and $\tilde\theta$ are constant along $\Sigma$, while $\theta_{z}dz$ and $\tilde\theta_{\bar{z}}d\bar{z}$ are holomorphic $(1,0)$ and antiholomorphic $(0,1)$-forms on $\Sigma$ respectively.   The $\kappa$ symmetry gauge condition (\ref{kap_cons}) implies
that all these modes have left-handed 4d chirality, that is, they transform as $(1/2,0)$ under the 4d rotation group $SU(2)_\ell\times SU(2)_r$.    Thus
$\theta$ and $\tilde\theta$ have 2 zero-modes each, and if $\Sigma$ has genus $g$, then $\theta_z$ and $\tilde\theta_{\bar z}$ each have $2g$ zero-modes.

To match the notation that we used in section \ref{schwpart}, we write the constant $(1/2,0)$ modes of  $\theta$ and $\tilde\theta$  as $\theta^A=\frac{1}{2}\psi_1^A$ and $\t\theta^A=\frac{1}{2}\psi_2^A$, respectively, where $A=1,2$ is a left-handed spinor index.  The fields $\psi_1^A$ and $\psi_2^A$ together make up the field
that in section \ref{schwpart} was called $\psi_i^A$, $i,A=1,2$.  Introduce a basis of holomorphic $(1,0)$-forms $\lambda_\sigma,\ \sigma=1\dots\g$ and a complex conjugate basis of antiholomorphic $(0,1)$-forms $\bar\lambda_\sigma,\ \sigma=1\dots\g$, such that:
\begin{equation}
i\int_\Sigma \lambda_\sigma\wedge\bar\lambda_\kappa = \delta_{\sigma\kappa}.
\end{equation}  We expand the $(1/2,0)$ parts of $\theta_z$ and $\t\theta_{\bar z}$ in this basis:
\begin{align}
\theta_z^A\d z &= \frac{1}{2}\sum_{\sigma=1}^\sg\rho^A_\sigma \lambda_\sigma,\cr
\tilde\theta^A_{\bar z}\d\bar{z} &= \frac{1}{2}\sum_{\sigma=1}^\sg\tilde\rho^A_\sigma \bar\lambda_\sigma.\cr
\end{align}
The fields $\rho^A_\sigma$ and $\t\rho^A_\sigma$ were introduced in section \ref{expform}.

If $\Sigma$ were not isolated and $\chi$, $\tilde\chi$ had some zero modes, then the $\kappa$ gauge-fixing condition (\ref{kap_cons}) would force them
to be of positive chirality in the 4d sense;  that is, they would satisfy $-i\Gamma_0\chi=+\chi$ and would transform as $(0,1/2)$ under the 4d rotation group.
The possible role of such modes was discussed in section \ref{expform}.

\subsection{Superparticle Action}

We now give a slow $t$-dependence to the fermionic zero-modes $\psi_i^A$ and $\rho^A_\sigma$, $\tilde\rho_\sigma^A$ and turn on the background gauge fields $V_M^I=h^I V_M$. The mass and charges of the wrapped M2-brane
\begin{align}
M &= \int_\Sigma \omega = \int_\Sigma \d^2z 2\mathcal{G}_{z\bar{z}}\cr
q_I &= \int_\Sigma \omega_I
\end{align}
are related by the usual formula\footnote{One might recall from eqn. (\ref{refo}) that in general $\omega=v^I\omega_I$, and so the mass of the BPS particle in M-theory units is $\sum_I q_Iv^I$. The 5d Einstein frame metric is related to the 11d metric by rescaling by a certain power of the Calabi-Yau volume, and the BPS mass in 5d Einstein frame is instead $M=q_I h^I$. For simplicity, in this appendix, one can just assume that the volume of Calabi-Yau is $1$ from the beginning, so the rescaling is unnecessary. Then $\omega=h^I\omega_I$, where $\CC_{IJK}h^Ih^Jh^K=1$.} $M=q_I h^I$. 

Starting from eqn. (\ref{act1}), it is not hard to write the action for an arbitrary spacetime with small and slowly varying curvature and for an arbitrary worldline $\gamma$ that has everywhere
a large radus of curvature. For an arbitrary worldline, the $\kappa$ gauge-fixing conditions look as follows:
\begin{align}
\label{ksgauge}
\frac{\dot{x}^M\Gamma_M}{i\sqrt{-\dot{x}^2}}\psi_i &= -\psi_i,\cr
\frac{\dot{x}^M\Gamma_M}{i\sqrt{-\dot{x}^2}}\rho_\sigma &= -\rho_\sigma,\cr
\frac{\dot{x}^M\Gamma_M}{i\sqrt{-\dot{x}^2}}\tilde\rho_\sigma &= -\tilde\rho_\sigma.
\end{align}
These conditions state that the fermions $\psi_i$, $\rho_\sigma$, and $\t\rho_\sigma$ all transform as $(1/2,0)$ under rotations of the normal plane to the worldline.  

The action takes the form:
\begin{align}
S=\int \d t \Big[ -M\sqrt{-\dot{x}^2} + q_I V^I_M\dot{x}^M + \frac{i}{2}M\varepsilon^{AB}\varepsilon^{ij}\psi_{Ai}\nabla_t\psi_{Bj} - \frac{i}{16}M\sqrt{-\dot{x}^2}\TT^-_{AB}\varepsilon^{ij}\psi_i^A\psi_j^B\cr + \sum_{\sigma=1}^\sg \left( i\varepsilon^{AB}\tilde\rho_{A\sigma}\nabla_t\rho_{B\sigma} + \frac{3i}{8}\sqrt{-\dot{x}^2}\TT^-_{AB}\tilde\rho_\sigma^A\rho_\sigma^B\right)\Big].
\end{align}
Here $\nabla_t$ is the pull-back to the particle world-line $\gamma$ of the Levi-Civita connection of $M$, projected onto the plane normal to the worldline.  And as usual,  $\TT^-_{AB}=\TT^-_{\mu\nu}\gamma^{\mu\nu}_{AB}$ is the anti-selfdual part of $\TT$ in the normal plane or equivalently in the local rest frame of the particle.   One interesting thing about this action is that the kappa-symmetry gauge (\ref{ksgauge}) ensures that only the projections of $\TT^-$ and of the Levi-Civita connection $\omega_M^{\bA\bB}$ to the plane normal to $\gamma$ enter this action, while the components along $\gamma$ drop out automatically. If we were writing corresponding equations of motion, we would have to impose this by hand.

To get the particle action used in section 3, one has to specialize this action to the graviphoton background and assume that the particle is almost at rest, i.e. do a non-relativistic expansion. In the graviphoton background, the spin-connection contribution  cancels the magnetic moment coupling of $\psi_i$ and  modifies it for $\rho$
and $\t\rho$. In the end, we get just the following familiar result (where we did not perform the non-relativistic expansion for the bosonic kinetic energy):
\begin{align}
S=\int \d t \Big[ -M\sqrt{-\dot{x}^2} + q_I V^I_M\dot{x}^M + \frac{i}{2}M\varepsilon^{AB}\varepsilon^{ij}\psi_{Ai} \frac{\d}{\d t}\psi_{Bj} + \sum_{\sigma=1}^\sg\left(i\varepsilon^{AB}\tilde\rho_{A\sigma}\frac{\d}{\d t}\rho_{B\sigma} + \frac{i}{2}\TT^-_{AB}\tilde\rho_\sigma^A\rho_\sigma^B \right)\Big].
\end{align}
All  fermions transform as $(1/2,0)$ under $SU(2)_\ell\times SU(2)_r$.

 \section{Do Short-Range Interactions  Affect the OV Formula?}\label{supercon}
 
In this appendix, we address a point that arose in section \ref{morch}.  In the derivation of the OV formula, one ignores short-range
interactions between superparticles in deriving the formula (\ref{meloxy}) that governs multiple winding.  (If there are long range
statistical interactions between BPS particles, one cannot ignore them and the OV formula will be modified.)  It is not immediately
obvious why this step is valid, since in general in nonrelativistic quantum mechanics in $D=2$ space dimensions, 
a short-range interaction may not be  irrelevant in the infrared.   Here we address this issue using
detailed properties of the relevant supersymmetric quantum mechanics.  But the answer we will get is not quite as simple as one might hope.

A preliminary point is that only two-body interactions have to be considered.  
In nonrelativistic quantum mechanics in $D$ space dimensions, a two-body short-range interaction $\delta^D(x-x')$ between particles with
positions $x$ and $x'$ has dimension
$E^{D/2}$ where $E$ is energy (we recall that as in section \ref{winding}, in nonrelativistic quantum mechanics, the position
coordinates have dimension $E^{-1/2}$).  Since the Hamiltonian $\hH$ has dimension $E$,  a contribution to $\hH$ proportional
to $\delta^D(x-y)$ is irrelevant only above $D=2$.    But a short-range three-body interaction between particles $x,x',x''$ 
would be something like
$\delta^D(x-x')\delta^D(x'-x'')$, with dimension $E^D$, and is irrelevant for $D>1$.  Interactions among more particles are only less
relevant.  So we only have to consider two-body interactions.

A second preliminary remark is that concretely, the meaning of the statement that a short-range two-body interaction is 
not irrelevant in $D=2$ is that scattering amplitudes may not vanish in the zero momentum
limit.  To show that in a specific problem, the interaction actually is irrelevant in the IR, one must show that the scattering becomes trivial
at zero momentum.  Of course, in a system with internal spin degrees of freedom, one must show that this is the case for every spin state.
We will do this by explicitly solving the Schrodinger equation at zero momentum.   

A final preliminary is that the  analysis is done in the absence of a graviphoton
field, though the application involves the graviphoton.  The derivation of the OV formula
or likewise the GV formula involves turning on a small\footnote{It is natural and sufficient to  consider  $\TT$  small since
the interactions one is computing, such as (\ref{mell}), are each of finite order in $\TT$.   And in any case $\TT$ small is the only region in which we have a hope to argue
that the interactions are unimportant. Finally, since the supersymmetric G\"odel solution is not real,
it is safer to consider $\TT$ small.} graviphoton field $\TT$.  
For small $\TT$, the computation is
dominated by contributions involving  distances of order $ 1/\sqrt {\TT}$.  So $\TT$ being small means that one has to know
how the unperturbed system would behave in the infrared.

\subsection{Basic Calculation}\label{basic}

A single BPS particle relevant to the OV problem is described by bosonic coordinates $x_1,x_2$ 
and  fermionic coordinates $\psi_+,$ $\psi_-$, with the free action
\begin{equation}\label{izz} I(x,\psi)=\int\d t \left(\frac{1}{2}\sum_{i=1,2}\left(\frac{\d x_i}{\d t}\right)^2
+i\psi_-\frac{\d}{\d t}\psi_+\right). \end{equation}
This is the projection of (\ref{uft}) to the $\eR'$-invariant sector, with the constant $-M$ dropped and $M$ scaled out.
Two widely separated BPS particles would be described by the sum of two such actions, with bosonic coordinates $x_i,\psi_\pm$
for one particle and $x'_i$, $\psi'_\pm$ for the second.  As usual for a two-body system with translation symmetry, it is convenient
to introduce center of mass coordinates $w_i=(x_i+x'_i)/2$, $\xi_\pm =(\psi_\pm+\psi'_\pm)/2$ and also difference coordinates
$y_i=(x_i-x'_i)/2$, $\eta_\pm=(\psi_\pm-\psi'_\pm)/2$.  The center of mass system is unaffected by the interactions and is simply
governed by a free action $I(w_i,\xi_\pm)$.  The relative motion is described asymptotically for large $y$ by another free action 
$I(y_i,\eta_\pm)$, but
we assume that this free action is modified by short-range interactions.  All that we will assume about these interactions is that
they are supersymmetric and conserve angular momentum.   

It is convenient to combine the spatial coordinates $y_1, y_2$ to complex variables $z=y_1+iy_2$, $\bar z=y_1-iy_2$.
Under rotations of the plane in which the superparticle propagates, $z$ and $\bar z$ transform with spin 1 or $-1$. 
We take 
$\eta_\pm$  to have spin $\pm 1/2$ and normalize them to satisfy $\eta_+^2=\eta_-^2=0$, $\{\eta_+,\eta_-\}=1$.
 It is convenient
to work in a basis consisting of two supercharges $\bar Q$ and $ Q$ of definite spin, namely $1/2$ and $-1/2$. This is
a complex basis, with  $Q$ the adjoint
of $\bar Q$.   The supercharges of definite spin are
\begin{align}\label{melfo} \bar Q& = i\eta_-\frac{\partial}{\partial\bar z} \cr
                                                 Q& =i\eta_+ \frac{\partial}{\partial z}. \end{align}
The supersymmetry algebra is
\begin{equation}\label{omvo} Q^2=\bar Q^2=0,~~~\{Q,\bar Q\}=\hH=-\frac{\partial^2}{\partial \bar z\partial z}. \end{equation}
We can think of the Hilbert space $\widehat\H_0$ of this system as the space of $(0,q)$-forms on the complex $z$-plane for $q=0,1$.
In this interpretation, we view $i\eta_-$ as the operator of multiplication by the $(0,1)$-form $\d\bar z$ (and $-i\eta_+$ as the
dual operation of contracting with the vector field $\partial_{\bar z}$).  Then $\bar Q$ becomes the usual $\bar\partial $ operator
$\bar\partial=\d\bar z\partial_{\bar z}$, and $Q$ is its adjoint.   The Hamiltonian commutes with a fermion number operator that
assigns the value $q-1/2$ to a $(0,q)$-form, for $q=0,1$. Separate conservation of angular momentum and fermion number reflects
the $R$-symmetry of the OV problem (the spin and $R$-symmetry generators are called $J_{12}$ and $J_{34}$ in section \ref{morch}).

How might  $Q$ and $\bar Q$ (and therefore  $\hH=\{Q,\bar Q\}$) be modified by interactions?  The
corrections are supposed to conserve angular momentum and vanish at large distances.  To preserve the fact that $Q^2=\bar Q^2=0$,
we do not add corrections to $\bar Q$ that are proportional to $\eta_+$, or corrections to $Q$ that are proportional to $\eta_-$.
We are not interested in irrelevant interactions, so we add to $Q$ or $\bar Q$ differential operators of order at most 1.

It is not interesting to add to $\bar Q$ a term proportional to $\partial_z$.  Indeed a perturbation $\Delta \bar Q=i\eta_- h_{\bar z}^z\partial_z$
(for some function $h_{\bar z}^z$ that vanishes rapidly at infinity -- often called a Beltrami differential) would represent a deformation of the complex structure of the $z$-plane.
Any such deformation is trivial, and can be removed by redefining the variable that we have called $z$.  Since $Q$
is the adjoint of $\bar Q$, this would simultaneously remove $\partial_{\bar z}$ from $Q$.

So we only consider in 
$\bar Q$ a perturbation of degree at most  1 and with no contribution proportional to $\partial_z$.  The general form is\footnote{By conjugating
by a suitable unitary transformation, one can make $A$ and $B$ real.  $A$ can then be interpreted in terms of a Kahler metric on the
complex $z$-plane, and $B$ in terms of a hermitian metric on a trivial complex line bundle over the $z$-plane.  We will not
make use of those facts.}
\begin{equation}\label{elop}\bar Q=ie^{A}\eta_-\left(\frac{\partial}{\partial \bar z}+B\right),\end{equation}
with functions $A$ and $B$ that vanish rapidly at infinity.  To conserve angular momentum, $A$ is a function only of $\bar z z$, and 
 $B(z,\bar z)=z C(\bar z z)$.  So                                    
\begin{equation}\label{zelop}\bar Q=ie^{A(\bar z z)}\eta_-\left(\frac{\partial}{\partial \bar z}+z C(\bar z z)\right).\end{equation}
$Q$ is the adjoint of this:
\begin{equation}\label{welop}Q=i\eta_+\left(\frac{\partial}{\partial z}-\bar z \bar C(\bar z z)\right)e^{\bar A(\bar z z)}.\end{equation}
The Hamiltonian is
\begin{equation}\label{belop}\hH=\{Q,\bar Q\}. \end{equation}

In a moment, we will have to solve the equation
\begin{equation}\label{yelop}\frac{\partial K}{\partial \bar z}=z C(\bar z z). \end{equation}
Taking $K$ to be a function of $\bar z z$ only, we see that the equation is $K'=C$ so that if $C$ vanishes rapidly at infinity,
we can take 
\begin{equation}\label{pozelo}K(\bar z z)=-\int_{\bar z z}^\infty \d u \, C(u). \end{equation}
This gives a solution $K(\bar z z)$ that is rotation-invariant and vanishes rapidly at infinity.  $K(\bar z z)$ does not necessarily vanish at $z=0$, but is regular there if $C$ is.

Now we have to solve the Schrodinger equation.
Even in the presence of the perturbations, the 
Hamiltonian commutes with a fermion number operator (or $R$-symmetry generator) $F=-\frac{1}{2}[\eta_+,\eta_-]$.  We write $|\neg\downarrow\rangle$
for a state annihilated by $\eta_+$ and $|\neg\uparrow\rangle$ for a state annihilated by $\eta_-$.  Such states have $F=-1/2$
and $F=1/2$, respectively.  (We have arranged this definition so that a $(0,q)$-form for $q=0,1$ has $F=q-1/2$.)
To solve the Schrodinger equation, we consider separately the two cases of states of $F=-1/2$ or $F=1/2$.

On a state $\Psi=\psi(z,\bar z)|\neg\downarrow\rangle$ of $F=-1/2$, $\hH$ reduces to $Q\bar Q$.  So we can find a zero energy
solution by requiring that $\bar Q\Psi=0$.  This gives
\begin{equation}\label{yurt}\Psi=\exp(-K(\bar z z))|\neg\downarrow\rangle. \end{equation}
Similarly, on a state $\Upsilon=\chi(z,\bar z)|\neg\uparrow\rangle$ of $F=+1/2$, $\hH$ reduces to $\bar Q Q$.  So we can find a zero energy solution
by requiring that $Q\Upsilon=0$.  This gives
\begin{equation}\label{zurt}\Upsilon=\exp(-\bar A(\bar z z)+\bar K(\bar z z))|\neg\uparrow\rangle.\end{equation}

Each of these solutions rapidly approaches the free particle solution $1\cdot |\neg\downarrow\rangle$ or $1\cdot |\neg\uparrow\rangle$
at infinity.  Thus there is no scattering in the zero-momentum limit and the interactions are irrelevant in the infrared.  

\subsection{A Critical Look}\label{critical}

Let us now take a more careful look at some of the above claims.  Suppose that instead of assuming that the function $C(\bar z z)$ vanishes rapidly at infinity,
we assume that $C(\bar z z )\sim c/\bar z z$, with a constant $c$. We set $A=0$, since a generalization of the discussion to include $A$ adds
little.    Then we have, at large $z$,
\begin{align}\label{werg} \bar Q&\sim i\eta_-\left(\frac{\partial}{\partial\bar z}+\frac{c}{\bar z}\right)\cr
                                            Q & \sim i\eta_+\left(\frac{\partial}{\partial z}-\frac{\bar c}{z}\right). \end{align}
We may be tempted to reject the possibility that $c\not=0$ on the grounds that $\bar Q$ and $Q$ do not approach the free model as quickly as we expect
for the BPS states relevant to the OV formula.  However, let us assume that $c$ is real.  We can try to conjugate $\bar Q$ and $Q$ to the free model
by a unitary transformation:
\begin{align}\label{onerg}\bar Q& \sim \left(\frac{z}{\bar z}\right)^c   i\eta_-\frac{\partial}{\partial\bar z}     \left(\frac{z}{\bar z}\right)^{-c} \cr
                                                        Q& \sim \left(\frac{z}{\bar z}\right)^{c}   i\eta_+\frac{\partial}{\partial z}     \left(\frac{z}{\bar z}\right)^{-c}. \end{align}
For generic $c$, this is not a well-defined unitary transformation, even in the region of large $z$, because the function $(z/\bar z)^c$ is not single-valued even
at large $z$.
But this multi-valuedness
 actually shows the physical meaning of the perturbation to $c\not=0$: the particles become anyons, which are free particles near $z=\infty$ except
that the wave-function acquires a $c$-dependent phase when one of them loops around the other.

For BPS particles that obey anyonic statistics, we expect the OV formula to require modification.  But what happens for $c\in \Z/2$?  In this case,
the function $(z/\bar z)^c$ is single-valued (away from $z=0$) so the model with $c\in\Z/2$ actually is unitarily equivalent to the free model near $z=\infty$.
Even so, we do not expect the OV formula to take its usual form in such a situation.           With $C(\bar z z)\sim c/\bar z z$, the integral (\ref{pozelo}) that was used
to define the function $K(\bar z z)$ diverges.    If one tries to modify the definition of the zero energy states $\Psi$ and $\Upsilon$ that 
approach $1\cdot |\downarrow\rangle$
and $1 \cdot |\uparrow\rangle$ at infinity, one finds that  this is possible for one of these two states but not the other (which one depends on the sign of $c$).   
Under these
conditions, we again expect the OV formula to require modification: for one of the two spin states, a pair of BPS particles at low energies
cannot behave as free particles.      

Thus, to ensure the validity of the OV formula, it is not sufficient to assume that the BPS particles have ordinary statistics.    The best that we can say
is that there is a universality class of non-anyonic models within which the interactions are irrelevant at low energies and thus the OV formula should hold.  

An interesting point is that invariance under a reflection $z\leftrightarrow \bar z$, which we did not assume in the above derivation, would force $c$ to
be imaginary and thus force us (if the model rapidly becomes equivalent to  the free model at infinity) to be in the universality class in which the OV fomula holds.       We observed in section \ref{morch} that the
basic example studied in \cite{OV} is in a universality class that can admit such a reflection symmetry.    So in that
universality class, it is reasonable to expect short-range interactions to be irrelevant at low energies. 
                                                
\vskip 1cm\noindent {\it Acknowledgments}
Research of EW was supported in part by NSF Grant PHY-1314311.   We thank N. Berkovits, J. Bryan, E. Diaconescu, D. Kaplan, J. Maldacena, R. Meyer, V. Mikhaylov, R. Pandharipande, N. Seiberg,
H. Ooguri,  V. Pestun, C. Vafa,  J. Walcher,  I. Yaakov, and M. Yamazaki for discussions.
\bibliographystyle{unsrt}

\end{document}